 \documentclass[11pt,a4paper]{article}
 \usepackage{amsmath}
\usepackage{cite}
\usepackage{graphicx}
\usepackage{amsfonts}
\usepackage{amssymb}
\input epsf
\begin{document}
\vphantom{0}
\vskip1.1truein
\renewcommand{\theequation}{\thesection.\arabic{equation}}
\newcommand{\eqal}[2]{\begin{eqnarray} #2 \end{eqnarray}}
\newcommand{\eqarr}[2]{\begin{array} #2 \end{array}}
\newcommand{\eqn}[2]{\begin{equation} #2 \end{equation}}
\newcommand{\newsec}[1]{\setcounter{equation}{0} \section{#1}}
\newcommand{\no}{\nonumber}
\newcommand{\tr}{{\rm tr~}}
\newcommand{\Tr}{{\rm Tr~}}

\def\yboxit#1#2{\vbox{\hrule height #1 \hbox{\vrule width #1
\vbox{#2}\vrule width #1 }\hrule height #1 }}
\def\fillbox#1{\hbox to #1{\vbox to #1{\vfil}\hfil}}
\newcommand{\ybox}{{\lower 1.3pt
\yboxit{0.4pt}{\fillbox{8pt}}\hskip-0.2pt}}
\makeatletter
\ifcase\@ptsize
 \font\tenmsy=msbm10
 \font\sevenmsy=msbm7
 \font\fivemsy=msbm5
\or
 \font\tenmsy=msbm10 scaled \magstephalf
 \font\sevenmsy=msbm8
 \font\fivemsy=msbm6
\or
 \font\tenmsy=msbm10 scaled \magstep1
 \font\sevenmsy=msbm8
 \font\fivemsy=msbm6
\fi
\newfam\msyfam
\textfont\msyfam=\tenmsy  \scriptfont\msyfam=\sevenmsy
  \scriptscriptfont\msyfam=\fivemsy
\def\Bbb{\ifmmode\let\next\Bbb@\else
 \def\next{\errmessage{Use \string\Bbb\space only in math 
mode}}\fi\next}
\def\Bbb@#1{{\Bbb@@{#1}}}
\def\Bbb@@#1{\fam\msyfam#1}
\newcommand{\IZ}{{\Bbb{Z}}}
\makeatother

\def\Sp{
\mathbb{S} }

\def\al{
\alpha }

\def\Ga{
\Gamma_{b} }

\def\up{\Upsilon_{b}}

\def\del{\delta_{x}}
\def\de{\Delta}
\def\P{\mathcal{P}_{\al}}
\def\p#1{\mathcal{P}_{\al _{#1}}}
\def\uq{\mathcal{U}_q(\mathfrak{sl}(2,\mathbb{R}))}
\def\uqtilde{\mathcal{U}_{\tilde{q}}(\mathfrak{sl}(2,\mathbb{R}))}
\def\fus#1#2#3#4#5#6{
C_{#5#6}\left[
\begin{array}
[c]{cc}%
#3 & #2\\
#4 & #1%
\end{array}
\right]}
  
\def\Fus#1#2#3#4#5#6{
F_{#5#6}\left[
\begin{array}
[c]{cc}%
#3 & #2\\
#4 & #1%
\end{array}
\right]}

\def\d{{\rm d}}
\def\Re{{\rm Re\,}}
\def\Im{{\rm Im\,}}
\def\inv{^{\raise.15ex\hbox{${\scriptscriptstyle -}$}\kern-.05em 1}}
\def\uq{\textstyle{1\over 4}}
\def\hf{{\textstyle{1\over 2}}}
 
\def\d{\partial}
\def\dbar{{\overline\partial}}
\def\inv{^{-1}}
\def\Tr{{\rm Tr}}
\def\cO{{\cal O}}
\def\cC{{\cal C}}
\def\cF{{\cal F}}
\def\cN{{\cal N}}
\def\cW{{\cal W}}
\def\cG{{\cal G}}

\def\<{\langle\,}
\def\>{\,\rangle}
 \def\({\left( }
\def\){\right) }
 \def\[{\left[ }
\def\]{\right] }

\font\zfont = cmss10 
\font\litfont = cmr6
\font\fvfont=cmr5
\def\bigone{\hbox{1\kern -.23em {\rm l}}}  
\def\ZZ{\hbox{\zfont Z\kern-.4emZ}}
\def\hf{{\litfont {1 \over 2}}}
\def\mx#1{m_{\hbox{\fvfont #1}}}
\def\gx#1{g_{\hbox{\fvfont #1}}}
\def\Re{{\rm Re ~}}
\def\Im{{\rm Im ~}}
\def\lfm#1{\medskip\noindent\item{#1}}
\def\p{\partial}
\def\a{\alpha}
\def\b{\beta}
\def\g{\gamma}
\def\d{\delta}
\def\e{\epsilon}
\def\th{\theta}
\def\vt{\vartheta}
\def\k{\kappa}
\def\l{\lambda}
\def\m{\mu}
\def\n{\nu}
\def\x{\xi}
\def\r{\rho}
\def\vr{\varrho}
\def\s{\sigma}
\def\t{\tau}
\def\z{\zeta }
\def\vp{\varphi}
\def\G{\Gamma}
\def\D{\Delta}
\def\T{\Theta}
\def\X{\Xi}
\def\P{\Pi}
\def\S{\Sigma}
\def\L{\Lambda}
\def\O{\Omega}
\def\oo{\hat \omega }
\def\ov{\over}
\def\o{\omega }
\def\bbox{{\sqcap \ \ \ \ \sqcup}}
\def\tria{$\triangleright $}
\def\dlr{\buildrel \leftrightarrow \over \partial }
\def\ddlr{\buildrel \leftrightarrow \over d }
\def\Gsl{{G }}
\def\Csl{{C }}
\def\partialsl{\partial \ \ \ /}
\def\sn{{\rm sn} }
 \def\cn{{\rm cn} }
 \def\dn{{\rm dn} }
 \def\z{\zeta }
 \def\uy{u_{\infty} }
 \def\hb{\hbar\ }
 \def\bh{{1\over \hbar}}
 \def\Im{{\rm Im}}
 \def\Re{{\rm Re} } 
 \def\CA {{\cal A}}
 \def\CB {{\cal B}}
 \def\CC {{\cal C}}
 \def\CD {{\cal D}}
 \def\CE {{\cal E}}
 \def\CF {{\cal F}}
 \def\CG {{\cal G}}
 \def\CH {{\cal H}}
 \def\CI {{\cal I}}
 \def\CJ {{\cal J}}
 \def\CK {{\cal K}}
 \def\CL {{\cal L}}
 \def\CM {{\cal M}}
 \def\CN {{\cal N}}
 \def\CO {{\cal O}}
 \def\CP {{\cal P}}
 \def\CQ {{\cal Q}}
 \def\CR {{\cal R}}
 \def\CS {{\cal S}}
 \def\CT {{\cal T}}
 \def\CU {{\cal U}}
 \def\CV {{\cal V}}
 \def\CW {{\cal W}}
 \def\CX {{\cal X}}
 \def\CY {{\cal Y}}
 \def\CZ {{\cal Z}}
 %

\chardef\tempcat=\the\catcode`\@ \catcode`\@=11
\def\cyracc{\def\u##1{\if \i##1\accent"24 i%
 \else \accent"24 ##1\fi }}
\newfam\cyrfam
\font\tencyr=wncyr10
\def\cyr{\fam\cyrfam\tencyr\cyracc}

 \def\R{\mathbb{R}}
 \font\cmss=cmss10 \font\cmsss=cmss10 at 7pt 
 \def\Z{\mathbb{Z}}
 \def\N{\mathbb{N}}

 \font\cmss=cmss10 \font\cmsss=cmss10 at 7pt 
 \def\p{\partial}
 \def\D{\slash\ \ \ \ D}
 \def\11{1\ \ 1}
 \def\Det{{\rm Det \,}}
 \def\llangl{\left\langle\ \left\langle }
 \def\rrangl{\right\rangle \ \right\rangle }
 \def\rangl{\right\rangle }
 \def\langl{\left\langle }
 \def\gst{\gamma _{\rm str}}

\def\gs{g_{_{\rm s}}}
\def\gym{ g_{_{\rm YM}}}
 \def\mb{ {\mu^{_{\ninepoint B}} } }
\def\mbt{\tilde \mu_{_{\ninepoint B}}}
 
 \def\Ga{\Gamma_b}


\def\cqg#1#2#3{{ Class. Quantum Grav.} {\bf #1} (#2) #3}
\def\np#1#2#3{{Nucl. Phys.} {\CB#1} (#2) #3}
\def\pl#1#2#3{{Phys. Lett. }{\CB#1} (#2) #3}
\def\prl#1#2#3{{Phys. Rev. Lett.}{\bf #1} (#2) #3}
\def\physrev#1#2#3{{Phys. Rev.} {\bf D#1} (#2) #3}
\def\ap#1#2#3{{Ann. Phys.} {\bf #1} (#2) #3}
\def\prep#1#2#3{{Phys. Rep.} {\bf #1} (#2) #3}
\def\rmp#1#2#3{{Rev. Mod. Phys. }{\bf #1} (#2) #3}
\def\rmatp#1#2#3{{Rev. Math. Phys. }{\bf #1} (#2) #3}
\def\cmp#1#2#3{{Comm. Math. Phys.} {\bf #1} (#2) #3}
\def\mpl#1#2#3{{Mod. Phys. Lett. }{\bf #1} (#2) #3}
\def\ijmp#1#2#3{{Int. J. Mod. Phys.} {\bf #1} (#2) #3}
\def\lmp#1#2#3{{Lett. Math. Phys.} {\bf #1} (#2) #3}
\def\tmatp#1#2#3{{Theor. Math. Phys.} {\bf #1} (#2) #3}
\def\hepth#1{{e-Print Archive:  hep-th/}#1}

\vskip -2cm

\rightline{SPhT-T03/102}
 
\rightline{hep-th/0307189}
 
\vskip 2cm

\centerline{\Large{\bf  Boundary  Liouville Theory and 
2D Quantum Gravity}}
 
\vskip 2cm
\centerline{Ivan K. 
Kostov\footnote{\texttt{kostov@spht.saclay.cea.fr}}\footnote{Associate 
member of {\cyr IYaIYaE -- BAN}, \ 
Sofia, Bulgaria }, 
B\'en\'edicte Ponsot\footnote{\texttt{ponsot@spht.saclay.cea.fr}},
 Didina Serban\footnote{\texttt{serban@spht.saclay.cea.fr}}}
\vskip 16pt
\centerline{\sl Service de Physique Th{\'e}orique,  CNRS -- URA 2306}
\centerline{\sl  C.E.A. - Saclay, }
\centerline{\sl  F-91191 Gif-sur-Yvette, France}
  
\vskip 2cm
 
\begin{abstract}
We study the boundary correlation functions in Liouville theory
 and in solvable  statistical  models of  2D quantum gravity. 
In Liouville theory we derive functional identities  for  all 
fundamental  boundary   structure constants, similar to the one
 obtained   for the boundary two-point function by Fateev,
Zamolodchikov and Zamolodchikov.   All these  functional identities
  can be written as  difference equations with respect to one of the
 boundary parameters.   Then we  switch to the  microscopic
realization of 2D quantum gravity  as a height model on a 
 dynamically triangulated disc and consider the boundary correlation
 functions of  electric, magnetic and  twist  operators.  By cutting
 open the sum over  surfaces along a domain wall, we derive
difference equations  identical to those obtained  in Liouville theory.  
 We conclude that there is a complete agreement  between  the
predictions of Liouville theory and the discrete approach. 
 \end{abstract}

  \newpage

\tableofcontents

\section{Introduction}

 \def\QG{2D QG}

  The two-dimensional quantum gravity (\QG) allows two complementary
  descriptions based on Liouville field theory \cite{PolyakovL} or on
  dynamical triangulations \cite{discreteqg}.  From the point of view of
  string theory, \QG\ is the theory of non-critical bosonic strings,
  which do not have a tachyon state and therefore admit an
  interpretation as statistical ensemble of embedded worldsheets. 
  (For  reviews see \cite{GDFZ, GM}.)  In this paper we will
  consider the  boundary problem in the two descriptions.
  We will present a systematic  approach to evaluate the 
  boundary correlation functions in Liouville and discrete \QG\ 
  based on a set of difference equations with respect to the boundary parameters. 
  Although derived by completely different methods, the
 difference equations  for the two- and three-point functions 
 obtained  in both approaches  become identical after rescaling with an appropriate factor.
 This factor  does not depend on the Liouville boundary parameters and 
 has a natural interpretation as  the contribution of the matter  fields.

   \subsection{The  continuum approach}
   
    The first success of the continuum  approach was to predict the 
    scaling exponents  of gravitationally dressed primary fields   \cite{kpz, DDK}.  
    Subsequently, the so-called resonance correlation functions 
    of vertex operators were found by doing the zero mode integration 
over the Liouville field and then using analytic continuation with
respect
 of the number of the Liouville interaction insertions \cite{GouLi, VDotsenko}. 
    These results showed  that the Liouville CFT, although
`noncompact', 
can be treated similarly to the standard  `compact' conformal field theories \cite{BPZ}.   
      Indeed,  in a series of impressive works  \cite{DO, ZZ,FZZb,
PTtwo, hosomichi},
 the Liouville CFT   has been practically solved.
          The exact expressions  obtained for  the 
        bulk \cite{DO, ZZ}   and boundary  \cite{FZZb, PTtwo,
hosomichi} 
 Liouville structure constants  
           were obtained as solutions of a set of identities 
    that follow from   the  OPE  of     vertex operators with 
    degenerate bulk or boundary Liouville fields. 

In Liouville theory with boundary the situation is more subtle because
of the presence,
 besides the cosmological constant $\mu$, 
 of a second dimensional parameter, the 
boundary cosmological constant  $\mu_B$.
The boundary Liouville structure constants thus depend on a new
dimensionless parameter,
 $\mu_B/\sqrt{\mu}$.  The observables are meromorphic functions 
of this parameter having a branch point singularity  at the point $- \mu_B^0$ where
 $\mu_B$ equals the boundary entropy density. The branch point can be resolved by 
 introducing a uniformization parameter  $\t$
 \eqn\deftau{
 \label{deftau}
 \mu_B = \mu_B^0 \cosh\t.}
In the  correlation function of $n$ boundary operators on the disc, one can define 
   $n$ independent boundary cosmological constants $\mu_{B_1},...,\mu_{Bn}$
   associated with the segments between two neighboring operators.  
 Fateev,  A. Zamolodchikov and Al. Zamolodchikov 
\cite{FZZb} found that   the boundary two-point function, or boundary
reflection  amplitude, satisfies  a  simple functional equation, which
can
 be given the form of a difference equation with respect to one of the boundary parameters, $\t_1$ or $\t_2$.
  One can show (see section 3 of this paper) that  this is a general property 
  of the    Liouville  boundary structure constants: all they 
  satisfy    recurrence  identities that have the form of difference
equations with respect  to one of the boundary parameters.
 In particular, using the pentagonal equation satisfied by the
boundary three point function \cite{PTtwo}, we find two difference  equations
for the latter,
one of them 
quite similar to the  equation obtained in \cite{FZZb}.  
 The shifts of the boundary parameters  come from the fusion  with the
lowest  
 degenerate boundary Liouville fields.

   The  two and three point functions in \QG\  are products of 
     the corresponding Liouville and matter structure constants, and therefore 
     have the same dependence on the boundary parameters.
    As a consequence,  they should satisfy   difference equations of the same form.
    
    In this paper we will concentrate in the simplest realization 
    of the matter field is as a gaussian field with background charge.
  Then the matter   three-point function  is trivial (no matter
screening charges)
 and the difference equation obtained in Liouville theory 
  should hold also for the three-point boundary correlator in 2D quantum gravity.

  In order to check this result experimentally, we need an adequate 
  lattice realization of the gaussian field  coupled to Liouville.
   This realization is given by the  SOS model  on a random surface
   and the related statistical models: the $O(n)$    \cite{Ion} and the 
   ADE     \cite{ADEold, Inonr, Idis}    models of quantum gravity. 
 In all these models the  matter fields are described as a  loop gas on the random surface.

            \subsection{The discrete approach}

 The    loop gas  realization of   \QG\  has been particularly 
   useful  in studying  problems involving two-dimensional geometrical
 critical phenomena, typically associated with non-unitary CFT (see,
 for example, the review \cite{duplan}).  
After being transposed to a random surface,  many of those problems
become 
easily solvable.  The   solutions  then can be used to  obtain information about the 
  same problems before coupling to gravity. 
   In particular, the scaling dimensions in the theory on the plane
are related to
 the gravitational dimensions  by KPZ scaling formula \cite{ kpz, DDK}.  The 
   KPZ rule was checked on various
 microscopic realizations of the 2D quantum gravity in terms of
 statistical models on random lattices and in the last years have been
 applied successfully to evaluate some difficult ``flat'' geometrical critical
 exponents associated with percolating clusters and polymers
 \cite{duplan}. 

  In the loop gas  model, the   matter fields  
   are represented by  non-intersecting
 loops (which can be also interpreted as domain walls) 
 on a randomly triangulated surface.  
 The corresponding CFT is non-unitary, with continuous spectrum 
of the central charge $-\infty<c\le 1$ \cite{nienhuis}.
 The matter correlation functions describe the critical behavior 
of networks of linear polymers \cite{dsds}\footnote{This geometrical 
description of  CFT was recently used in the analysis of stochastic 
 conformal maps \cite{bernbau}.}. 
 Bulk
 correlation functions of linear polymers have been first calculated in
 \cite{BKpol} using a factorization property of the measure over random
 surfaces.  The same method is readily  generalized to the case of
 boundary correlation functions.  More complicated examples of boundary
 correlators have been calculated in \cite{KKloop}, including the
 2-point function for the boundary twist operators intertwining between the
 free and fixed boundary conditions.

  The exact results  presented in  \cite{FZZb} 
   inspired the work \cite{Ibliou}, where it was
shown that the boundary two-point function
of star polymers   on a random disc 
satisfies, when appropriately normalized, the same   difference equation  
as the Liouville boundary reflection amplitude for
  degenerate fields of type $(r,1)$ or $(1,s)$. 
   It was also  noticed  that the  
   two-point function of the boundary twist operator calculated previously 
in\cite{KKloop}  agrees with  the general formula
presented in \cite{FZZb}  with  the corresponding  identification  of the 
twist operator  as a boundary  conformal field. 

 Here  we  apply the  approach  of \cite{Ibliou} to  derive difference equations for 
  the  boundary three-point function and the bulk-boundary correlation
 function.   Comparing the results with those obtained in Liouville
theory we   find  that  there is perfect  agreement  between the continuum and
the discrete approaches.

%
   
 The layout of this paper is as follows.  In sect.  2 we sketch the
 world sheet description of 2D QG as a theory of a gaussian matter
 field $\chi$ with background charge $e_{0}$, and a Liouville field
 $\phi$ with background charge $Q$ and exponential interaction.  We
 introduce the KPZ primary fields in the bulk and on the boundary. 
 Sect.  3 contains a review of the boundary Liouville problem and the recent results
 \cite{FZZb, PTtwo, hosomichi}.  There
 we summarize the finite-difference equations satisfied by the boundary
 structure constants.  In sect. 4 we interpret the difference equations obtained 
 in pure Liouville theory from the point of view of \QG\ with gausian matter field.
  In sect.  5 we  give the microscopic definition of the 2D QG as a non-restricted
 height model on the dynamically triangulated disc.  First we define
 the height model on a  given  triangulation and establish the
 equivalence with a gas of loops.  We give
 the loop-gas formulation of the Dirichlet and Neumann boundary
 conditions as well as  that of the electric, magnetic and twist operator of
 the  gaussian field.  In sect.  6 we derive the difference equations
 for the basic boundary correlation functions in 2D QG.  First we remind
 how the loop equation for the disc amplitude can be solved by being 
 transformed into finite-difference  equation.  Then, using
  the knowledge of the disc  amplitude, we derive  another type of functional
 identities using a factorization property of the functional measure
 of 2D QG.  These identities are most easily obtained for the correlation
  functions involving degenerate matter operators 
  with Dirichlet boundary condition. These operators are described 
geometrically as sources of open lines starting at the boundary.  
We also consider the case  where both Dirichlet and Neumann  
boundary conditions are present.  Here we derive a difference equation
for 
the correlator of two  intertwiners between Dirichlet and
 Neumann boundary conditions, which we call boundary twist operators,
and  a  vertex operator with arbitrary charge. 
  Our conclusions are presented in sect.  7.

 \section{The world-sheet description of 2D QG}

 \subsection{Liouville Quantum Gravity on a disc.}

  Let $\CM$ be the world-sheet manifold, which we take with the
  topology of a disc.  We denote by $\hat R$ and $\hat K $
  correspondingly the local gaussian curvature in the bulk and the
  geodesic curvature along the boundary\footnote{We normalize the two
  curvatures so that $ \int _{\CM} R ^{(2)}+2\int _{\p \CM} K = 4\pi. 
  $ }.  In the conformal gauge $g_{ab}(x)\sim \d_{ab}$ the
  fluctuations of the metric are described by a local field $\phi$
  (the Liouville field) with exponential interaction, which couples to
  the matter field through the conformal anomaly \cite{PolyakovL}. 
 The conformal gauge necessarily introduces the Faddeev-Popov
reparametrization
 ghosts with central charge $-26$.   We will assume that the matter
is realized
  as a scalar field $\chi$ with background charge $e_0$ and central
charge
 $c= 1- 6 e_0^2$.  Such a field is described by
  an effective action\footnote{Here we choose the boundary interaction
  that supports the quasiclassical description.  The other possible
  interaction is
$\tilde \mu_B \ e^{\phi/b}$.}%

\eqal\actg{
\label{actg}
\CA [\chi,\phi]&& = \int \limits_\CM
\( {1\over 4\pi}
 [ (\nabla \phi)^2 + (\nabla \chi)^2+ 
 (Q \phi -i e_0\chi ) \hat R 
 ]
+\mu 
e^{ 2b\phi }\)
 \no\\
&& +
\int\limits_{\p\CM}   \( \frac{1}{2\pi} ( Q\phi-i e_0\chi )\hat K
+ \mu_B \ e^{b\phi}\) + {\rm ghosts}}
where $\mu $ and $\mu_B$ are correspondingly the bulk and the boundary
cosmological constants and the background charges are expressed in
terms of the Liouville coupling constant $b$ as %
 \eqn\bgcharges{ 
 \label{bgcharges}
 Q= {1\over b}+b, \qquad e_0= {1\over b}-b .  } 
 With the choice
 (\ref{bgcharges}) the two background charges satisfy $Q^2 - e_0^2=4$,
 which is equivalent to the vanishing of the central charge $ c_{\rm
 tot}\equiv c_\phi +c_\chi+ c_{\rm ghosts}=(1+6 Q^2)+(1- 6e_0^2) -26
 =0.  $ 
 The bulk and boundary cosmological constants  are coupled 
 correspondingly to the area $A=\int d^2 z \ e^{2b\phi(z, \bar z)}$  
 and the boundary length $\ell =\int dx \ e^{b\phi(x)}$.

 After conformally mapping the disc $\CM$ to the upper half plane, the
 curvature term disappears
\eqn\actUHPQG{ \CA[\phi, \chi] =\int \limits_{\Im z\ge 0}d^2 z\(
{1\over 4\pi} [ (\nabla \phi)^2 + (\nabla \chi)^2 ] +\mu e^{ 2b\phi 
}\)
+\int\limits_{-\infty}^\infty dx \ \mu_B\ e^{b\phi} + {\rm ghosts} }
and the background charges are introduced through the asymptotics of
the fields at spatial infinity 
$$\phi(z, \bar z ) \sim - Q \log |z|^2
,\quad \chi (z, \bar z ) \sim - e_0\log |z|^2.
$$
The partition function on the disc is defined as the functional
integral 
over the fields $\chi$ and $\phi$ with action (\ref{actg})
\eqn\ptfdisc{
\label{ptfdisc}
\Phi(\mu, \mu_B)=
\int [d\chi][d\phi] \  e^{-\CA[\chi,\phi]}.
}
To define the  functional measure 
we have to fix the boundary conditions. In order to have correspondence 
with the sum over triangulated surfaces one should impose  non-homogeneous 
Neumann boundary 
condition for the Liouville field\footnote{Very recently, the
Dirichlet boundary condition for the Liouville field studied in \cite{ZZ-Pseudo}
was also found to have interpretation in terms of a sum over random
surfaces
 \cite{ZZ-D}.  It describes the non-perturbative effects related  to 
D-instantons and D-particles.
The  role  of the Dirichlet  Liouville boundary condition in  
two-dimensional string theory has been previously discussed in  \cite{PolyakovDB}.}
$$ i(\p-\bar \p) \phi  =4\pi \mu_B \ e^{b\phi(x)},$$
 while  the matter field $\chi$  can  satisfy  either   Dirichlet or Neumann
 boundary conditions.
   
\subsection{Bulk and boundary matter fields in the Coulomb gas 
description}

 \paragraph{ \bf  Bulk matter fields:}
 There is a continuum of holomorphic  bulk primary   fields 
 realized either as  electric  or as magnetic operators.
 The electric, or   {\it vertex}, operators  
 $$V_e(z, \bar z)=e^{2i e\chi(z, \bar z)}$$
 are
primaries with respect to the energy momentum tensor
\begin{eqnarray}
T(z)=-(\partial\chi)^{2}-ie_0\ \partial^{2}\chi  , \quad 
\bar{T}(\bar{z})&-(\bar{\partial}\chi)^{2}-ie_0\ \bar{\partial}^{2}\chi \; 
\nonumber
\end{eqnarray}
and have conformal weights
\eqn\Deltae{ \Delta_e= \bar
 \Delta_e= e(e- e_0) =(e- e_0/2)^2 - e_0^2/4 . } 
The   magnetic, or {\it vortex}, operators $\tilde V_{m}(z, \bar
 z)$ are associated with discontinuities of the field $\chi$. %
 The vortex operator with magnetic charge $m$ describes a
 discontinuity $\delta\chi= 4\pi m $ along a line starting at the
 point $(z, \bar z)$ and has left and right conformal dimensions  
\eqn\Deltam{ \Delta_m=m(m- e_0),\qquad \bar \Delta_m=
 m(m+ e_0).} 
 In presence of a background charge the purely magnetic
 operators have spin.  On the other hand, only the spinless fields
 survive in 2D QG, since the local rotations can be compensated by
 coordinate transformations.  Therefore we are led to consider the
 mixed electric-magnetic operator $\CO_{e, m}$, which have left and
 right conformal dimensions
 \eqn\Deltamn{ \Delta_{e,m}=(e+m)(e+m -e_0) \qquad \bar \Delta_{e,m}=
 (e-m)(e-m -e_0).} The condition that the spin $ \Delta -\bar \Delta $
 of such an operator vanishes fixes the electric charge to $e= e_0/2$
 or the magnetic charge to $m=0$.  Thus the spinless operators
 carrying magnetic charge are of the form $ \CO_{ e_0/2,m} $, with
 conformal dimensions
\eqn\DeltaS{
 \Delta_{\e_0/2, m}= \bar \Delta_{ e_0/2, m}= m^2-e_0^2/4. }

  An important set among the primaries are the fields 
  degenerate with respect to the conformal symmetry algebra
\cite{BPZ} with  scaling dimensions
\eqn\Deltars{\Delta_{rs} = {(r/b-sb)^2 - e_0^2\over 4} \qquad 
(r,s\in \N).}
These fields can be constructed either as vertex operators 
with charges 

\eqn\ers{e_{rs}= \hf(e_0 - {r / b} +{s} b)} or alternatively as
spinless operators with electric charge $e_0/2$ and magnetic charge
\eqn\mrs{m_{rs} = \hf (-{r/b+sb}).} The reflection symmetries $e \to
e_0-e$ and $m\to -m$ act as $\{r,s\}\to \{-r,-s\}$.  Nontrivial
correlation functions of these operators can be constructed by adding
a finite number of screening operators $V_{1/b}$ and $V_{-b}$ with
conformal dimension one.
  
 We will be interested only in the magnetic charges of the form $m= Lb/2, L\in \Z,$
 which have a nice geometrical meaning in the microscopic approach.
  This spectrum of magnetic operators  is compatible with the
assumption  that the gaussian field is compactified   at radius
$b$: $\chi \sim \chi + 2\pi b$.

\paragraph{\bf  Dirichlet and Neumann boundary conditions:}
For the matter field, we will consider Neumann ($N$) boundary condition
\eqn\neumann{
\label{neumann}
(\partial -\bar{\partial})\chi =0\qquad (\rm Neumann)}
 or Dirichlet ($D$) boundary 
condition 
\eqn\dirichlet{
\label{dirichlet}
(\partial +\bar{\partial})\chi =0\qquad (\rm Dirichlet).}
These are the only boundary conditions that 
can be imposed on a gaussian field. 

In order to define completely the D and N boundary conditions, 
we have to fix the global modes.
 In the case of  Dirichlet boundary condition, eqn. \ref{dirichlet}
 means that the field has  constant value of $\chi(x)=\chi_B$ along the boundary.  
 Thus there is a continuum of Dirichlet boundary conditions labeled by the
 value of $\chi_B$.
 Similarly, in the case of Neumann boundary condition we should specify the increment of 
 $\chi$ along the boundary:  $\delta\chi =
 \int \chi(x) dx$.
      The Neumann boundary condition is related to the Dirichlet one by duality
      transformation
      \eqn\chiduality{
      \label{chiduality}
       b\leftrightarrow 1/b, \qquad 
      \chi\equiv  \chi(z)+\bar\chi(\bar z)\quad \leftrightarrow \quad
      \tilde \chi \equiv \chi(z)-\bar\chi(\bar z).
      }
%

\paragraph{\bf  Boundary matter  fields:}
 Boundary fields interpolate between different boundary conditions.
 To completely characterize a   boundary operator, one should specify the
  boundary conditions on the left and on the right \cite{cardy}.    
  The  boundary primary fields  can be constructed  out of vertex and 
vortex  operators.  
There are three  types of 
boundary operators for
the gaussian field, $N/N, D/D$ and $N/D$.
 
 \smallskip
 
 \noindent 
 $\bullet$ \   \ \ The  {\it N/N } boundary fields are the boundary 
vertex operators
 $$ B_e(x)= e^{ie\chi(x)},
 $$
 where $x\in \R$ is the coordinate on the boundary.  The boundary
 scaling dimension of this field is given by the same formula as for
 the corresponding bulk operator:
\eqn\DeltaBe{
\label{DeltaBe}
\Delta^B_e = e (e - e_0)= (e-e_0/2)^2 - e_0^2/4.
 }
 
  \smallskip
  
 \noindent $\bullet$ \ \ \ The $D/D$ boundary fields are boundary
 magnetic operators $\tilde B_m$, which introduce a discontinuity
 $\delta \chi = 2\pi m$ of the boundary value of the field.  
 The magnetic operator changes the  boundary value of the
  matter field from $\chi_B$ to $\chi_B+2\pi m n$.
 The
 dimension of such operator is 
 \eqn\DeltaBm{
 \label{DeltaBm}
 \Delta^B_m = m (m - e_0)= (m-e_0/2)^2
 - e_0^2/4.
 }
 Unlike the spinless magnetic operator in the bulk, the boundary
 magnetic operator does not carry electric charge.  The degenerate
 electric and magnetic boundary operators have the spectrum of charges
\eqn\emrsB{e_{rs}= \hf(e_0 - {r / b} +{s} b),\qquad
  m_{rs} =  \hf (e_0 -r/b+sb).
  } 
   We will be concerned only by the magnetic charges of the form $m=
L/2$
 with $L$ integer, which can be  realized microscopically as $L$
domain lines ending at the boundary (see sect. 5).  
   
    \smallskip
    
  \noindent $\bullet$ \ \ \ Finally, there are the {\it D/N} and {\it N/D} boundary
 operators.  
 We call the operators  of lowest dimension intertwining
 between Dirichlet and Neumann boundary conditions  {\it boundary twist
 operators}.
 For $b=1$ ($c=1$) the boundary twist operator is well known \cite{Hashimoto, AOS}
  and its  dimension is  $\Delta_T=1/16$.  The boundary twist operator
for the
 Coulomb gas with $b<1$  will be introduced in sect. 5.4.   Its dimension is that of 
 a ``half-degenerate" field with $\{r,s\}=\{\hf, 0\}$.

\subsection{Gravitationally dressed bulk and boundary fields}

 \paragraph{\bf  Bulk  operators:}

     We will restrict  ourselves to the vertex operators since the 
magnetic operators are vertex operators for the dual field $\tilde 
\chi$.
     In 2D QG the vertex operators of the matter field are dressed by 
Liouville 
 vertex operators
\eqn\grve{V_{e,\a }(z,\bar z) =
e^{2i e \chi (z,\bar z)} \ e^ {2\a
\phi(z,\bar z) }}
 in such a way that the composite operator is a density, {\it i.e.} 
its conformal dimensions are $\Delta=\bar \Delta =1$ \cite{DDK}
 \eqn\dimbal{ e(e- e_0) + \a(Q-\a)=1.
 \label{dimbal}} 
  In the language of the 2D string theory, the shifted matter and 
Liouville charges
 \eqn\PandE{
 P=e_0-2e, \qquad E=i(Q-2\a)
 }
 are the components of the momentum in the two-dimensional Euclidean 
  target space \\ $(X^0, X^1)= (\phi, \chi)$.
  The  balance  of dimensions (\ref{dimbal}) then  takes the form 
of the mass-shell condition\footnote{Here we assume that the Liouville
field describes the time direction of the target space of the
Euclidean 2D string theory.}
\eqn\masshellPE{
\label{masshellPE}
 P^2 +E^2 = 0.} 
  The states with real Liouville energy $E$   appear as normalizable 
intermediate states,  while  
the non-normalizable on-shell fields 
(\ref{defVP})  describe massless particles in the target space, 
the closed string  `tachyons', which correspond to 
  local operators on the world sheet.
  The two solutions  $E=\mp i P$ of the 
  mass shell  condition (\ref{masshellPE})   describe closed string tachyons 
  with left and right chirality\footnote{Since we will be  eventually  interested 
  in the comparison with the discretized theory, we choose a  normalization
  which    eliminates  the ``leg factors"  in the correlation functions \cite{GDFZ, GM}.  Below we will
normalize  the boundary fields following the same principle.}
  \eqn\defVP{
   \label{defVP}
  \CV_P^{(\pm)}=\mp {1\over \pi}
 {\G(\pm b^{\pm 1} P)\over
 \G(1\mp  b^{\pm 1} P)}\ 
 V _{e, \a_\pm} , \qquad e=\hf (e_0- P), \ \
 \  \a_\pm = \hf (Q  \mp  P). 
  }
  In this way, there are   two possible gravitational dressings
   of the matter field vertex operator with  charge $e$. 
  For  instance,  for the identity operator $ e=0$ we have two possible 
Liouville
 dressings: the Liouville interaction $\a=b$ and dual one $\a=1/b$. 
   The solution 
 ${\CV }_P$
 with the smallest 
Liouville exponent describes 
 physical a physical   state,  
 \eqn\KPZ{
 \label{KPZ}
  {\CV}_P  
   =   
 \begin{cases}
 \CV^{(+)}_P & \qquad  ( P>0)\cr
 \CV^{(-)}_{P} & \qquad  (P<0)
 \end{cases}
}
 while the larger exponent is usually
 referred as `wrongly dressed' state,
  \eqn\WRDS{
  \label{WRDS}
    {\check  \CV }_P= 
 \begin{cases}
 \CV^{(-)}_P &  \qquad ( P>0)\cr
 \CV^{(+)}_{-P} &  \qquad (P<0).
 \end{cases}
 }
   The condition $\a<Q/2$ 
for the physical fields is also known as  Seiberg bound \cite{seiberg}. 
The wrongly dressed states are not realized in Euclidean 2D QG, but
they will be important when comparing the microscopic quantum gravity
with Liouville theory.  The correlation functions of the operators
${\bar \CV }_{P}$ are obtained from the correlation functions of the
`physical` operators ${\CV }_{P}$ by analytic continuation with
respect to $P$ beyond the point $P=0$.

The gravitational dressing of the identity operator ($e=0$ or $P=e_0$)
 is   the Liouville interaction ${\CV }_{e_0}= e^{2 b\phi}$;
this implies that we have chosen $b\le1$ in the action
 \ref{actg}.

 \paragraph{ \bf Boundary operators:}

  Similarly to the bulk fields, the boundary fields in  \QG\  are constructed as 
  matter and  Liouville vertex operators 
    \eqn\Bbound{{B}_{e,\b}(x) =e^{ ie \chi(x) +\b\phi (x)}.} 
 In the string-theoretical context,   the vertex operators with complex momenta
\eqn\parambetae{ e= \hf {e_0 } -P , \ \ \ \b = \hf {Q}-iE} 
appear as normalizable  intermediate states of the open string 
amplitudes while the on-shell fields
 \eqn\onshellB{
\CB  _P^{(\pm)}
   = {\G(\pm 2b ^{\pm 1}P)\over\pi} \ 
  B_{e, \b_\pm},\qquad e={e_0\over
 2} -P, \ \ \ \b_\pm = {Q\over 2}\mp  P.
 \label{onshellB}
 } 
are the open string tachyons.
The  physical boundary operators  
 \eqn\BKPZ{
 \label{BKPZ}
  {\CB}_P  
   =   
 \begin{cases}
 \CB^{(+)}_P & \quad ( P>0)\cr
 \CB^{(-)}_{P} &\quad  (P<0)
 \end{cases}
 }
 are related to the  fields with  the  `wrong' dressing 
 \eqn\wrongTB{
 \label{wrongTB}
  {\check \CB }_P= 
 \begin{cases}
 \CB^{(-)}_P & \quad ( P>0)\cr
 \CB^{(+)}_{P} & \quad (P<0)
 \end{cases}
 }
by analytic continuation through the point $P=0$.

 We  will denote the  degenerate matter fields dressed by gravity as
\eqn\degene{
\label{degene}
\CB_{rs} = \CB_{P_{rs}}, \ \ \ P_{rs} = r {1\over 2b} - s {b\over 2}.
}
As we will see later, the  degenerate fields $\CB_{L+1,1}$
 and $\CB_{1,L+1}$   play  special role in the microscopic approach.

 In order to define completely a boundary operator,  we have to specify  the
 left and right boundary conditions both for the matter and Liouville
 fields.  The Liouville boundary conditions are  labeled  by the
 values of the cosmological constant on both sides.

Let us remark that  here both dressings can  
have physical meaning.  In particular,
 the boundary Liouville interaction can be $e^{b\phi}$ as well as
 $e^{\phi/b}$, the two exponents corresponding to the two possible 
dressings of the identity operator. 
   In the Liouville literature only the first case (which
 was chosen  as the boundary term in  the action \ref{actg}) has been
 considered.  In this case the dimension of the boundary is one, {\it
 i.e.} half the dimension of the bulk: $\mu_B\sim \mu^{1/2}$ and the
 boundary can be treated quasiclassically. 
 In the second case
 $\mu_B\sim \mu^{1/2 b^2}$, which corresponds to the 
 ``wrong" dressing  of the identity
 boundary operator,  the boundary has
 anomalous fractal dimension larger than one.
It has been shown that both boundary  Liouville interactions 
  can be realized microscopically \cite{ ADEold, 
Idis}.   This is why it is useful to 
  introduce a new scaling exponent $\nu$ having the meaning of  
inverse
 fractal dimension of the boundary \eqal\numu{
  {\mu_{_B}}^{2\nu} \sim \mu}
  where
  \eqal\nunu{
  \label{nunu}
   \nu =  1\ \  &&{\rm for\ boundary\ interaction}\  \ e^{b\phi}\no\\
 \nu =  b^2\ \ 
 && {\rm  for\  boundary \ interaction }  \ \ e^{\phi/b}.\no}
 The two possible values of $\nu$ are  dual to each other 
 in the sense that if the
classical boundary ($\nu=1$) corresponds to a Dirichlet boundary
condition, then  the fractal boundary ($\nu= b^2$) corresponds to
Neumann boundary condition for the matter field $\chi$, and {\it vice 
versa}
\cite{KKloop}.

\paragraph{\bf  Correlation functions and KPZ scaling:}
  Consider the correlation function of $n$ boundary and $m$ bulk fields
 as\footnote{To completely define the correlation function, 
 we should also specify the matter boundary conditions, which is understood below.}
 \eqn\defcorW{W_{ P_1,...,P_n}^{( K_1,...,K_m)}(\mu_{_{B1}}, ... 
 , \mu_{_{Bn}};\mu)= \left\langle {\CB}_{P_{1}}^{\mu_{_{B1}},
 \mu_{_{B{2}}}} ...  {\CB}_{P_{n}}^{\mu_{_{Bn}}, \mu_{_{B{1}}}} \ 
{\bf
 \CV}_{K_{1}}...  {\CV }_{K_{m}} \right\rangle _{\rm disc},
 \label{defcorW}} 
 where  $\<\ ... \>$ denotes  the functional integral with respect to 
the matter and Liouville fields.   By the conformal
 invariance   this correlation function has the form
 \eqn\scfactor{\mu^{\gamma}\ F\left( \mu_{_{B1}} \mu^{-1/2\nu_1},
..., \mu_{_{Bn}}  \mu^{-1/2\nu_n} \right) ,
}
where the exponent $\g$ depends linearly on the target space momenta
\eqn\expgamma{\gamma = 1- \hf \gst
 - \sum_{j=1}^m (1- \delta^{^{\rm bulk}}_{K_j})-
{1\over 2} \sum_{k=1}^n (1- \delta^{^{\rm bound} }_{P_k} ) .}
 Here 
 $\nu_1,...,\nu_n$ are the exponents (\ref{nunu}) for  the $n$ segments of the boundary,
\eqn\gssg{\gst = 1- 1/b^2<0} 
is the string susceptibility exponent,
\eqn\grdim{\delta_{K} ^{^{\rm bulk}}= { |K|\over  2b}+\hf \gst 
\ \qquad \delta_{P} ^{^{\rm bound} }= { |P|\over  b}+\hf \gst } 
are  correspondingly the 
gravitational dimensions of the bulk and boundary fields, and $F$ is a
scaling function.  The gravitational dimensions $\d$ are related to
the flat dimensions $\Delta$ and the conformal anomaly $c$ of the
matter field by the KPZ formula
\eqn\KPZsc{\Delta = {\delta(\delta-\gst)\over 1-\gst},
\qquad c =1-6 {\gst ^{2} \over 1-\gst}.
} 

Using the results obtained in boundary Liouville theory and reviewed
in the next section, we can evaluate instantly the boundary
correlation functions in 2D QG that involve   three or less fields. 
In this case there is no moduli integration and the correlation
functions of the matter and Liouville fields factorize and the
correlation function in 2D QG is equal (up to a numerical factor) to
the corresponding Liouville structure constant.  There is a subtlety
related to the fact that only physical fields are present in 2D QG. It
is resolved by using the reflection symmetry to express the
correlation functions of wrongly dressed fields (whenever they appear)
in terms of the correlation functions of physical fields.

 \newsec{Correlation functions in   boundary Liouville theory}

\subsection{Review of boundary Liouville theory  }

In this subsection  we present a brief review of some of the
exact results for Liouville on a disc, obtained recently in
\cite{ZZ,FZZb}.  
In  the upper half plane  the Liouville theory is defined by the action  
\begin{equation}
A=\int\limits_{\mathrm{UHP}}\left(  \frac{1}{4\pi}(\partial_{a}%
\phi)^{2}+\mu e^{2b\phi(z,\bar{z})}d^{2}z\right)
+\mu_{B}\int\limits_{\mathbb{R}}e^{b\phi(x)}  dx
\label{bound}
\end{equation}
 with asymptotics   at infinity  $\phi(z,\bar{z})\simeq -Q\log(z\bar{z})$ and 
 Neumann  boundary condition
  on the
real axis 
 $i(\partial -\bar{\partial})\phi = 4\pi \mu_B e^{b\phi}$.
 The theory is conformal  invariant with central charge   
$c_L=1+6Q^2$.

\paragraph{\bf Bulk fields:}
The fields $V_{\al}(z,\bar{z})= e^{2\a\phi(z,\bar z)}$ are
primaries with respect to the energy momentum tensor
\begin{eqnarray}
T(z)&=&-(\partial\phi)^{2}+Q\partial^{2}\phi \; ,\nonumber \\
\bar{T}(\bar{z})&=&-(\bar{\partial}\phi)^{2}+Q\bar{\partial}^{2}\phi 
\nonumber
\end{eqnarray}
and have conformal weight
$\Delta_{\alpha}=\bar{\Delta}_{\alpha}=\alpha(Q-\alpha)$.
Because of the invariance $\al \to Q-\al$, one identifies the
operator $V_{\al}$ with its reflected image $V_{Q-\al}$:
\eqal\reflblc{
\label{reflblc}
V_{\al}(z,\bar{z})=S(\al)V_{Q-\al}(z,\bar{z})
}
where we  used  the bulk reflection amplitude \cite{ZZ}
\begin{eqnarray}
\label{LiouvR}
S(\al)=\frac{(\pi \mu
\gamma(b^{2}))^{(Q-2\al)/b}}{b^{2}}\frac{\gamma(2\al
b-b^{2})}{\gamma(2-2\al/b+1/b^{2})}, \label{ra}
\end{eqnarray}
and $\gamma(x)=\Gamma(x)/\Gamma(1-x)$. The reflection amplitude
satisfies the unitarity condition:
 \eqal\unitblc{
 \label{unitblc}
 S(\al)S(Q-\al)=1.
 }
 An important set among the
primaries are the fields $V_{-nb/2},\;n \in \mathbb{N},$ which are
degenerate with respect to the conformal symmetry algebra and
satisfy linear differential equations \cite{BPZ}. For example, the
first non trivial case consists of $\al=-b/2$, and the
corresponding operator satisfies
\begin{equation}
\left(\frac{1}{b^{2}}\partial^{2}+T(z)\right)V_{-b/2}=0, 
\end{equation}
as well as the complex conjugate equation.
It follows from these equations that when one performs the
operator product expansion of one of these degenerate operators
with a generic operator, then the OPE is truncated \cite{BPZ}. For
example:
\begin{eqnarray}
V_{-b/2}V_{\al}=c_{+}V_{\al-b/2}+c_{-}V_{\al+b/2} \, .\label{OPE}
\end{eqnarray}
The structure constants $c_{\pm}$ are special cases of the bulk
three point function, and can be computed pertubatively as Coulomb
gas (or screening) integrals \cite{FF,DF}. One can take $c_+=1$,
as in this case there is no need of insertion of interaction,
whereas $c_-$ requires one insertion of the Liouville interaction
$-\mu\int e^{2b\phi}d^2z $, and
\begin{eqnarray}
c_-&=&-\mu\int d^2z \left\langle V_{\alpha}(0)
V_{-b/2}(1)e^{2b\phi(z,\bar{z})}V_{Q-\alpha-b/2}
(\infty)\right\rangle\nonumber \\
&=& -\mu\int d^2z |z|^{-4b\al}|1-z|^{2b^2}\nonumber \\
&=&
-\mu\frac{\pi\gamma(2b\alpha-1-b^2)}{\gamma(-b^2)\gamma(2b\al)} \,
.\nonumber
\end{eqnarray}
In the first line, we used the property of invariance under global
transformations to set $z_1=0,\, z_3=1,\, z_4=\infty$, and in the
second line
$<\phi(x)\phi(y)>=-\log |x-y|$.\\
Similarly, there exists also a dual series of degenerate operators
$V_{-m/2b}$ with the same properties.\\

\paragraph{\bf Boundary fields:}
There is a one-parameter family of conformally invariant boundary
conditions characterized by different values of the boundary
cosmological constant $\mu_B$. For later convenience we will
parametrize
 $\mu_B$ in terms of a dimensionless angular variable 
$\s$ defined as follows \cite{FZZb}:
\begin{eqnarray}
\text{cos}\left(2\pi
b(\sigma-\frac{Q}{2})\right)=\frac{\mu_{B}}{\sqrt{\mu}}\sqrt{\text{sin}(\pi
b^{2})}.  \label{relation mu-sigma}
\end{eqnarray}
The meaning of this parametrization is that it unfolds the branch point of the 
observables at $\mu_B = - \mu_B^0\equiv \sqrt{\mu/\sin \pi b^2} $.
We shall denote the boundary
operators as $B_{\beta}^{\sigma_{2}\sigma_{1}}(x)$; they have
conformal weight $\Delta_{\beta}=\beta(Q-\beta)$ and are labeled
by two left and right boundary conditions $\sigma_{1}$ and
$\sigma_{2}$ related to $\mu_{B_1}$ and $\mu_{B_2}$ by (\ref{relation mu-sigma}).
Again, because of the invariance $\beta \to Q-\beta$, one should
identify the field $B_{\beta}^{\sigma_{2}\sigma_{1}}(x)$ with its
reflected image $B_{Q-\beta}^{\sigma_{2}\sigma_{1}}(x)$:
\eqn\Brefl{
B_{\beta}^{\sigma_{2}\sigma_{1}}(x)=D(\beta,\sigma_{2},\sigma_{1})
\ B_{Q-\beta}^{\sigma_{2}\sigma_{1}}(x)
}
 The  boundary reflection amplitude  $D(\beta,\sigma_{2},\sigma_{1})$ 
satisfies the   unitary condition
\eqn\unitaryD{
\label{unitaryD}
D(\beta,\sigma_{2},\sigma_{1})D(Q-\beta,\sigma_{2},\sigma_{1})=1
}
and gives the nontrivial piece of the boundary two-point 
correlation function, see the next section.

As it was argued in \cite{FZZb}, the degenerate boundary fields
$B_{-nb/2}^{\sigma_{2}\sigma_{1}},\; n\in \mathbb{N},$  have
truncated operator product expansion with all  primary fields 
if the left and right boundary parameters are related by
$\s_2-\s_1=
-\frac{nb}{2},  -\frac{(n-2)b}{2} ,,, \frac{nb}{2} $.
 For example, for $\s_2=\s_1\pm \frac{b}{2}$
\begin{eqnarray}
B_{\beta}^{\sigma_{3}\sigma_{2} }
B_{-b/2}^{\sigma_{2}   \sigma_{1}}
= c_{+}^{\pm}B_{\beta-b/2}^{\sigma_{3}\sigma_{1}}+c_{-}^{\pm}
B^{\sigma_{3}\sigma_{1}}_{\beta+b/2}\; , \quad \sigma_2=\sigma_1
\pm b/2  \label{OPE -b/2}
\end{eqnarray}
where, as in the bulk situation, $c_{\pm}^{\pm}$ are structure
constants and are obtained as certain screening integrals.  In the
first term of (\ref{OPE -b/2}) there is no need of screening insertion
and therefore $c_{+}^{\pm}$ can be set to 1, whereas the computation
of $c_{-}^{\pm}$ requires one insertion of the boundary interaction
$-\mu_B\int e^{b\phi(x)} dx$ and was explicitly computed in
\cite{FZZb} with the following result
\begin{eqnarray}
\lefteqn{c_{-}^\pm=
2\left(-\frac{\mu}{\pi\gamma(-b^2)}\right)^{1/2}
 \Gamma(2b\beta -b^2-1)\Gamma(1-2b\beta)\times} \nonumber \\
&& \sin\pi b(\beta\pm(\sigma_1-\sigma_3) -b/2) \sin\pi b(\beta
\pm(\sigma_1+\sigma_3-Q)-b/2).  
\end{eqnarray}
Similar properties have the fields $B_{-n/b}^{\s_1\s_2}$ with $\s_2-\s_1=
-\frac{n}{2b},  -\frac{n-2}{2b} \dots  \frac{n}{2b} $.

\paragraph{\bf Self-duality:}
The observables are invariant with respect to the duality transformation
 $b\to 1/b$ 
provided the dual cosmological constant $\tilde{\mu}$ is related
to $\mu$ as
\begin{eqnarray}
\pi\tilde{\mu}\gamma(1/b^2)=(\pi\mu\gamma(b^2))^{1/b^2},\label{mu}
\end{eqnarray}
and the dual boundary cosmological constant is defined as follows
\eqn\mubs{
\label{mubs}
\cos\left(\frac{2\pi}
{b}(\sigma-\frac{Q}{2})\right)=\frac{\tilde{\mu}_{B}}{\sqrt{\tilde{\mu}}}
\sqrt{\sin\frac{\pi} {b^2}}\;.
}

\subsection{Correlation functions}
In order to characterize LFT on the upper half plane, one needs to
know the following structure constants:
\begin{enumerate}
\item
bulk one-point function
\begin{equation}
\left\langle V_{\alpha}(z,\bar{z})\right\rangle_{\sigma}
=\frac{U_{\sigma}(\alpha)}{\left| z-\bar{z}\right|
^{2\Delta_{\alpha}}} \label{onepoint} \nonumber
\end{equation}
\item
boundary two-point function
\begin{equation}
\left\langle
B_{\beta_1}^{\sigma_{1}\sigma_{2}}(x)B_{\beta_2}^{\sigma_{2}
\sigma_{1}}(0)\right\rangle
=\frac{\delta(\beta_2+\beta_1-Q)+D(\beta_1,\sigma_{2},
\sigma_{1})\delta(\beta_2-\beta_1)}{\left|
x\right|^{2\Delta_{\beta_1}}} \nonumber
\end{equation}
\item
bulk-boundary two-point function \footnote{The bulk one point
function is a special case of the bulk-boundary coefficient with
$\beta=0$.}
\begin{equation}
\left\langle
V_{\alpha}(z,\bar{z})B_{\beta}^{\sigma\sigma}(x)\right\rangle
=\frac{R_{\sigma}(\alpha,\beta)}{\left|  z-\bar{z}\right|
^{2\Delta_{\alpha}-\Delta_{\beta}}\left| z-x\right|
^{2\Delta_{\beta}}} \label{bbound} \nonumber
\end{equation}
\item
boundary three-point function
\begin{eqnarray}
\left\langle
B_{\ \beta_{1}}^{\sigma_{2}\sigma_{3}}(x_{1})
B_{\ \beta_{2}}^{\sigma_{3}
\sigma_{1}}(x_{2})
B_{\ \beta_{3}}^{\sigma_{1}\sigma_{2}}(x_{3})
\right\rangle
=\frac{C_{\beta_1 \beta_{2}\beta_{3}}^{(\sigma_{2}\sigma_{3}
\sigma_{1})}}{\left|
x_{21}\right|  ^{\Delta_{1}+\Delta_{2}-\Delta_{3}}\left|
x_{32}\right|
^{\Delta_{2}+\Delta_{3}-\Delta_{1}}\left|  x_{31}\right|  
^{\Delta_{3}%
+\Delta_{1}-\Delta_{2}}}  \nonumber
\end{eqnarray}
 By its definition, this function is invariant under cyclic
permutations.
 In the following we will also use the notation
$$C_{\b_1\b_2}^{(\sigma_{2}\sigma_{3}
\sigma_{1})\beta_{3}}
\equiv C_{ \beta_1 \beta_{2}, Q-\beta_{3}}^{(\sigma_{2}\sigma_{3}
\sigma_{1})} .
$$
All correlation functions considered above are invariant w.r.t.
the duality transformation $b \to 1/b$.
\end{enumerate}

\subsection{Explicit expressions and functional relations}

Below we list the explicit representation for the structure constants,
and the functional equations they satisfy.  Except for the boundary
three point function, they have been computed by considering an
auxiliary correlation function containing a degenerate bulk or
boundary field: this leads to solvable functional relations for the
correlation function sought for.  This trick was first introduced in
\cite{Teschner}, and used extensively in \cite{T,FZZb,Z,hosomichi}. 
For each of the equations listed below, there exists a dual equation,
obtained by replacing $b$ by $1/b$.  It is important to note that the
functional relations are of two types: one type involves shifts on
boundary parameters and momentum, the other type involves shifts of
boundary parameters only.

\begin{enumerate}
\item { bulk one-point function}
\paragraph{\it Expression\cite{FZZb,T}:}
\begin{eqnarray}
\label{Upsigma}
\lefteqn{U_{\sigma}(\alpha)=\frac{2}{b}(\pi \mu
\gamma(b^2))^{\frac{(Q-2\alpha)}{2b}}\times}\nonumber \\
&&
\times\Gamma(2b\alpha-b^2)\Gamma(\frac{2\alpha}{b}-\frac{1}{b^2}-1)
\cos[\pi(2\alpha-Q)(2\sigma-Q)]
\end{eqnarray}
\paragraph{\it Reflection property:}
$$
U_\sigma (\alpha)=S(\alpha)U_\sigma (Q-\alpha),
$$
where $S(\alpha)$ is the bulk reflection amplitude (\ref{ra}).

\paragraph{\it Functional equation:}
\begin{eqnarray}
U_{\sigma-\frac{b}{2}}(\alpha)+U_{\sigma+\frac{b}{2}}(\alpha)=2\cos\pi
b(2\alpha-Q)U_{\sigma}(\alpha). \label{fa}
\end{eqnarray}

\item boundary two-point function
\paragraph{\it Expression\cite{FZZb}:}
\begin{eqnarray}
\label{Dbeta}
\lefteqn{D(\beta,\sigma_{2},\sigma_{1})
 =\(\pi\mu\gamma(b^{2})b^{2-2b^{2}}\)^{\frac{1}{2b}(Q-2\beta)}\times}
\nonumber \\
& & \times \frac{\Gamma_{b}(2\beta-Q)}{\Gamma_{b}(Q-2\beta)}
\frac{S_b(\sigma_{2}+\sigma_{1}-\beta)S_b(2Q-\beta-\sigma_{1}-\sigma_{2})}
{S_b(\beta+\sigma_{2}-\sigma_{1})S_{b}(\beta+\sigma_{1}-\sigma_{2})}
\end{eqnarray}

\paragraph{\it Reflection property:}
it takes here the form of the unitarity condition
\eqal\Duni{
\label{Duni}
D(\beta,\sigma_{2},\sigma_{1})\ D(Q-\beta,\sigma_{2},\sigma_{1})=1.
}

\paragraph{\it Functional equations:}
The first of these equations can be seen as a consequence from the
different possible factorizations for an auxiliary boundary three
point function containing a degenerate boundary field
$<B_{\beta+b/2}(x_3)B_{-b/2}(x_2)B_{\beta}(x_1)>$, see equation (4.3)
of \cite{FZZb}.
\begin{eqnarray}
\lefteqn{D(\beta,\sigma_{2},\sigma_{1}-b/2)-D(\beta,\sigma_{2},
\sigma_{1}+b/2)=
\left(-\frac{\mu}{\pi\gamma(-b^2)}\right)^{1/2} }\nonumber
\\
&& \times \frac{-2\pi\Gamma(1-2b\beta_1)}{\Gamma(2+b^2-2b\beta_1)}
\sin
 \pi b(2\sigma_1-Q)D(\beta+b/2,\sigma_{2},\sigma_{1}).
 \label{1fe}
\end{eqnarray}
One can also find
\begin{eqnarray}
\lefteqn{\sin^2\pi
b(\sigma_1-\sigma_2-\beta+\frac{b}{2})D(\beta,
\sigma_2+\frac{b}{2},\sigma_1)}\nonumber
\\
&& -\sin^2\pi
b(\sigma_1+\sigma_2-\beta-\frac{b}{2})D(\beta,
\sigma_2-\frac{b}{2},\sigma_1)\nonumber
\\
&& =\sin\pi b(2\sigma_1-Q)\sin\pi
b(2\sigma_2-Q)D(\beta,\sigma_2,\sigma_1+\frac{b}{2}), \label{2fe}
\end{eqnarray}
and when the two boundary parameters are equal
$\sigma_1=\sigma_2\equiv \sigma$:
\begin{eqnarray}
\sin\pi b(2\sigma-Q-\beta) D_{\sigma-b/2}(\beta) + \sin\pi
b(2\sigma-Q+\beta) D_{\sigma+b/2}(\beta)\nonumber
\\
 =2\sin\pi b(2\sigma-Q)\cos\pi b(\beta-Q)
D_{\sigma}(\beta) .\nonumber
\end{eqnarray}

\item bulk-boundary two-point function
\paragraph{\it Expression\cite{hosomichi, Z}:}
\begin{eqnarray}
\lefteqn{R_{\sigma}(\alpha,\beta)=2\pi (\pi\mu\gamma(b^{2})
b^{2-2b^{2}})^{\frac{1}{2b}(Q-2\alpha-\beta)}}\nonumber
\\
&&\frac{\Gamma_b^3(Q-\beta)\Gamma_b(2\alpha-\beta)
\Gamma_b(2Q-2\alpha-\beta)}
{\Gamma_b(Q)\Gamma_b(Q-2\beta)\Gamma_b(\beta)
\Gamma_b(2\alpha)\Gamma_b(Q-2\alpha)}\nonumber \\
&& \int_{- i\infty}^{+i\infty}  dp \; e^{2i\pi
p(2\sigma-Q)}\frac{S_b(p+\beta/2+\alpha-Q/2)S_b(p+\beta/2-\alpha+Q/2)}
{S_b(p-\beta/2-\alpha+3Q/2)S_b(p-\beta/2+\alpha+Q/2)}.\nonumber
\end{eqnarray}
\paragraph{\it Reflection properties:}
$$
R_{\sigma}(\alpha,\beta)=S(\alpha)R_{\sigma}(Q-\alpha,\beta),\quad
R_{\sigma}(\alpha,\beta)=D(\beta,\sigma,\sigma)R_{\sigma}
(\alpha,Q-\beta)\footnote{The
first of these relations is trivially verified. We checked the
second one is also true, as it was not done in \cite{hosomichi}.}.
$$
\paragraph{\it Functional equations:}
The two functionals relations written below are nothing but the
Fourier transform of the equations (21) found in \cite{hosomichi}.
We recall that these equations arise from the different possible
factorizations for the auxiliary three point function containing a
degenerate boundary field
$<V_{\alpha}(0)B_{\beta}(1)B_{-b/2}(\eta)>$. We find:
\begin{eqnarray}
\lefteqn{R_{\sigma-\frac{b}{2}}(\alpha,\beta)-
R_{\sigma+\frac{b}{2}}(\alpha,\beta)=\sin\pi
b(2\sigma-Q)R_{\sigma}(\alpha,\beta+b)\times}\nonumber \\
&& -2\pi \left(-\frac{\mu}{\pi \gamma(-b^2)}\right)^{1/2}
\frac{\Gamma(1-2b\beta)\Gamma(1-b^2-2b\beta)}{\Gamma^2(1-b\beta)
\Gamma(1-b\beta-2b\alpha)
\Gamma(1-b\beta+2b\alpha-bQ)}, \nonumber\\
 \label{toto}
\end{eqnarray}
as well as
\begin{eqnarray}
\sin\pi b(2\sigma-Q-\beta) R_{\sigma-b/2}(\alpha,\beta) + \sin\pi
b(2\sigma-Q+\beta) R_{\sigma+{b\over 2}}(\alpha,\beta)\nonumber
\\
 =2\sin\pi b(2\sigma-Q)\cos\pi b(2\alpha-Q)
R_{\sigma}(\alpha,\beta).\nonumber\\
\label{rfe}
\end{eqnarray}
One notices that the special case $\beta=0$ in the latter equation
reproduces equation (\ref{fa}).
\paragraph{\it Particular case:}
When $2\alpha=\beta$, one can show the following relation that
links the bulk-boundary structure constant to the boundary
reflection amplitude:
\begin{eqnarray}
\label{RvD}
\lefteqn{R_{\sigma}(\beta)\equiv \mathrm{res}_{2\alpha=\beta}
R_{\sigma}(\alpha,\beta)
}\nonumber\\
&&
=2\pi\frac{\Gamma_b^2(Q-\beta) }{\Gamma^2_b(\beta)
\Gamma_b(Q)}
D(\beta,\sigma,\sigma)\mathrm{res}_{x=0}\Gamma_b(x),
\end{eqnarray}

\item boundary three-point function
\paragraph{\it Expression\cite{PTtwo}:}
 \begin{eqnarray}
\lefteqn{C_{\beta_{2}\beta_{1}}^{(\sigma_{3}\sigma_{2}
\sigma_{1})\beta_{3}}
=
 \bigl(\pi \mu \gamma(b^2) 
b^{2-2b^2}\bigr)^{\frac{1}{2b}(\beta_3-\beta_2-\beta_1)}} \nonumber \\
&&
\times\frac{\Gamma_b(Q+\beta_2-\beta_1-\beta_3)
\Gamma_b(Q+\beta_3-\beta_1-\beta_2)}
{\Gamma_b(Q-2\beta_1)\Gamma_b(Q)}\nonumber \\
&& \times
\frac{\Gamma_b(2Q-\beta_1-\beta_2-\beta_3)\Gamma_b
(\beta_2+\beta_3-\beta_1)}{\Gamma_b(2\beta_3-Q)\Gamma_b(Q-2\beta_2)}
\nonumber \\ &&
\quad\times\frac{S_b(\beta_3+\sigma_1-\sigma_3)
S_b(Q+\beta_3-\sigma_3-\sigma_1)}{S_b(\beta_2+\sigma_2-\sigma_3)
S_b(Q+\beta_2-\sigma_3-\sigma_2)} \nonumber \\
&& \quad \times \frac{1}{i}\int\limits_{-i\infty}^{i\infty}ds \;\;
\frac{S_b(U_1+s)S_b(U_2+s)S_b(U_3+s)S_b(U_4+s)}
{S_b(V_1+s)S_b(V_2+s)S_b(V_3+s)S_b(Q+s)}  
 \label{ff3p}
\end{eqnarray}

and the coefficients $U_i$, $V_i$ and $i=1,\ldots,4$ read
$$
\begin{array}{ll}
 U_1 =\sigma_1+\sigma_2-\beta_1            &  V_1 = 
Q+\sigma_2-\sigma_3-\beta_1+\beta_3 \\
 U_2 = Q-\sigma_1+\sigma_{2}-\beta_1       &  V_2 = 
2Q+\sigma_2-\sigma_3-\beta_1-\beta_3 \\
 U_3 = \beta_2+\sigma_2-\sigma_3           &  V_3 = 2\sigma_2 \\
 U_4 = Q-\beta_2+\sigma_2-\sigma_3         \\
\end{array}
$$
\paragraph{\it Reflection properties:}
It was shown in \cite{PTtwo} that
\begin{equation}
C_{\beta_{1}\beta_{2}\beta_{3}}^{(\sigma_{2}\sigma_{3}\sigma_{1})} 
\equiv
C_{\beta_{1}\beta_{2}}^{(\sigma_{2}\sigma_{3}\sigma_{1})Q-\beta_{3}} 
=%
D(\beta_{3},\sigma_{2},\sigma_{1})
C_{\beta_{1}\beta_{2}}^{(\sigma_{2}\sigma_{3}\sigma_{1})\beta_{3}}.\nonumber
\end{equation}
as well as two other similar equations  obtained by cyclic permutations.
  
 \paragraph{\it Functional equations:}
It is shown in the appendix B that the following two equations are
a consequence of the 
pentagonal equation  satisfied by the boundary
three point function\footnote{We recall that this fact follows
from the consistency condition that expresses the associativity of
the product of four boundary operators \cite{Runkel, BPPZ} .} 
\cite{PTtwo}:
\begin{eqnarray}
\label{tripfe}
\lefteqn{C_{\beta_{1},\beta_{2},\beta_3}^{(\sigma_{2},\sigma_{3},
\sigma_{1}-
\frac{b}{2})}-
C_{\beta_{1},\beta_{2},\beta_3}^{(\sigma_{2},\sigma_{3},
\sigma_{1}+
\frac{b}{2})}
=C_{\beta_{1}, \beta_2 +\frac{b}{2} ,\beta_{3}+\frac{b}{2}}^{(\sigma_{2},\sigma_{3},
\sigma_{1})}}
\nonumber \\
&& \times 
-2\pi \left(-\frac{\mu}{\pi\gamma(-b^2)}\right)^{1/2}
 \sin\pi b(2\sigma_1-Q)\times \nonumber \\
&& \times \frac{\Gamma(1-2b\beta_2) \ \Gamma(1-2b\beta_3)}{\Gamma(1-b(\beta_2+\beta_3- \beta_1))\ 
\Gamma(2-b(\beta_1+\beta_2+\beta_3-b))}
\nonumber
\\
\end{eqnarray}
and
 \begin{eqnarray}
\lefteqn{\sin\pi b(\sigma_2-\sigma_1-\beta_3+\frac{b}{2})\sin\pi
b(\sigma_3-\sigma_1-\beta_2+\frac{b}{2})\
C_{\beta_1,\beta_2,\beta_3}^{(\sigma_2,\sigma_3,
\sigma_1+\frac{b}{2})}}\nonumber
\\
&& - \sin\pi b(\sigma_2+\sigma_1-\beta_3-\frac{b}{2})\sin\pi
b(\sigma_3+\sigma_1-\beta_2-\frac{b}{2})\
C_{\beta_1, \beta_2,\beta_3}^{(\sigma_2,\sigma_3,\sigma_1-\frac{b}{2})}
\nonumber
\\
&& =\sin\pi b(\sigma_2+\sigma_3-\beta_1)\sin\pi b(2\sigma_1-Q)\
C_{\beta_1,\beta_2,\beta_3}
^{(\sigma_2+\frac{b}{2},\sigma_3+\frac{b}{2}, \sigma_1
)} \quad. \label{3fe}
\end{eqnarray}
These equations reproduce equations (\ref{1fe}) and (\ref{2fe}) in
the case $\beta_1=0,\; \sigma_2=\sigma_3,$ and $\beta_2=\beta_3$,
where the structure constant
$C_{0,\beta,\beta}^{\sigma_2,\sigma_2,\sigma_1}$ reduces to the
boundary reflection amplitude $D(\beta,\sigma_2,\sigma_1)$.
\end{enumerate}

\noindent

\subsection{Important cases}

    To compare the Liouville approach with the matrix models approach, it
is useful to compare the dependence with respect to the boundary
parameter $\sigma$ of some important correlation functions, such as
the one point function of the boundary cosmological operator
$B_b^{\sigma\sigma}$, the bulk-boundary structure constant with one
boundary cosmological operator inserted $R_{\sigma}(\alpha,b)$, and
the boundary three point function with three boundary cosmological
operators inserted.
\begin{itemize}
\item
$<B_b^{\sigma\sigma}>:$\\
It amounts to evaluate the quantity $R_{\sigma}(0,b)$ directly
from its integral representation. In sending $\alpha \to 0$, some
of the poles of the $S_b$ functions will cross the imaginary axis,
so one has to deform the contour accordingly. Picking up the
residues at poles leads to the following expression:
\begin{eqnarray}
\label{Bbss}
<B_b^{\sigma\sigma}>\sim \mu^{\frac{1}{2b^2}} \cos\pi
b^{-1}(2\sigma-Q).
\end{eqnarray}
In other words, the one point function of the boundary
cosmological operator $B_b^{\sigma\sigma}$ is proportional to the
dual boundary cosmological constant $\tilde \mu_B$.
\item
$R_{\sigma}(\alpha,b), \alpha \neq 0:$\\
Setting $\beta=0$ in (\ref{toto})
 provides a relation between $R_{\sigma}(\alpha,b) $
and the one
 point function $R_{\sigma}(\alpha,0)=U_{\sigma}(\alpha)$.For the
 time being, we are only interested in the dependence of 
 $R_{\sigma}(\alpha,b)$  and $U_\sigma(\a)$ on
$\sigma$, which is found to be:
\begin{eqnarray}
\label{Rabs}
R_{\sigma}(\alpha,b)\sim \mu
^{\frac{1}{2b}(b^{-1}-2\alpha)}\frac{\sin\pi
[(2\alpha-Q)(2\sigma-Q)]}{\sin \pi b(2\sigma -Q)} \label{11PT}
\end{eqnarray}
\item
$C_{bbb}^{(\sigma_{2}\sigma_{3}\sigma_{1})}:$ \\
It is not straightforward to derive it directly from (\ref{ff3p}), but one can check that
  the following expression
\begin{eqnarray}
\label{Csss}
C_{bbb}^{(\sigma_{2}\sigma_{3}\sigma_{1})}
\sim  \frac{(\mu_{B_{2}}-\mu_{B_{3}})
\tilde{\mu}_{B_1}+(\mu_{B_{3}}-\mu_{B_{1}})\tilde{\mu}_{B_2}+
(\mu_{B_{1}}-\mu_{B_{2}})\tilde{\mu}_{B_3}}{(\mu_{B_{3}}-\mu_{B_{1}})
(\mu_{B_{1}}-\mu_{B_{2}})(\mu_{B_{2}}-\mu_{B_{3}})} \label{3PT}
\end{eqnarray}
 is a solution of the functional equation (\ref{3fe}). 
\end{itemize}

   \section{Boundary correlation functions in  Liouville quantum gravity}

In order to carry out the comparison with the discrete approach, we
need to
 take into account the matter field and also normalize the fields and
the 
coupling constants in  a way to match with their microscopic realizations.  
Among the results listed in the previous section it is sufficient to 
consider eqn. (\ref{tripfe}) together with  its dual since it contains all the information 
we need about the  boundary correlators. 
We will show that eqn. (\ref{tripfe}) together with the duality and   reflection  
properties  is sufficient to determine, up to a normalization constant, the 
one-, two- and three-point functions in \QG.

Since we are considering the simplest case of a free field with no
screening
 charges, the  role of the matter field is reduced to imposing the
neutrality
 condition for the charges of the vertex operators.  We   assume Neumann  
boundary conditions for the  matter field along all segments of the boundary.
Thus  the difference equations in Liouville theory hold also for the
physical
 fields  in \QG.     

The only subtle point here is that the equations do not close on the
set of the
 physical fields. Repeated application of the difference operator
produces eventually  a ``wrongly dressed" field. In this case we will need the
 boundary reflection amplitude to express the ``wrongly dressed" field through  a physical one.
 
 Once   closed  equations for the three-point functions are found, 
 we  will  formulate them entirely in terms of the observables that
can be measured in the microscopic theory.  For this purpose we evaluate 
the  boundary one- and two-point functions  in \QG.

\subsection{The rescaled bulk and boundary cosmological constants}

  We will  redefine the bulk and boundary  cosmological constants
   $\mu$ and $\mu_B$ according to the  normalizations (\ref{defVP})and  (\ref{BKPZ})
for the bulk and boundary operators. The new bulk and boundary  cosmological constants\footnote{Here 
we use the conventions  of ref.  \cite{Idis}.} $\Lambda$ 
and $z$  are defined as the couplings of the operators $\CV^{(+)}_{e_0}$ and $\CB  ^{(+)}_{e_0/2}$ respectively:
$$\mu  \int e^{2b\phi} =
\Lambda\  \CV^{(+)}_{e_0},
\quad \mu_B \ e^{b\phi}= z\  \CB  ^{(+)}_{e_0/2}.  $$
This gives
 \eqn\Mmu{
\label{Mmu}
{\Lambda\over \mu}  =\pi {\G(b^2)\over \G(1-b^2)}   ,
\qquad {z\over \mu_B} =  {\pi\over \G(1-b^2)}.
} 
We  will  also introduce the boundary parameter $\tau$ of \cite{Idis}
 related to $\s$ by
    \begin{equation}
  \qquad  \s=  \frac{Q}{2}  - i  {\tau \over 2\pi b} .
  \end{equation}
Now  the boundary cosmological constant is  parametrized as 
$$ z = M\cosh\t, \qquad M= \sqrt{\Lambda}.$$

\subsection{On the duality property of \QG}

It follows from (\ref{mu})-(\ref{mubs}) that the
 duality $b\to \tilde b = 1/b$ transforms 
the new bulk and boundary cosmological constants  into 
 \eqn\DUALTY{
\tilde \Lambda =  \Lambda^{1/b^2}, \qquad 
\tilde z =M^{1/b^2}\cosh  {\t/ b^2}
}
where the dual  boundary cosmological constants are   defined  by
 $$\tilde\Lambda\  \CV_{-e_0}=
  \tilde\mu \ e^{2\phi/b},
 \quad\tilde z\  \CB  ^{(-)}_{-e_0/2}= \tilde\mu_B\ e^{\phi/b}.
 $$
If we denote
$$ \tilde z = \tilde M \cosh \tilde \t,$$
then the duality  transformation   
$\tilde b = 1/b$  is formulated as 
  \eqn\DUALITY{
  \label{DUALITY}
    \tilde \Lambda  =  \Lambda^{2b^2} \ \ {\rm or} \ \ \tilde M = M^{1/b^2},
  \quad  \tilde\t = \t/b^2.
  }
 The duality 
 acts on 
  left and right chiral tachyons  as
  \eqn\TDU{
\label{TDU}
\tilde \CV ^{(\pm)}_{P} = \overline{ \CV ^{(\mp)}_{-P}} , \qquad
[\tilde \CB ^{(\pm)}_{P} ]^{\tilde \t_1\tilde \t_2} 
=[  \overline{  \CB ^{(\mp)}_{-P}}]^{\t_1\t_2},
}
where the bar means   complex conjugation.
In  the (non-unitary) realization of \QG\ based on the gaussian field  the 
duality $b\to \tilde b =1/b$ is not
 a symmetry of the Hilbert space of the theory. It acts in a pair of CFT
  characterized by  
matter background charge: $e_0$ and $\tilde e_0=-e_0$.

The duality property of the correlation functions thus reads
$$
\tilde W_{P_1,...,P_n}^{(K_1,...,K_m)} (\tilde \t_1,...,\tilde \t_n)
=\overline{
W_{-P_1,...,-P_n}^{(-K_1,...,-K_m)} (\t_1,...,\t_n)}.
$$

\subsection{The boundary  reflection  amplitude in \QG}

The open string tachyons of same momenta but opposite chiralities 
 are related by the reflection amplitude
 \eqn\reflB{
\label{reflB} [\CB^{(+)}_P]^{\t_1\t_2} =  D_P^{+-}  (\t_1, \t_2) 
\ [\CB^{(-)}_P]^{\t_1\t_2}, \quad
 [\CB^{(-)}_P]^{\t_1\t_2} =   D_P^{-+} ( \t_1,  \t_2) 
\ [\CB^{(+)}_P]^{\t_1\t_2}
}
The latter is obtained from the   Liouville reflection amplitude (\ref{Dbeta})
by  taking into account  the   leg factors in the definition (\ref{BKPZ}).
 Indeed,  the conventions used in sect. 2 are such that the
reflection amplitude for
 the matter  gaussian field $e^{e\phi}\to e^{(e_0-e)\phi}$  is equal to one.
  The expression of  the rescaled reflection amplitude is
   \eqal\DsubP{
 \label{DsubP}
 & &D_P^{+-}( \t_1,\t_2) =
  {1\over b}  { M^{2 P/b}  \over S_b(2P+b)
    }\ 
   \frac{S_b\({Q\over 2} +P - i {\t_1+\t_2\over 2\pi b}\)
  \ S_b\({Q\over 2} +P - i {\t_1-\t_2\over 2\pi b}\)}{
  S_b\({Q\over 2} -P - i {\t_1+\t_2\over 2\pi b}\)
  \ S_b\({Q\over 2} -P - i {\t_1-\t_2\over 2\pi b}\)}
 \cr
 & =&  {1\over b}
  M^{{2 P/ b}}   {\G_b(-2P+1/b)\over \G_b(2P+b)}
 \exp\left( 
{1\over 2}  { \int\limits_{0}^{\infty}}
 {dy\over y} \left[{  \sinh( \pi P y )\ \cos {y\t_1\over b}\ \cos {y\t_2\over b}
 \over 
\sinh (\pi y/b) \sinh
(\pi  by )}  -{2P\over \pi y}\right] \).
}
 The    amplitude  $D_P^{+-} = (D_P^{-+} )^{-1}$   has
 the    symmetries
   
 \eqn\REFLD{
 \label{REFLD}
 D_P^{+-} D_{-P}^{+-} = {1\over b^2}  {\sin {2\pi\over b}  P
 \over\sin 2\pi b P}, \qquad
 D_P^{-+} D_{-P}^{-+} = b^2 {\sin 2\pi b P
 \over\sin {2\pi\over b}  P}
}
and 
\eqn\REFB{
\label{REFB}
D_{-P}^{-+} (\t_1,\t_2) =
\tilde D^{+-} _{P}(\tilde \t_1, \tilde \t_2),
}
where the function 
$\tilde D_P^{+-} (\t_1,\t_2)$ is defined by (\ref{DsubP})
with $b$  replaced by   $\tilde b = 1/b$, and $\tilde\t_{1,2}=\t_{1,2}/b^2$.
Remarkably, the function (\ref{DsubP})
satisfies   difference equation (\ref{1fe}) and its dual  without the 
$\G$-functions  on the r.h.s.: 
   \eqal\DEDD{
   \label{DEDD}
 \sin \(\pi b^2 {\p\over\p\t_1}\)  D_P^{+-}  (\t_1,\t_2) 
 = M 
  \sinh \t_1 \  D^{+-}_{P-{b\over 2}}(\t_1,\t_2) .
  }
     \eqal\DED{
   \label{DED}
 \sin \(\pi  {\p\over\p\t_1}\)    D^{-+} _P(\t_1,\t_2) 
 =   M^{{1/ b^2}}  
  \sinh \({ \t \over b^2}\) D^{-+}_{P+{1\over 2 b}}(\t_1,\t_2)
  }
  
  Note that the reflection amplitudes  for  the degenerate momenta 
  $ P_{mn} = \hf(m/b - nb)$  are given by rational functions of
   $z=M\cosh\t$ and $\tilde z = M^{1/b^2}\cosh(\t/b^2).$
    They can be obtained from the reflection amplitudes  for $P=0$ 
  $$ D^{-+}_0  =  b S(b)= b^ 2, \ \ \ 
  D^{+-}_0  =  {1\over b }S(1/b)= {1\over b^ 2}
   $$
  by applying $m$ times eq. (\ref{DED}) and $n$ times eq. (\ref{DEDD}).
   Let us calculate the reflection amplitude for the Liouville
interaction,
 which will play important role in the sequel.  We have
  \eqal\expls{\label{expls}
  D^{+-}_{b/2}&=&  {b^{-2}   \over  \sin( \pi b^2)}( z_1+ z_2)
  \cr
  & & \cr
    D^{-+}_{-1/2b + b/2}&=&- {b^2\sin( \pi b^2)\over \sin(\pi/b^2)}  
  {\tilde z_1- \tilde z_2\over z_1-z_2}\cr
    & & \cr
    D^{+- }_{1/2b - b/2}&=& 
 - {\tilde z_1- \tilde z_2\over z_1-z_2}
  }
  where in the last line we used the property (\ref{REFLD}).

\subsection{Difference  equations for the three-point function}

Consider the boundary three-point function, which factorizes to a
 product of a matter and Liouville components 
\eqn\Wtrip{
\label{Wtrip}
 W_{P_1,P_2,P_3}(\t_1,\t_2,\t_3) = 
\< [\CB_{P_1}]^{\t_2\t_3}[\CB_{P_2}]^{\t_3\t_1}[\CB_{P_3}]^{\t_1\t_2} \>_{\rm disc} 
= C^{\rm matter} _{e_1e_2e_3} C_{\b_1\b_2\b_3}^{\s_2\s_3\s_1}.
}
 In absence of screening operators\footnote{The argument that follow
can
 be of course carried out  within the Coulomb gas formalism in its
full generality,  which involves  an arbitrary number of screening
charges. Then the  difference equations will  depend also on the  
boundary three point function of the matter boundary fields.
 }, 
  the gaussian matter field correlator is non-zero only if  the three
  electric charges satisfy $e_1+e_2+e_3= e_0$ or, in terms of momenta, 
 \eqn\neuttr{
 \label{neuttr}
 P_1+P_2+P_3=\hf e_0.
 }
  In this case the matter  structure constant  does not depend on the matter 
  boundary conditions and is equal to one.
   Indeed, the difference of the matter boundary parameters adjacent to a 
  vertex operator is fixed by the charge of the operator. The overall shift of the boundary parameter is just  a  global mode and  does not change the three-point function.

  Let us assume  that  
      $P_1 >b/2, \quad P_2 <-b/2,$
 in which case $\CB_{P_1}= \CB_{P_1}^{(+)} $ and $\CB_{P_2}= \CB_{P_2}^{(-)} $.
    Taking into account  the normalization (\ref{onshellB})
    and using the neutrality condition (\ref{neuttr}),  we write eqn. (\ref{tripfe})
     as
    \eqn\TRIPFE{
    \label{TRIPFE}
    W_{P_1,P_2,P_3}(\t_1,\t_2,\t_3+i b^2\pi ) -
      W_{P_1,P_2,P_3}(\t_1,\t_2,\t_3-i b^2\pi) 
   = 2i M \  \sinh \t_3\ W_{P_1-b/2,P_2+b/2,P_3}(\t_1,\t_2,\t_3). 
   }
 \def\TB{ {\Delta} }
 Remarkably, all $\G$-functions disappeared. This allows to write
 the difference equation in the  operator form
 \eqn\DOE{
 \label{DOE}
 \TB  _{P_2, P_1} (\t_3) \  W_{P_1,P_2,P_3}(\t_1,\t_2,\t_3 )=0
 }
 where $\TB  _{P_2, P_1} (\t_3) $ is a  difference operator associated with the
 segment  $\<12\>$  with boundary parameter $\t_3$ and  
   momenta $P_2$ and $P_2$ at  the ends:
 \eqn\OPFE{
 \label{OPFE}
\TB  _{P_2,  P_1} (\t_3)  =   \sin \(b^2\p{\t_3}\)   -
 M  \sinh (\t_3)\ e^{{b\over 2} (\p_{P_2 }
 - \p_{P_1})} \qquad \qquad \(P_1  >{b\over 2},  \ P_2 < -{b\over 2}\)
 }
 In a similar way, the equation dual to (\ref{tripfe}) can be reformulated 
 as an equation for the three-point  function  (\ref{Wtrip}) with 
 $\CB_{P_1}= \CB_{P_1}^{(-)} $ and $\CB_{P_2}= \CB_{P_2}^{(+)} $
 under the condition that $P_1 <-{1/2b}$ and $P_2 >{1/ 2b}$.
 The resulting difference equation has again the form 
  (\ref{DOE}) with difference operator given by 
 \eqn\OPFED{
  \label{OPFED}
   \TB _{P_2,  P_1} (\t_3)=
\sin\( \p_{\t_3}\)   - M ^{1/b^2}  
 \sin \({\t_3/b^2}\)\ e^{{1\over 2 b} (\p_{P_2} 
 - \p_{ P_1})} \qquad\qquad \(P_1 <-{1\over 2b}, \ P_2 >{1\over 2b}\)
 }
 
 The  difference operator in  (\ref{DOE})  involves translations of the momentum 
  by $b/2$ or  $1/2b$,  and  after  its repeated application 
  on gets on the r.h.s.   a correlation function in which one or two
operators get wrong Liouville dressing. In order to obtain closed  
set of equations  we should express the ``wrongly dressed" fields
(\ref{WRDS}) in terms of the physical ones using the  boundary reflection
amplitude. This gives again an equation of the form (\ref{DOE}), with 
  a   difference operator which depends on all three boundary parameters:

  \bigskip
   
 \noindent
-- if $P_1\in  [{1\over 2b},\infty]$ and 
  $ P_2\in[ -{1\over 2b},0]$:
  \eqal\DEWa{
   \label{DEWa}
   \TB _{P_2,  P_1} (\t_3)=
  \sin \(\pi  \p_{\t _3}\) 
 -
 M^{{1/ b^2}}  \sinh \({ \t_3/ b^2}\)  \
 D^{-+}_{P_2+{1\over 2b}}( \t_1 ,
 \t_2)\  e^{{1\over 2b} (\p_{P_2 }
 - \p_{P_1})}
}   
\bigskip

\noindent
if  $P_1 \in [0, {1\over 2b}]$ and 
 $P_2\in [ -{1\over 2b}, 0]$:
  \eqal\DEWb{
   \label{DEWb}
    \TB _{P_2,  P_1} (\t_3) = \sin \(\pi \p_{\t _3}\) 
   -
      M^{{1/ b^2}}  \sinh\({ \t _3/ b^2} \)
        D^{+-}_{P_1-{1\over 2b}}(\t_1 ,\t_3)
     D^{-+}_{P_2+{1\over 2b}}(\t _3,\t_2)\
      e^{ {1\over 2b}(\p_{P_2 }
 - \p_{P_1})} .}
        Similarly we get the equations for the  intervals 
  $P_1\in[-{b\over 2},0], P_2\in[{b\over 2}, \infty]$ 
  and $P_1\in [0, {b\over 2}],   P_2\in[-{b\over 2}, 0].$

It is possible  to express these equations entirely in terms of the
observable 
in the microscopic theory.  Indeed, the hyperbolic sine in the numerator 
of the second term  is   proportional to the 
one-point function (\ref{Bbss}) and the reflection amplitudes in the
denominator
 are proportional to the boundary two-point function.
In order to fix the coefficients, we will need the exact expressions of the 
boundary one- and two-point functions in \QG.

  \subsection{The boundary two-point function}
  
  The    boundary  two-point function in  \QG\  
  $$W_{P_1,P_2}(\t_1, \t_2)
      =        \< [\CB   _{P_1}]^{\t_1\t_2}[\CB_{P_2}]^{\t_1\t_2}\>_{\rm disc}$$
      can be obtained from the  boundary three-point function  
      using the relation
   \eqn\TPTP{
   \label{TPTP}
  {\p W_{P_1,P_2}(\t_1, \t_2)
  \over \p z_2} =   
   \< [\CB   _{P_1}]^{\t_1\t_2}[\CB_{\hf e_0}]^{\t_2\t_2}
   [\CB_{P_2}]^{\t_2\t_1}\>_{\rm disc}=
     W_{P_1,P_2, \hf e_0}(\t_1, \t_2, \t_2).
  }
  As a consequence, it satisfies the difference equation
  (\ref{TRIPFE}),   (\ref{DEWa})  as well as the dual equations in the
corresponding momentum intervals.   
   By the neutrality condition (\ref{neuttr})  the two-point function
vanishes unless  $P_1+P_2$ is equal to zero
\footnote{In a more general setting when screeneng charges are allowed, 
  this condition   will  be    relaxed to  $P_1+P_2 = m/2b -
nb/2$.}.
  We will denote
   \eqn\WP{
   \label{WP}
    W_P(\t_1,\t_2) =W_{-P}(\t_1,\t_2)=W_{P, -P}(\t_1,\t_2).
    }
    This  function   is proportional to the boundary reflection amplitude, but the
    evaluation of the exact coefficient is not easy   in   CFT because 
    of the zero mode associated with the residual dilatation symmetry.
    This is why we will evaluate it indirectly, by solving the difference equation.
    
    The function $W_P(\t_1,\t_2)$ should satisfiy  (\ref{DOE})   
    for any $P$, with the corresponding identification of the difference operator. 
    This gives two independent equations for each value of $P>0$.
      For sufficiently large $P$  the function $W_P(\t_1,\t_2)$
      satisfies 
        \eqn\OPF{
 \label{OPF}
\(  \sin \(b^2\p/\p\t_1\)   -
 M   \sin( \t_1) \ e^{-{b\over 2}  \p/\p P }  \)    W_{P}(\t_1,\t_2)=0 \qquad (|P|>-b/2)
 }
 and 
 \eqn\OPFd{ \label{OPFd}
\(  \sin \( \p/\p\t_1\)   -
 M  ^{1/ b^2}\  \sin (\t_1/b^2) \ e^{-{1\over 2b}  \p/\p P }
  \)  W_{P}(\t_1,\t_2)=0 \qquad (|P|>1/2b).
 }
These are exactly the equations  (\ref{DED}) and (\ref{DEDD}) for the 
       boundary reflection  amplitudes $D^{+-}$ and $D^{-+}$.
       Therefore $W_P$ should be proportional to both amplitudes, with $P$-dependent coefficients.
   The   common solution is determined up to a multiplicative constant  $\k= \k(b)$:
   \eqn\WPD{
   \label{WPD}
   W_{P}(\t_1,\t_2)= \k_b{ b\over | \sin ( 2\pi  P/b)|}
   D_{|P|}^{+-}(\t_1,\t_2) =\k_b
   {1/ b\over  |\sin (2\pi Pb)|}
     D_{-|P|}^{-+}(\t_1, \t_2) .}
    The duality transformation acts, according to (\ref{REFB}),   as
 \eqn\WPDD{\label{WPDD}
  \tilde W_{P}(\tilde \t_1,\tilde \t_2)
   =  {\k_{1/b} \over \k_b}
   W_{P}(\t_1,\t_2) .
}

    For   $|P|<1/2b$ or $|P|<b/2$ the two-point function satisfies
    non-linear  difference equations, which follow from (\ref{DEWa}) and (\ref{DEWb}):   
  \eqal\WEW{
   \label{WEW}
     \sin ( i \pi \p_{\t_1}) W_{ {1\over 2b}-|P|} (\t_1 ,\t_2) 
   =
     {\k_b ^2  \over    \left|  \sin (2\pi bP) \sin \(  { 2\pi P \over b}\)\right| }
 \   { M^{{1/ b^2} }\sinh\({ \t_1 / b^2} \)
 \over W_{P} (\t_1,\t_2) } \quad \(|P|\le{1\over 2b}\)
 }
 and
   \eqal\WEW{
  \label{WEWd}
   \sin ( i \pi b^2 \p_{\t_1})    W_{{b\over 2}-|P|}(\t_1 ,\t_2)    =
     {\k_b ^2 \over  \left|   \sin( 2\pi bP)\sin \(\  {2\pi P \over b}  \)\right|}
 \   { M^{}\sinh\({ \t_1 } \)\over W_{P}(\t_1,\t_2)} \quad \(|P|\le{b\over 2}\).
 }
  
 

 \subsection{The boundary one-point function (the disc loop amplitude)}

The boundary one-point function, or the disc loop amplitude, is
defined as 
     $$W(\t)= \<[ \CB   _{e_0}]^{\t\t}\>_{\rm disc}=
      - {\p\Phi\over \p z}, \qquad z = M\cosh\t  
  $$
  where $\Phi$ is the partition function of the disc
  with fixed boundary condition. It   is equal, up to a constant factor, to the Liouville amplitude  
  (\ref{Bbss}). 
  
  As in the case if the two-point function, the direct calculation 
  of $W(\t)$ from CFT is  difficult  because of the zero modes associated with the residual global conformal symmetry.
  We  can alternatively  evaluate $W(\t)$ using its relation with the 
  boundary two-point function\footnote{The easiest way 
  to prove that is to
  perform a Laplace transformation and 
  express both sides in terms of the disc amplitude with fixed length 
  $\ell = \int dx e^{ b\phi(x)} $.}
  \eqn\Weo{\label{Weo}
  W_{e_0/2}(\t_1, \t_2)=- {W(\t_1) - W(\t_2) \over z(\t_1) - z(\t_2)}.
  }
The explicit expression for
 $ W_{e_0/2}(\t_1, \t_2)$ follows from  (\ref{expls}):  
 \eqn\Weeo{\label{Weeo}
 W_{e_0/2}(\t_1,\t_2)=-    b \k_b \  {  M^{1/b^2-1} \over | \sin (\pi/b^2)|} 
{\cosh( \t_1/b^2)-\cosh( \t_2/b^2)\over
\cosh \t_1 -\cosh \t_2}.
}
 Comparing (\ref{Weeo}) and (\ref{Weo}) we get the expression for the
 disc loop amplitude
 \eqn\Wtau{
\label{Wtau}
 W(\t	) =  b \k_b \ {  M^{1/b^2} \cosh (\t/b^2)  \over |\sin \pi/b^2|}
.
}
In this section we are considering pure Neumann boundary conditions for the 
 matter field. The loop amplitude is proportional to the volume of the
momentum space,
 which is $2b$.  If we normalize the amplitude  by  the  volume of the 
 momentum space, the factor $b$ in the denominator will disappear.
Anticipating the result of the comparison with the discrete 
theory we will fix the value of the constant $\k$:
\eqn\kapp{
\label{kapp}
\k = {1\over b}.
}
 Note that the normalized amplitude is not self-dual.
 The T-dual amplitude is  given by
\eqn\TdW{
\label{TdW}
\tilde W(\tilde \t) = { 1\over b^2}\ 
{  \tilde M^{b^2} \cosh (\tilde \t b^2)    \over| \sin \pi b^2|}
=    {  1\over b^2}\  {M \cosh \t\over  |\sin \pi b^2|}
 .
 }

  \subsection{The difference equations  in 
   terms of  loop observables}
  
 To compare with the microscopic approach  we  should 
 translate   the 
 finite-difference equations 
 in terms of   boundary observables.

\bigskip

   \noindent
   a) when $ P_1\in [{1\over 2b},\infty]$
    and  $P_2\in [-\infty, -{1\over 2b}]$:
   \bigskip
     \eqal\FDEa{
   \label{FDEa}
  & W_{P_1,P_2,P_3}(\t_1, \t_2, \t_3 + i\pi)-W_{P_1,P_2,P_3}(\t_1, \t_2, \t_3 - i\pi)
   =\cr
   &
   \cr
   &
    \[ W(\t_3+i\pi) -  W(\t_3-i\pi)\] 
     W_{P_1-  {1\over 2b}  ,P_2+ {1\over 2b},P_3}(\t_1, \t_2, \t_3 )
 }

  \bigskip 

 \noindent
 a1) when $P_1\in  [{1\over 2b},\infty]$ and 
  $ P_2\in[ -{1\over 2b},0]$:
  \bigskip
  
\begin{eqnarray}
   \label{FDEaa}
 & & W_{P_1,P_2,P_3}(\t_1, \t_2, \t_3 + i\pi)
 -W_{P_1,P_2,P_3}(\t_1, \t_2, \t_3 - i\pi)
   =\cr
   & &
   \cr
    & &
   \cr
   & & 
 {1 \over   \left|\sin \(  {\pi\over b^2} - 2{\pi \over b} |P_2| \)\right|  }\
  {  W(\t_3+i\pi) -  W(\t_3-i\pi)\over
    W_{{1\over 2b}-|P_2|}( \t_3 ,\t_1)}\ 
     W_{P_1-  {1\over 2b}  ,P_2+ {1\over 2b},P_3}(\t_1, \t_2, \t_3 )
 \end{eqnarray}
 \bigskip 
 
   \noindent
 a2) when $P_1\in  [0, {1\over 2b}]$ and 
  $ P_2\in[ -\infty, -{1\over 2b}]$:

\begin{eqnarray}
   \label{FDEaab}
 &  &W_{P_1,P_2,P_3}(\t_1, \t_2, \t_3 + i\pi)
 -W_{P_1,P_2,P_3}(\t_1, \t_2, \t_3 - i\pi)
   =\cr
   & &
   \cr
    & &
   \cr
   & & {1 \over   \left|\sin \(  2\pi   b P_1\)\right| 
   }\ 
   {  W(\t_3+i\pi) -  W(\t_3-i\pi)  \over
    W_{{1\over 2b}-|P_1|}( \t_3 ,\t_2)}  \ 
     W_{P_1-  {1\over 2b}  ,P_2+ {1\over 2b},P_3}(\t_1, \t_2, \t_3 )
  \end{eqnarray}

\bigskip

\noindent
 a3) when $P_1 
 \in [ 0, {1\over 2b}]$ and $P_2\in [-{1\over 2b} ,0]$:

  \begin{eqnarray}
  &  &W_{P_1,P_2,P_3}(\t_1, \t_2, \t_3 + i\pi)
 -W_{P_1,P_2,P_3}(\t_1, \t_2, \t_3 - i\pi)
   =\cr
   & &
   \cr
    & &
   \cr
   & &
 {  W(\t_3+i\pi) -  W(\t_3-i\pi)  \over
    \left|\sin \(  2\pi   b P_1\)   \sin \(  {\pi\over b^2} - 2{\pi \over b} |P_2| \) \right| }\
  {1\over   W_{{1\over 2b}-|P_1|}( \t_3 ,\t_2)  
  W_{{1\over 2b}-|P_2|}( \t_3 ,\t_1)} \times\cr
  & & \cr
   & & \cr
  & &
  \times \ 
     W_{P_1-  {1\over 2b}  ,P_2+ {1\over 2b},P_3}(\t_1, \t_2, \t_3 )
     \label{FDEab}
      \end{eqnarray}
      \bigskip

  \bigskip 
  
 \noindent
   b) when   $ P_1
   \in [-\infty, -{b\over 2}]$
    and $P_2 \in [{b\over 2}, \infty]$:
   \bigskip
   \eqal\FDEb{
   \label{FDEb}
   & \tilde W_{P_1,P_2,P_3}( \tilde \t_1,  \tilde \t_2,  \tilde \t_3 + i\pi )
   - \tilde W_{P_1,P_2,P_3}( \tilde \t_1,  \tilde \t_2,  \tilde \t_3 - i\pi )
   =\cr
   &\cr
   &b^2 \  \[  \tilde W( \tilde \t_3+i\pi) -   \tilde W( \tilde \t_3-i\pi)\]
  \tilde  W_{P_1+  {b\over b}  ,P_2- {b\over 2},P_3}( \tilde \t_1,  \tilde \t_2, \tilde  \t_3 ).
 }

The cases b1), b2) and b3)  are related to a1), a2) and a3)  by duality.

\newsec{Microscopic formulation of 2D quantum gravity}

\subsection{SOS model on a dynamically triangulated disc}

The  functional integral of 2D Quantum gravity 
(\ref{ptfdisc})  is discretized by  considering the
statistical ensemble of all triangulations of the 
 world manifold $\CM$.
 Simultaneously, the matter field defined on $\CM$ 
 should be   discretized  as a 
 statistical  system  defined on   such a dynamical  triangulation\cite{discreteqg}. 
    
    Two-dimensional conformal field theories with central charge less than
one can be constructed as the continuum  limits of solvable two dimensional
statistical models on regular lattices. 
There are several microscopic formulations of the $c\leq 1$ conformal
field theories: as vertex models \cite{Baxter}, as height RSOS \cite{abf} and $ADE$
models \cite{vincent}, as loop models or as the $q$-state Potts model
or $O(n)$ models \cite{nienhuis}. 
Correspondingly, we have several solvable microscopic formulations of 
 the 2D quantum gravity with the  matter
central charge less than one.  The statistical models are in this
case formulated on a random lattice, the sum over geometries being
replaced by the sum over all the lattice realizations.
 As a result we obtain solvable models of quantum gravity, which can be 
 formulated also as  matrix models  \cite{isingi, Ion, adem}.  
 
   All these models  are  mapped onto 
 a special solid-on-solid (SOS) model with discrete unbounded
height variable $h\in \hf \Z$ and a Hamiltonian depending only on
height difference of nearby sites. 
The map is called Coulomb gas picture \cite{kkn, nienhuis}. 
The SOS model in question depends on a continuous parameter $g$, 
the Coulomb gas
coupling constant, and can be considered as a solvable microscopic
realization of the gaussian field with background charge $e_0\sim
g-1$.  This correspondence is very convenient since it allows to give
the operator identification of the fields already of the microscopic
level. 
Here we will    restrict ourselves to  the 
 microscopic formulation of 2D QG based on the SOS model.

We  start with  the definition of the SOS model on an arbitrary
triangulated surface. (The case of lattice paved  by squares   has been considered in \cite{ADEold}.) 
 The target
space of the SOS model is an infinite chain, the Dynkin graph of
$A_\infty$, whose nodes are labeled by half-integer {\it heights} $ h
\in \hf \Z$.  The structure of the target space is determined by its
adjacency matrix,
 $A_{hh'}$, whose matrix elements are one if $h$ and $h'$ are nearest 
neighbors and zero otherwise:
\eqn\adjm{ A_{h h'}=\delta_{h ,h '+\hf }+\delta_{h ,h'-\hf }.}
The adjacency matrix  is diagonalized in the momentum space 
 by the plane waves $S_h ^{(p)}=e^{2i \pi ph}$
\eqn\diagnaj{
 A_{hh'}= \int\limits_{-1}^1 d p \ e^{2i \pi p(h-h')}\ A_p, 
 \ \ A_p= 2\cos\pi p.} 
We will consider a non-unitary version of the SOS model, which is
characterized by a background target-space momentum $p_0$.  Consider a
triangulation $\CT$ of the disc that is dual to a   -trivalent planar
graph.  To each node $i\in \CT $ we associate a height $h _i\in \Z$. 
The partition function of the SOS model on $\CT$ is defined as the sum
over all height configurations $\{ i\to h_i| i\in \CT\}$. 

On the  flat lattice   the   
Boltzmann weight of a height configuration is 
a product of   factors
associated with the the triangles 
$\Delta_{ijk} $
\eqn\Wtr{
W_{\Delta} (h _1,h_2,h _3)=
T \ \delta_{h_1h_2} \delta_{h_2h_3}\delta_{h_3h_1} 
+ \delta_{h_1h_2} A_{h_2h_3} A_{h_3h_1} 
\( {S_{h_3} \over S_{h_1} }\) ^{1/6}
 +{\rm cyclic },
 \label{Wtr}
}
where  
\eqn\Shash{ S_h= e^{-2i\pi p_0
h}} 
 is the plane wave associated with the
 background  momentum  $p_0\in [0,
2]$.
The coupling   $T$    is  usually called  temperature.
  In presence of    curvature defects one should 
      also associate Boltzmann weights  with the nodes $j\in \CT$:
      \eqn\Wpt{
\label{Wpt}
W_{_{\bullet}}(h_i)=
  (S_{h_j})^{  \hat R_j/4\pi } } 
 where $\hat R_j$ is the curvature at the point $j\in\CT$.
As we will  see below, this factor is
 necessary to preserve the loop gas representation of the
partition and correlation functions.

The local curvature  is defined  as  the
deficit angle 
\eqn\currb{ \hat R_i= {2\pi\over 3} (6-c_i).
\label{currb}}
where $c_i$ denotes the coordination number at the point $i$ (the
number of triangles having this point as a vertex).  Similarly, the
boundary curvature at the point $j\in\p\CT$ is defined as
\eqn\bcurrb{
\label{bcurrb}
\hat K_j= {\pi\over 3} (3-c_j).} 
 By the Euler relation the total curvature
of the disc (including the curvature along the boundary) is 
\eqn\diseul{
\label{diseul}
\sum_{i\in\CT} \hat R_i + 2\sum_{i\in \p\CT} \hat K_i = 4\pi. 
}

   The partition function of the SOS model on the triangulation $\CT$ is
given by the sum over all height configurations:
\eqn\pfcmt{\CZ (\CT) =
\sum_{\{h_i| i\in \CT\}} \prod _{ i} W_{_{\bullet}}(h_i)
\prod_{\Delta_{ijk} } W_{\Delta}(h_i,h_j,h _k).
\label{pfcmt}} 
To define completely the partition function we should fix the boundary
conditions.  There are two natural choices of the boundary conditions
in the height model.  The fixed, or Dirichlet type boundary condition
prescribes a constant height $h$ along the boundary.  
The fixed boundary conditions where the hight takes a value $a$ on the
Dynkin graph  are  described by the type $(a,1)$ conformal boundary
conditions 
 \cite{BPPZ}\footnote{In a regular lattice the general $(a,r)$ boundary condition
is realized by attaching $r$ times fused weight to height $a$. 
This amounts to restricting the height variable to the 
 interval  $[a-r/r, a+r/2]$.  The condition type $(a,r)$ interpolates between the 
 fixed (D) and free (N)  boundary conditions.
  }
The free, or
Neumann type boundary condition prescribes to sum over all heights
along the boundary with weight one.

The  loop gas representation is obtained by  expanding the partition function (\ref{pfcmt})  as a sum of
monomials, each monomial corresponding  to choosing one of the four terms
in (\ref{Wtr}) for each triangle.  A neat geometrical
interpretation of the monomials can be achieved if we represent
graphically the r.h.s. of eqn.  (\ref{Wtr})\ as

\epsfxsize= 250pt
\eqn\tttrfig{\epsfbox{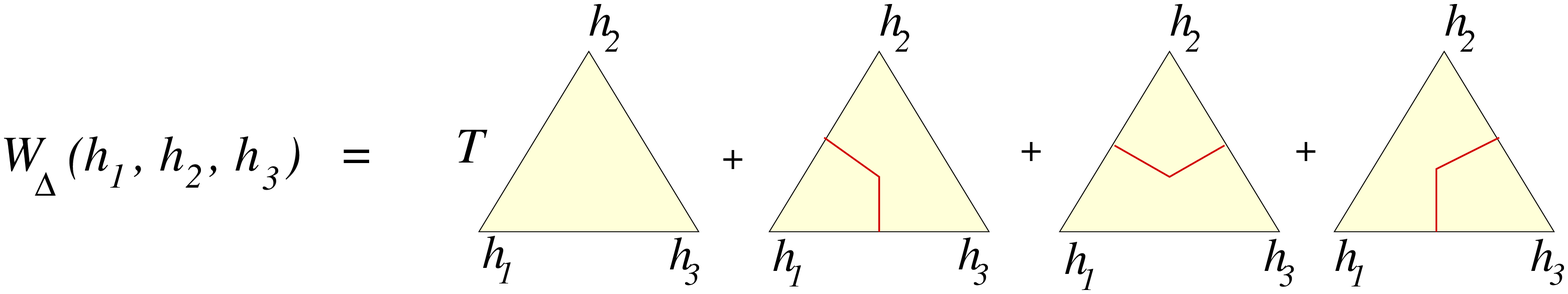}}

\noindent In the last three terms the lines separate the two vertices
with the same height from the vertex with different height.  Then a
monomial in the expansion of the partition function corresponds to a
collection of closed non intersecting loops on the dual trivalent graph
$\tilde \CT$:
 \bigskip \bigskip 

\centerline{\epsfxsize= 300pt\epsfbox{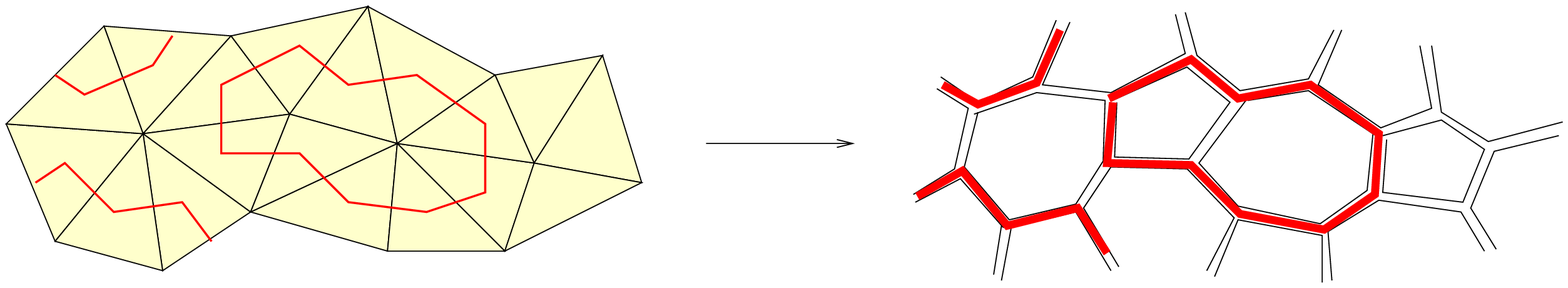}}
 \bigskip \bigskip
\noindent  The
loops can be considered as domain walls separating the domains of
constant height. The Dirichlet boundary condition means that 
the height   of the domain  adjacent to  the boundary   
 of $\CT$ is set to the value  $h_B$.

Let us consider  the case of a Dirichlet boundary condition $h=h_B$ at the boundary. 
The weight of a loop configuration is obtained by summing over all
possible domain heights.  The Boltzmann weight of a domain $\CD$ with
height $h=h_{\CD}$ is the product of the Boltzmann factors associated
with its nodes and triangles\footnote{We will assume that a triangle
belongs to given domain of constant height if at least two of its
three vertices are inside the domain.}.
By the Euler relation (\ref{diseul}) applied to  the  domain $\CD$,
%
%
%
the
total power of $S_h$ depends only on the number of boundaries of
$\CD$, which we denote by $N_B(\CD)$.  As a consequence, the partition
function (\ref{pfcmt}) with Dirichlet boundary condition $h_B$ can be
reformulated as a sum over the domain configurations
\eqn\lpgz{
\label{lpgz}
\CZ_{h _B} (\CT)=T^{\CA_{\rm tot}}\!\!\!\sum_{\ \ \ \
^{\rm loop \ configu-}_{\rm rations\ on\ \CT}} \ T^{-{\ell} _{\rm tot}} \
\sum_{ h_{\CD}\in \hf \Z} \ \ \prod_{\CD} \Big( S_{h_{\CD}}\Big)^{2-
N_B(\CD)} }
 where $ \ell_{\rm tot}$ is the total length of the loops, equal to the
 number of decorated triangles, and $\CA_{\rm tot}$ is the total number
 of triangles of $\CT$.  We remind that in the sum the heights of
 neighboring domains differ by $\pm \hf$.
 
 For any given domain configuration the sum over the heights can be
 performed using the fact that $S_x$ is an eigenvector of the
 adjacency matrix with eigenvalue $n = 2\cos\pi p_0$
\eqn\pby{
\label{pby}
\sum_{h'} A_{hh'} S_{h'} = 2\cos\pi p_0\  S_h  .}
 The summation starts with the innermost (simply connected) domains and
it generates a factor $n$ for each domain boundary.
After the summation over all the domain heights, one obtains
\def\xinfty{ {x(\infty)} } $n^{\# {\rm loops}} S_{h_B}$, where $h_B$
is the (fixed) height at the boundary.  Thus we obtain a representation of the
partition function as that of a gas of mutually- and self-avoiding
loops on the triangulation $\CT$, with fugacity $n = 2\cos p_0$ and
tension $ \log T$
\eqal\prlgf{
\label{prlgf}
 {\CZ _{h_B} (\CT) =  S_{h_B}\   T^{\CA_{\rm tot}} \ \sum_{\
 \ \ \ ^{\rm loop \ configu-}_{\rm rations\ on\ \CT}} T^{-\ell_{\rm
 tot} } \ ( 2\cos\pi p_0)^{ \# {\rm loops}} }\equiv 
 S_{h_{B}} \ \CZ(\CT) .}
If we fix, instead of the height, the momentum flowing through the
boundary, we have\footnote{A boundary state characterized by a fixed
height is also referred as a Cardy state, while the state with fixed
momentum $p$ flowing through the boundary is called Ishibashi state. 
The two kinds of states are related by Fourier transformation
$|p\rangle = \sum_{h\in\hf\Z} e^{2\pi i p h} |h\rangle$.}
 $$  \CZ ^{(p)}  (\CT) \equiv\sum_{h}e^{-2i\pi p  h_{B}}
  \CZ _{h_B} (\CT)= \delta(p- p_0)  \ \CZ (\CT)
  $$
Thus the ground state of the SOS model is translationally 
invariant in the target space only if $p_0=0$; otherwise there is a 
non-zero momentum flowing through the boundary.  

  The loop gas representation for Neumann  boundary 
condition has been considered in \cite{KKloop}.  
The free boundary condition means that the height can change by $\pm \hf$
at each point of the boundary.  To define a height configuration we
need to specify all heights along the boundary.  In this case, the
domain boundaries form not only closed loops, but also open oriented
lines with ends on the boundary.  There is an open line starting or
ending at each boundary link.  
After summing over the heights using eqn. (\ref{pby}), one obtains the
partition function as 
  the sum of all configurations of self- and
mutually avoiding loops and lines.  The closed loops   acquire  as before 
 factors  $n= 2\cos \pi p_0$ while  the factors associated with the open lines   
ending at the boundary are $2\cos( \pi p_0/2)$.  
We can also consider Neumann and Dirichlet
boundary conditions on different parts of the boundary.

%

%

The microscopic theory of gravity is constructed as the statistical
ensemble of all triangulations with given topology \cite{discreteqg}. 
The trivalent planar graphs dual to triangulations can be interpreted
as the Feynman graphs for the `t Hooft limit of a matrix model.  In
the case of the SOS and related models, such matrix model has been
constructed in \cite{adem}.

    Consider first the ensemble of all triangulated discs $\CT_{\ell,\CA}$ made
    of $\CA$ triangles and whose boundary has $\ell$ edges.  The disc SOS
    partition function with fixed height $h$ at the boundary is given
    by the average of the matter partition function (\ref{pfcmt}) \
\eqn\PhiSOS{\hat  W_h(\ell, \CA)=\sum_{\ \ \CT_{\ell,\CA}} 
\CZ_h(\CT)
 }
where we assumed that the triangulations have a marked point at the
boundary, so that there is no symmetry factor. 
 Introducing the bare
bulk and boundary cosmological constants 
$\Lambda   $ and $ z $, we
define the disc  partition function with Dirichlet boundary  condition  $h$ as
\eqn\barePhi{ \Phi _h(\Lambda, z  ) 
=\sum _{\ell=0}^\infty\
\sum_{\CA=1}^\infty \ e^{-\Lambda  \CA-z  \ell }
 \ { \hat W_h(\ell, \CA)\over \ell}. 
}
Using the loop representation (\ref{prlgf}), we can relate the disc
amplitude with fixed height to the partition function of the loop gas
on a random graph
 \eqn\Phih{\Phi_h(\Lambda, z  )=e^{2i\pi p_0 h} \
\Phi(\Lambda, z ),}
\eqn\PhiSOSb{ 
\Phi (\Lambda, z )= \sum_{\CT} {1\over \ell }\ e^{-\Lambda \CA- 
z  \ell}
\sum\limits_{{\rm loops\ on }\ \CT} (2\cos\pi p_0)^{[\#\ {\rm loops}]} 
\ \ e^{- 2M\ell _{\rm loops}} 
}
where  $\CA$ is the area of  the triangulation $\CT$, $\ell$ is the
length 
of its boundary  and   $M  \equiv  \hf\log T$.
 The factor $1/\ell$ takes into account  the cyclic symmetry.

  \subsection{Continuum limit and mapping to Liouville gravity}

  In the case when $\CT$ is the regular triangular lattice, the loop
model (\ref{prlgf}) has been solved by Bethe Ansatz technique
\cite{nienhuis, Baxter}.  The loop model is critical in the interval $0<T<T_c
$ where its continuum limit is described by a conformal theory with
central charge (we  assume that $p_0>0$)
\eqn\cdense{ 
\label{cdense}
c _{_{\rm dense\ phase }}
=1-6p_0^2/(1-p_0).} 
 This is the dense or low-temperature phase of the loop gas.  In this
 phase the loops occupy almost all the space, $\ell^{\rm tot} \sim \CA_{\rm
 tot}$.  At the critical temperature\footnote{The exact value of the
 critical temperature is $T_c=2\cos (\pi p_0/4) $} the loop gas
 exhibits a different critical behavior (the dilute phase) described
 by a larger central charge
\eqn\cdilute{
\label{cdilute}
c_{_{\rm dilute\ phase
}}=1-6p_0^2/(1+p_0).} 
In the dilute phase both the total length of the loops and the empty
space diverge correspondingly as $ {\CA_{\rm empty }}\sim {\CA_{\rm tot}}$
and $\ell^{\rm tot} \sim \sqrt{\CA_{\rm tot}}$.  Above the critical
temperature the  loop gas is not critical, {\rm i.e.} there is a
finite correlation length determined by the typical size of the loops.
A more
detailed identification of the continuum limit has been made in \cite{
kkn , nienhuis}, where it has been argued that the renormalization
group trajectories of the SOS model flow to a gaussian field with
electric charge
at infinity.

With Boltzmann weights defined by 
(\ref{Wtr})-(\ref{Wpt}) this statement can
be generalized to the case of a lattice with curvature defects \cite{ADEold}.  
More precisely, the SOS model renormalized at large distances to 
the gaussian field   (\ref{actg}),  with the identification %
\eqn\fieldschih{
\label{fieldschih}
e_0\chi =
 2\pi p_0 h.} 
Indeed,  the effect of the curvature is to associate  an additional
weight  factor
\eqn\latR{ \exp
 (-{i\over 2\pi } \sum_j R_j \pi h_jp_0)} 
  with each  height configuration.  The logarithm of
 this factor represents a discretization of the linear term of the
 gaussian action
 (\ref{actg}), which leads to the identification %
(\ref{fieldschih}). From (\ref{cdense}) and (\ref{cdilute})  one gets, assuming 
that $0<b<1$,
\eqal\bversp{
\label{bversp}
 &e_0= p_0/b, \ \ \chi = 2\pi b h \qquad 
 &{\rm in \ the \ dense \ phase}; \no \\ 
 &e_0= b p_0, \ \ \chi =2 \pi h
 /b\qquad &{\rm in \ the \ dilute  \ phase. }}
Thus  the gaussian field $\chi$  with background charge $e_0$ gives 
 the effective field theory for the large-distance behavior of the SOS model
with background  target-space momentum $p_0$, with the identifications 
(\ref{fieldschih}) and (\ref{bversp}) . 

In the same way as all $c \le 1$ conformal field theories are mapped
onto the gaussian field, one can map the underlying solvable spin
models ($ADE$, $O(n)$, Potts, ...)  onto the SOS model in the sense
that the partition and correlation functions can be interpreted in
terms of distributions of electric and magnetic charges.  This mapping
is known as Coulomb gas picture \cite{ kkn, nienhuis}. 
Traditionally one uses the Coulomb gas coupling constant
\eqal\cgcoupl{g=  1- p_0 =\ 
b^2&& \ \ \  {\rm in\  the\  dense\  phase};\no \\
 g=1+p_0= 1/b^{2} && \ \ {\rm in \ the\  dilute\  phase.}}
Then (\ref{bversp}) reads in both phases
\eqn\chihash{
\label{chihash}
 e_0 = p_0/\sqrt{g}, \quad \chi =2 \pi \sqrt{g} \ h.
 }
  
%

The SOS model  and related models on a random graph have been solved in
\cite{ADEold} \ (for the dense phase) and in \cite{Idis} (for the
dilute and the multicritical phases).  
The bulk critical   behavior due to    triangulations with divergent area
is achieved along a line
in the two-dimensional space of coupling constants $\{ \Lambda, M\}$ ending at 
 a tricritical point   $\{\Lambda^*, M^*\}$.    The
critical line describes the dense phase of the SOS model coupled to
gravity, and the tricritical point -- the  dilute phase.

From now on we will consider only the scaling regime and will denote by 
the same symbols $\Lambda$,
$M$ and $z$   correspondingly the  renormalized  cosmological
constant,
 loop tension and boundary cosmological constant, correspondingly.
%
 The scaling of the bulk and boundary cosmological constants
with the renormalized loop tension $2M$ is \cite{ Idis, KKloop}

\eqal\BBscbis{
\label{BBscbis}
 \Lambda^{\rm dilute} \sim M^2, &&\qquad 
\Lambda^{\rm dense} \sim M^{2g}\no \\
z^{\rm Dirichlet} \sim M, &&\qquad
z^{\rm Neumann } \sim M^g \qquad (\rm in \ both\ phases).}

   \noindent
 The Dirichlet boundary can be considered itself as a domain wall,
 which explains why $\mu _B^{\rm Dirichlet} $ scales as the domain
 wall tension.   
The scaling (\ref{BBscbis})  leads to the following identification of
the bulk and boundary Liouville interactions: \smallskip

\eqn\SSscaling{
\begin{matrix}
& &   _{ \rm Bulk}   & _ { \rm Dirichlet  \ boundary}  &_{\rm Neumann \ 
boundary}\\
    & & & \\
       {\rm Dilute\ phase}: &\ g= 1/b^2>1: & \mu\ e^{2 b\phi } & 
\mu_B \ 
e^{b\phi} & \tilde\mu_B \
e^{\phi/b}\\
 {\rm Dense\ phase}: &\ g=b^2 <1: & \mu\ e^{2b\phi} &\tilde \mu_B \ 
e^{\phi/b} & \mu_B \ 
e^{b\phi}
\end{matrix}
 }
 \\
 \\

  \noindent  
Each value of the central charge in the interval $-2\le c\le 1$ is
realized both in the  dense and in the dilute phase of the SOS
model.  The phase that connects to the continuum path integral is the
dilute phase with Dirichlet boundary or the dense phase with Neumann
boundary.  The dilute and dense phases are related by a T-duality
transformation for the matter field.
In order to have identical conventions with the continuous approach,
 we normalize the bulk cosmological  constant so that  
$\Lambda^{\rm dilute} = M^2$ and $ \Lambda^{\rm dense} = M^{2g}$.

The correlation functions of the boundary operators are meromorphic
 with respect to the boundary cosmological constants $z$, with a
 cut $[-\infty, - M]$ along the real axis. 
%
     When the boundary cosmological constant 
 approaches the value $-M$, the partition function diverges
 because of the dominance of world sheets with infinite boundary
 length or, in terms of the Liouville path integral, because of the
 diverging integral with respect to the zero mode of the Liouville
 field at the boundary. The branch point 
 is resolved by  by an
uniformization map 
\eqn\mubt{ z= M \cosh \tau.} 
 The complex $z$-plane is mapped
to the strip $| \Im \t | < \pi$ with the points $\t$ and $-\t$
identified.  The two sides of the strip parametrize the two edges of
the cut.  The branch point corresponds to the points $\t= i \pi \equiv
-i\pi $.

  \subsection{Bulk and boundary operators in the SOS model}

 \bigskip

  \paragraph{\bf Bulk electric  and magnetic operators:}
  The electric charge in the SOS model  corresponds to the  target space momentum
  $p$.  Since the momentum space is periodic, $p+2\equiv p$, we can
  restrict $p\in[0, 2]$.  The electric charge is carried by the electric, or {\it vertex},
  field
\eqn\vertSOS{
\label{vertSOS}
V^{(p)}(j) =
  e^{2 i \pi (p_0-p) h_j}.} %
  In terms of the
  loop gas the electric operator (\ref{vertSOS}) changes the weight of
  the non-contractible loops, {\it i.e.} those encircling the point
  where the operator is inserted.  Instead of $n=2\cos\pi p_0$, each
  non-contractible loop gets a factor $2\cos\pi p$ \cite{nienhuis}.
  It follows from the matching of charges
  (\ref{bversp})  that the bulk dimensions of the SOS vertex operators
  (\ref{vertSOS})  are
  \eqn\DimvSOS{ \Delta_{V^{(p)}}= {p^2- p_0^2\over 4g}.}

Unlike the electric operators, the
  magnetic, or {\it vortex}, operators  $S^{(L)}$   
  have infinite but discrete spectrum $L\in \Z$.
    In terms of the loop gas these operators  are the
   ``star polymers" \cite{dsds, Dstar },   representing
    configurations of $L$
open lines starting at  the same 
 point. Strictly speaking, the magnetic 
 operator $S^{(L)}$  on the lattice 
is associated with a  hole with boundary  of length $L$, with 
an open line starting at each edge.
For example,  the configurations 
associated with  the operator 
$S^{(3)}$  look as follows:

 \epsfxsize= 150pt
 \centerline{\epsfbox{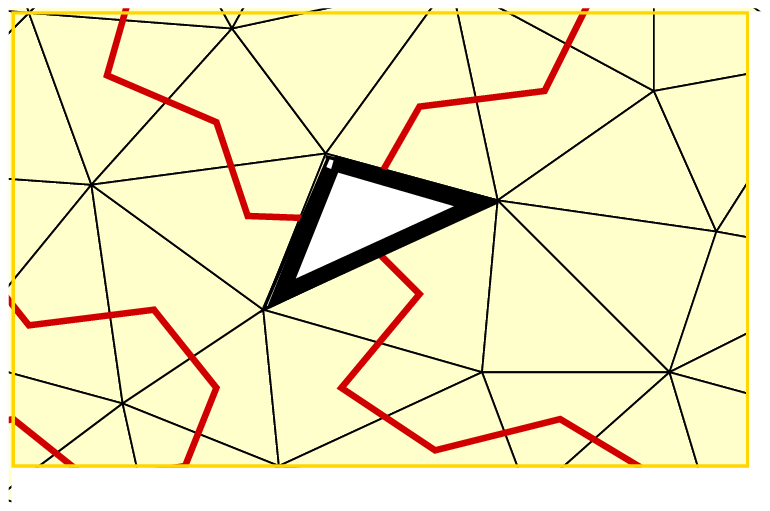} }

 Since in the SOS model each line represents a domain wall separating two 
 neighboring heights, the  magnetic  operator  creates a  
 a discontinuity $\delta h$  around the small loop, which can take values
 $\delta h = -{L\over 2} , -{L\over 2}+1, ..., {L\over 2}$.
 When $\delta h = \pm {L\over 2}$, the operator  $S^{(L)}$
 can be identified with the  spinless magnetic
operator $\CO_{e_0, m}$ for the gaussian field, with $m= \pm L\sqrt{g}/4$ and conformal
dimension\footnote{In the rational points $b^2=p/(p+1)$ the star
operators correspond to the magnetic operators $\phi_{N-1-n, N}$ of
the minimal models, with $N= {p+1\over 2}$.}
\eqn\DflatM{\Delta _{S^{(L)}}= \bar \Delta _{S^{(L)}}
= {g^2 L^2/4 - p_0^2 \over 4g}=
{  \begin{cases}
 \Delta_{L/2,0} \ \  \ {\rm in \ the\  dilute\  phase}\   (g>1)\\
  \Delta_{0, L/2} \ \ \ {\rm in\ the\ dense  \ phase} \ (g<1).
  \end{cases}}}

   \paragraph{\bf Boundary electric   operators:}   The boundary
 vertex operator is defined by inserting the vertex field
\eqn\BvxSOS{ V_p(j) = e^{2\pi i( \hf p_0 -p)h_j}}
  at a point on the boundary 
  with Neumann boundary conditions  on both sides.
%
%
In terms of the loop gas the boundary vertex operator changes the
 weight of the noncontractible lines, {\it i.e.} the lines that
 connect the boundary on its left and the right.  These loops get
 additional weight factor $ \cos (\pi p)/\cos (\pi p_0/2)$.
 Its conformal dimension  follows from (\ref{DeltaBe}):
\eqn\DbvxSOS{ \Delta_{V_p}= {4p^2 - p_0^2\over 4g.} 
  }

 \paragraph{\bf  Boundary magnetic operators:}

   The boundary star operator $S_L$ creates $L$ lines at some point of
  the boundary with Dirichlet boundary conditions:
  
  \bigskip
  \epsfxsize= 150pt
 \centerline{\epsfbox{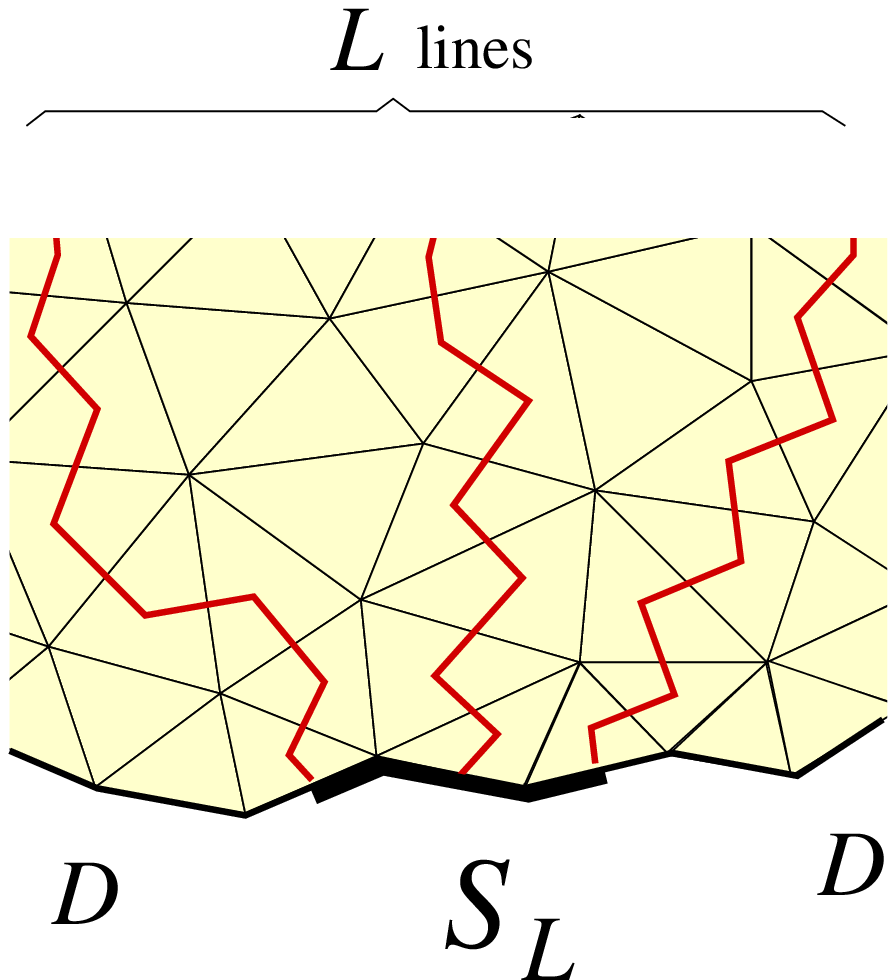} }
 \noindent 
  The operator $S_L$ implies a discontinuity of the boundary  height
$\delta h = h_{\rm right}-h_{\rm left}$, which can take values 
 $\delta h = -{L\over 2} , -{L\over 2}+1, ..., {L\over 2}$.
  Again, when  $\delta_h = \pm {L\over 2} $ the operator $S_L$
  can be identified with the boundary magnetic operator for
  the gaussian field with magnetic charge
  $m= \pm L\sqrt{g}/2$. 
 Therefore its  conformal dimension is given by  (\ref{DeltaBm}):   
\eqn\DflatMb{
 \Delta_{S_L}= { (L g- p_0 )^2 - p_0^2\over 4g}=
  \begin{cases}  \Delta_{L+1,1} \ \  {\rm in\  the\  dilute\  phase}  
\ (g>1)\\
  \Delta_{1, L+1} \ \ {\rm in\ the\ dense\ phase} \ (g<1).
  \end{cases}
  }

  \bigskip
  
   \noindent
  \paragraph{\bf  Boundary  twist operator:}
  The boundary twist operators  are  associated with the points 
separating the Dirichlet and Neumann type boundaries:
  \vskip 1cm
  
   \epsfxsize= 150pt
 \centerline{\epsfbox{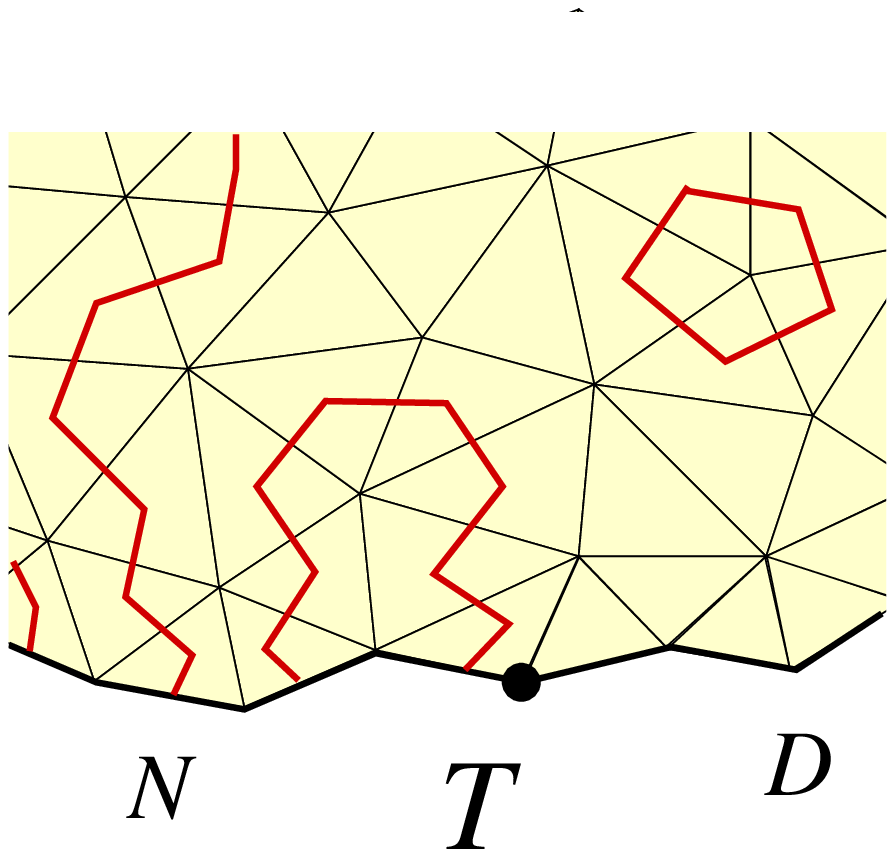} }
 \bigskip
 \noindent 

To our knowledge,  the boundary twist operator operator has not been
discussed
 in the context of the loop gas on a flat lattice. 
It seems to have a simple description in terms of the gaussian field only for 
$b=3/4$ and $b=1$. For generic $b$ we    found  its flat  scaling
dimension from 
  its gravitational  dimension   using   the KPZ scaling formula:
\eqn\gtwist{
\label{gtwist}
\Delta_T = \({1\over 4b } \)^2     - {e_0^2\over 4} 
= {(3- 2b^2)(2b^2-1)\over 16b^2}.  }  
%
%
 Thus the boundary twist operator is not a degenerate field,
 but can be classified as a ``half-degenerate" field with $r=\hf,
 s=0$.  
 Note however that in  the case of  a minimal model $b={p\over p+1}$ with
 $p$ odd, the boundary twist operator can be identified with the
diagonal
 operator $\{r,s\}=\{ p+1, p+1\}$ of the Kac table. For example, in the Ising model
  ($b=\sqrt{3/4}$) this is the order operator $\s$ ($r=2,s=2$)  with  
   dimension  $1/16$.
    
We will also consider excited boundary twist operators $T_L$ 
obtained by  the fusion of a  twist operator $T$ and a star operator $S_L$:
\bigskip

   \epsfxsize= 100pt
 \centerline{\epsfbox{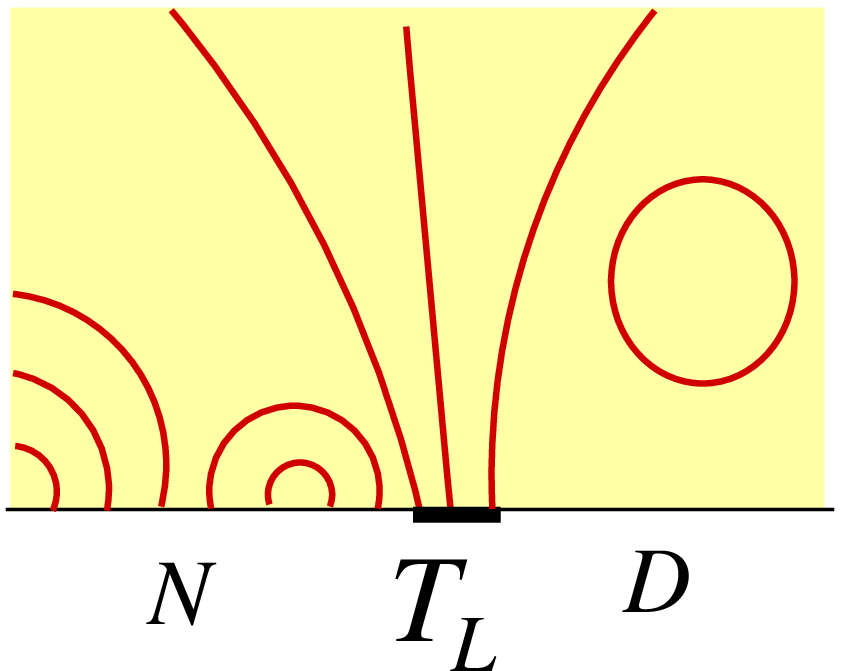} }
 \bigskip
 \noindent 
The  dimension of such operator is
 \eqn\dimTL{
  \Delta_{T_L}= \frac{  (L+\hf)^2/b^2 -  ( b-1/b)^2 }{4}=
  \Delta_{L+\hf, 0}.}
 Here again  we   extracted the conformal 
 dimensions   from the two-point correlation function 
 in 2D QG using the KPZ relation. 
 
 %

\newsec{Difference equations    for the boundary correlators 
in   the  microscopic  approach}

\subsection{Factorization property of the sum over geometries}

The boundary correlators in the SOS model satisfy a set of loop
equations, which can be restated as functional equations for the
boundary $\t$-parameters.  These equations are consequence of the
remarkable factorization property of the functional measure of the
loop gas on a random surface.

Let $\G$ be a self-avoiding line on the disc $\CM$ connecting two
points on the boundary.  It splits the disc $\CM$ into two pieces
$\CM_1$ and $\CM_2$ such that $\CM=\CM_1\cup \CM_2 $.  We can first
perform the integral with respect to all configurations of the
self-avoiding line $\G$, and then integrate with respect to all
metrics $g_{ab}$ on $\CM$.  Alternatively, cutting open the path
integral along the line $\G$, we can integrate first with respect to
the metrics on $\CM_1$ and $\CM_2$ and then with respect to the
configurations of the separating curve $\G$.  Due to the general
covariance of the measure, the result of the integration depends only
on the length $\ell$ of the curve $\G$.  Therefore the path integral
over the configurations of the curve $\Gamma$ reduces to
one-dimensional integral along its length $\ell$.  Thus we have the
following factorization formula \eqn\factrz{\int \CD g_{ab} (\CM)
\int\limits_\CM \CD \G= \int _0^\infty d\ell\ \CD g_{ab} (\CM_1) \int
\CD g_{ab} (\CM_2) } where $\CD\G$ is the measure over self-avoiding
lines $\G\in \CM$ with fixed endpoints.  This formula is quite obvious
from the microscopic realization of the measure as a sum over
triangulations (see \cite{ADEold} for details).

  Now we use the fact that the SOS partition function on $\CM$ in
  presence of such line factorizes into a product of two SOS partition
  functions on $\CM_1$ and on $\CM_2$ with Dirichlet boundary
  conditions along $\CL$: \eqn\factrSOS{ \CZ(\CM_1\cup \CM_2) =
  \CZ(\CM_1)\CZ(\CM_2).} As a consequence, we can represent the
  partition function of the SOS model in presence of a domain wall as
  an integral with respect to the length of the domain of the product
  of the SOS partition functions, for which the domain has the meaning
  of a Dirichlet boundary.

\subsection{ Disc loop amplitude }

The most basic observable is the disc loop amplitude $\hat W_h (\ell)$
, defined as the partition function of the loop gas on a random
surface with the topology of a disc with fixed  height $h$ on the boundary
(Dirichlet boundary condition),
  boundary  length $\ell$  and a marked point at the boundary \cite{ADEold}.  
 Since we are concerned only by the scaling limit, by
length we understand the renormalized length of loops or Dirichlet
boundaries.  The length is related to the constant mode of the
Liouville field on the boundary as $$\ell = \begin{cases}e^{b\phi_0} \
\ { \rm in\ the\ dilute\ phase},\\
  e^{\phi_0/b}  \ \ {\rm in \ the \ dense\ phase.}
  \end{cases}
  $$
 It follows from   eq. (\ref{lpgz})  that  $\hat W_h (\ell )$
  and its  Laplace transform 
  \eqal\partfBz{
W(z)& = \int_0^\infty d\ell \ e^{-z\ell}\
\hat W(\ell)= {\p\Phi_h(z)\over \p z}
}
 depend on the  boundary height  as
 $$
 \hat W_h (\ell )= S_{h} \  \hat W(\ell ), \qquad   W_h (z)= S_{h}\  
W(z),
$$
(where $\hat W(\ell)$ and $W(z)$ are the loop gas disc partition functions   
with a  marked point at the boundary).

The disc amplitude satisfies a quadratic equation of motion
\cite{ADEold, Idis}.  Here we give a heuristic  derivation  of this
equation in the continuum limit.  Consider first the ensemble of empty
world sheets, called pure gravity.  It corresponds to the values
$g=1/2\ (c=-2)$ and $g=3/2\ (c=0)$ of the SOS model.  It is known that
in this case the loop amplitude $W(z)$ satisfies a quadratic equation
\footnote{We have shifted $W(z)$ by a polynomial, which is irrelevant
in the continuum limit.}
$$ 
W^2(z)=  P(z),  
$$
where $P(z)$ is a quadratic polynomial in $z$, whose coefficients
depend on the bare couplings $M_0$ and $ \mu_0$.  Written in the
$\ell$-representation\footnote{Here we require that $\ell$ is strictly
positive, because we are not interested in the singular at $\ell=0$
terms.  } \eqn\loopeqCG{
\label{loopeqCG}
  \int_0^\ell d\ell' \ \hat W(\ell') \hat W(\ell-\ell') =0 \qquad
  (\ell >0) } this equation has the following meaning: a small
  variation of the world sheet metric at the marked point produces a
  contact term associated with the degenerate world-sheet geometries,
  for which another point of the boundary is at zero distance from the
  marked point.  The integral can be regarded as the total
  contribution of these degenerate geometries, for which the integral
  over metrics factorizes into a product of two disc amplitudes.  We
  can write (\ref{loopeqCG}) symbolically as \epsfxsize= 80pt
  \eqn\pictW{
\label{pictW}
 \epsfbox{   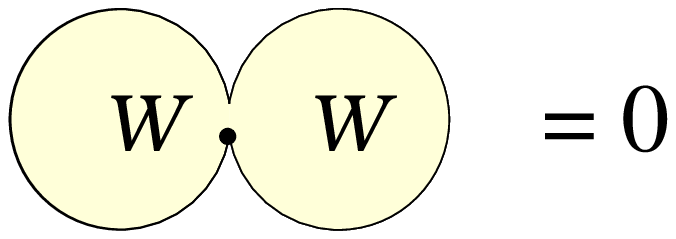 }}

Now consider the general case of matter coupled to gravity.  Apart
from the contact term that comes from the degenerate geometries, there
is a second contact term associated with a loop coming close to the
marked point\footnote{This contact term is in fact the boundary
one-point function of the energy operator of the loop gas.  The energy
operator is primary in the bulk, but appears as the descendent of the
identity operator when classified according to the irreducible
representations of the boundary Virasoro algebra \cite{cardy}.  }.  By
the factorization property we mentioned above, the contribution of
such loops is given by an integral over their length times the loop
fugacity $n= -2\cos\pi g$
\eqn\lopeqellW{
   \int_0^\ell d\ell' 
 \hat W(\ell')  \hat W(\ell-\ell') 
 - 2\cos \pi g \ \int_{0}^\infty d\ell' \ \hat W(\ell')\hat W(\ell + 
\ell')=0 \qquad (\ell >0}
  or pictorially
  \epsfxsize= 170pt 
%
%
\eqn\pictW{ \epsfbox{   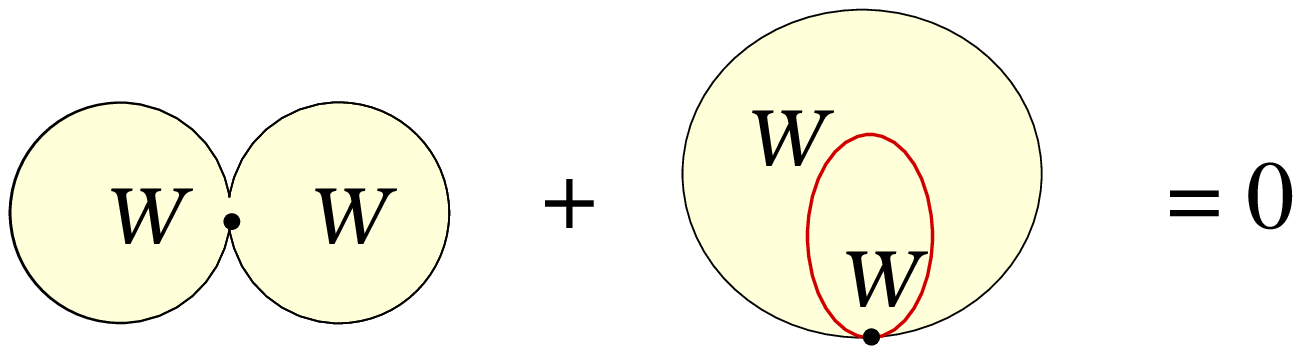 }}
\noindent The Laplace-transformed loop amplitude $W(z)$ is a rational
function of $z$ with a cut extending between the simple branch point
at $z=-M$ and a branch point of (generically) infinite order at
infinity.  The position of the right branch point is related to the
tension at the boundary, $ \hat W(\ell) \sim e^{-M\ell}$.  It is
consistent to assume that the tension at the boundary is equal to half
of the tension of the loops in the bulk.  Then weight factor
$e^{-2M\ell}$ of the loop in the second term of (\ref{pictW}) gives
the correct boundary tension for the loop amplitudes associated with
the internal and the external domains.  In terms of boundary
cosmological constants, eq.  (\ref{pictW}) reads\footnote{The integral
is divergent at infinity, but we assume that the amplitude $W(z)$ is
regularized so that the cut ends at a second branch point at large but
finite distance.  The integration then can be done using the Cauchy
formula, and then the cutoff may be lifted.  The rigorous procedure is
explained in \cite{ADEold}\ and \cite{Idis}.}
\eqn\loopqW{
\label{loopqW}
  W^2(z) - 2\cos\pi g\ \int {dz'\over 2\pi i }\ {W(z) -W(z') \over
  z-z'} W(-z')=0 } where the contour of integration encircles the cut
  $-\infty<z<-M$.  Then equation (\ref{loopqW}) implies the following
  condition relating the real and imaginary parts of $W(z)$ along the
  real axis
\eqn\cutD{
\label{cutD}
 \Im W(z)[W(z+i0) +W(z-i0)- 2\cos\pi g \ W(-z)]=0 , \ \  (z\in \R).}

In the parametrization $z=M\cosh\t$,  which unfolds  the branch point at
$z=-M$, the upper and the lower sides of the cut correspond to the
lines $\tau+i\pi$ and $\tau-i\pi$, $0<\tau<\infty$.  Therefore
\ref{cutD} can be written as the following finite-difference equation
for the analytic function $W(\tau) \equiv W(z(\tau))$
\eqn\eqW{
\label{eqW}
 W(\t+i\pi) +W(\t-i\pi) - 2\cos \pi g\ W(\t)=0}
which is solved  up to a constant factor by
\eqn\Wottau{
\label{Wottau}
W(\t)= {1\over   \sin \pi g }\ M^g \cosh (g\tau) .} 

 \subsection{Bulk one-point function  }

The bulk one-point function $U^{(p)}_h(z)$  is the partition function
of the loop gas on the punctured disc with a vertex operator $V_p$
 inserted at the puncture and Dirichlet boundary condition
characterized by a height $h$ and boundary cosmological constant $z=
M\cosh\t$.  It has the form 
$$ U ^{(p)}_h(z) =
e^{i\pi p h} U ^{(p)}(z)
 $$
where $U ^{(p)}(z)$ is the loop gas partition function on the disc in which 
the   fugacity of the loops encircling the puncture is  changed to $2\cos \pi p $
\cite{ADEold}.
In order to write the loop equation, we should first differentiate
with respect to $z$ in order to create a marked point on the boundary. 
The derivative 
\eqn\dUdz{R^{(p)}(z)=\p U ^{(p)}(z)/\p z,} which is
also a bulk-boundary correlation function, satisfies the following
integral equation \cite{ADEold}
\eqal\loopqzU{
 2 R ^{(p)}(z) W(z) &+&
 2\cos \pi p_0 \  \int {dz'\over 2\pi i }\ { R^{(p)}(z) - R 
^{(p)}(z') 
\over z-z'} W(-z') \cr
&+ &2\cos \pi p \  \int {dz'\over 2\pi i }\ {W(z) - W(z') 
\over z-z'}   R^{(p)}(-z')
=0
 }
  or pictorially
  \epsfxsize= 300pt 
\eqn\pictU{ 
\label{pictU}
\epsfbox{   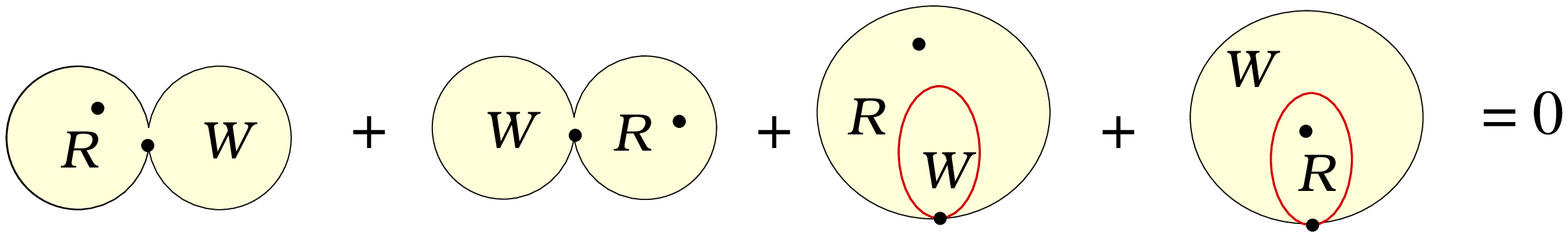 }}
From here we obtain,  for $z\in \R$,  
\eqal\cutDU{ & \Im W(z)[ R ^{(p)}(z+i0) + R ^{(p)}(z-i0)+2\cos\pi p
\ R ^{(p)}(-z)]\cr 
& + \Im R ^{(p)}[W (z+i0) + W(z-i0)+2\cos\pi p_0 \
W(-z)] =0.} 
The second line of vanishes due to (\ref{eqW}) and one
finally obtains the following functional equation in the
$\tau$-parametrization
\eqn\eqUtau{
 R ^{(p)}(\t+i\pi) +R ^{(p)}(\t-i\pi) +2\cos \pi p\ R^{(p)}(\t)=0.}
The solution of this equation  with the correct scaling is given by
\eqn\Rtau{R ^{(p)}(\t)=
\p_zU ^{(p)}\sim M^{|p|-1}  {\sinh  p\tau\over \sinh \tau}.
}
This expression is to be compared  with the expression (\ref{Rabs}) 
found from Liouville theory.
Integrating with respect to $z$, we find
\eqn\Utau{U ^{(p)}(\t) \sim M^{|p|}\cosh p\t  .
}
  in accordance with (\ref{Upsigma}).

One can easily write loop equations like (\ref{pictW}) and
(\ref{pictU}) for any boundary correlation function of electric and
magnetic operators under the condition that there should be at least
one with Dirichlet boundary condition for the SOS field.  The r.h.s.
will contain a contact term due to a pinching of the disc, another
term associated with the energy operator at the boundary, and a number
of contact terms, each associated with a line or a non-contractible
loop touching the marked point.
 
   In the next section we will consider another class of
functional equations, which 
  follow only from the  factorization property of the  
measure. These equations  will be shown to be equivalent to the 
 difference equations in Liouville gravity.

 \subsection{The boundary two-point function of magnetic operators 
}

In this subsection we recall  the  calculation  of the 
 boundary two-point function of magnetic, or star,  
 operators presented in \cite{Ibliou}.
To begin with, consider the  simplest example of   the disc
amplitude with two marked points and two different boundary
cosmological constants, 
\eqn\Wtwocc{
\label{Wtwocc}
W(z_1, z_2)= \int_0^\infty d\ell \ e^{-z_1\ell_1 -z_2\ell_2}\
\hat W(\ell_1+\ell_2)={W(z_1)-W(z_2)\over z_1-z_2}.}
This is a particular case ($L=0$) of the correlation function of two
gravitationally dressed star operators ${ \bf S}_L $
 \eqn\SScorf{D _L(\t_1,\t_2)=  \<{ \bf S}_L^{\t_1\t_2}{\bf  
S}_L^{\t_2\t_1}\>.
    } 
In the loop gas representation, this correlation function is the loop
gas partition function in presence of $L$ open non-intersecting lines
connecting two points on the boundary:
\eqn\SLeps{
 \epsfxsize=200pt
\epsfbox{  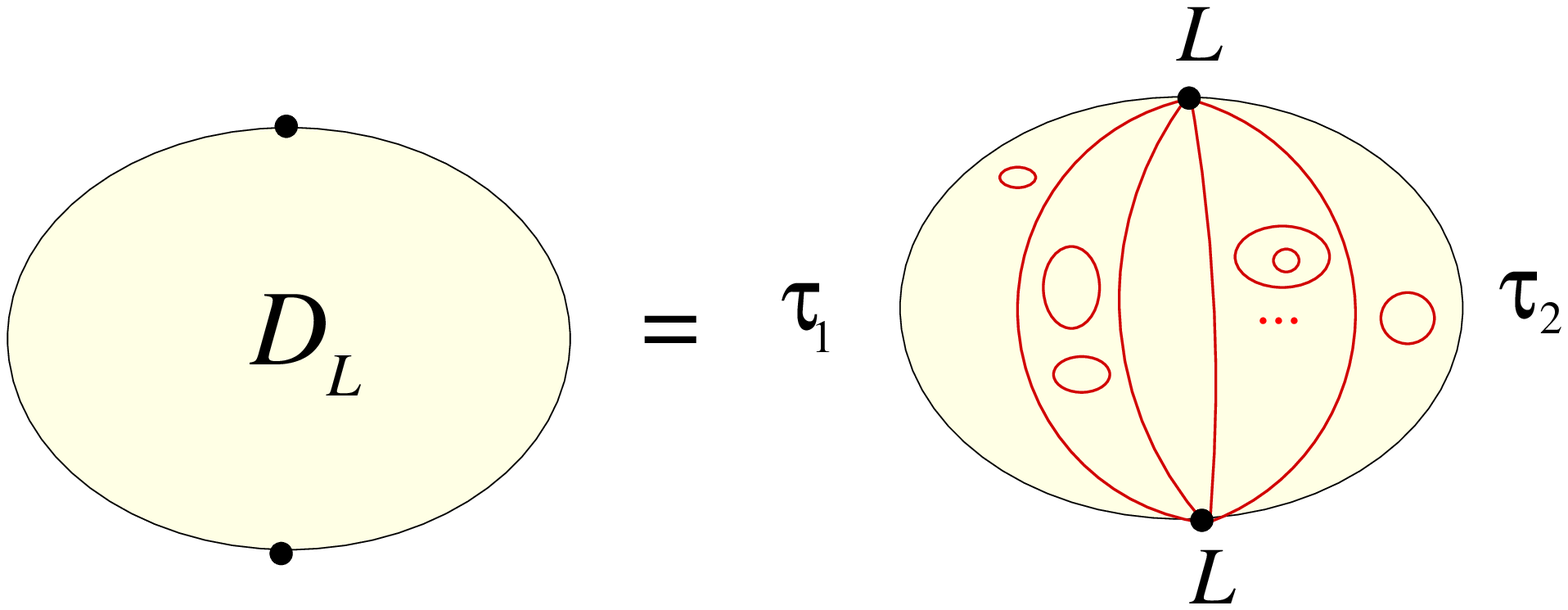  }
  }
 The inverse Laplace transform of $D_L$, which is the partition
 function for fixed lengths $\ell_1$ and $\ell_2$ of the two
 boundaries
\eqn\WlW{ D _L(\t_1,\t_2) = \int_0^\infty d\ell_1 d\ell_2 \
e^{{-z_1\ell_1 -z_2\ell_2}}\  \hat D _L(\ell_1, \ell_2)} 
can be
expressed through the disc partition function $W(\ell)$ as follows. 
Consider the sum over all loop configurations with fixed lengths
$\ell_1',..., \ell_{L}'$ of the branches of the star polymers, which
is given by the product of $L+1$ disc partition functions, and then
integrate with respect to the lengths
\eqn\IntWL{
\label{IntWL}
 \hat D _L(\ell_1, \ell_2)= \int_0^\infty
 d\ell_1'  ... d\ell_{L}' \
\hat W(\ell_1+\ell_1')\hat W(\ell_1'+\ell_2')...
\hat W(\ell_{L}'+\ell_2).}

The integral representation (\ref{IntWL}) is equivalent to the
recurrence relation between $\hat D _L$ and $\hat D _{L-1}$
\eqn
\SLeps{ \epsfxsize=190pt \epsfbox{ 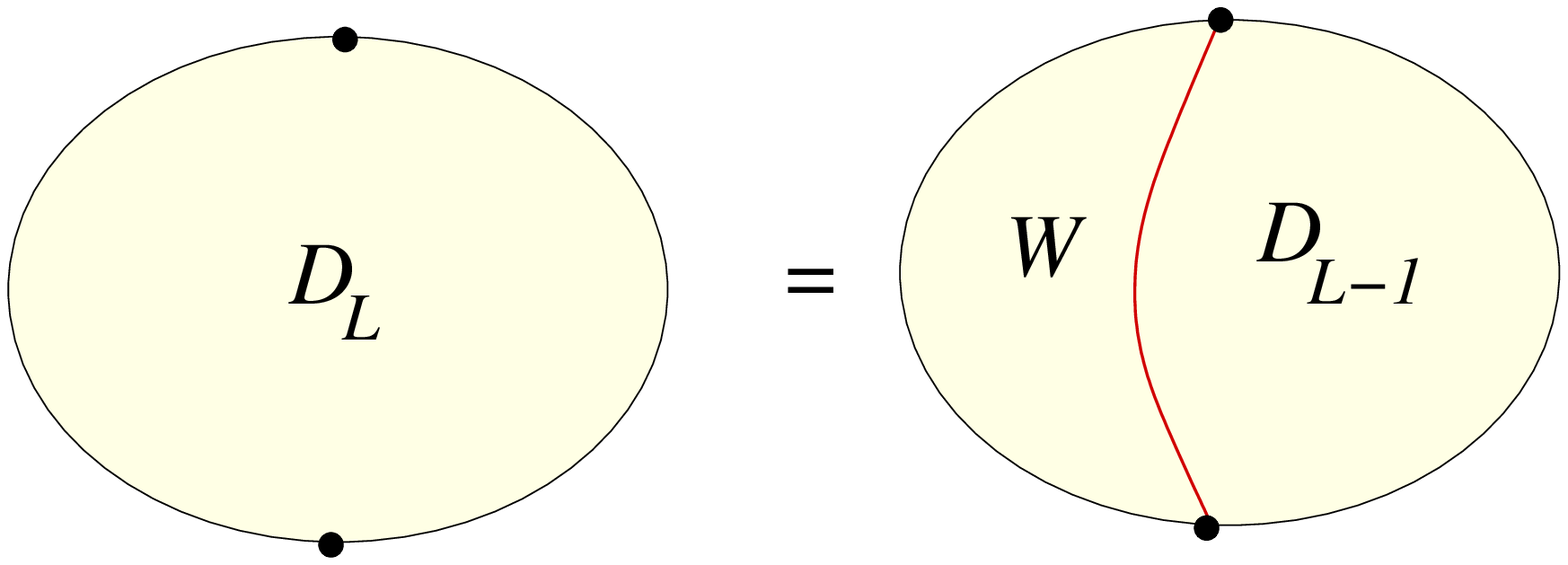 } }
\eqn\recInt{
\hat D _{L}
(\ell_1, \ell_2)=\int_0^\infty d\ell\  
 \hat W(\ell_1+\ell)\hat D _{L-1}(\ell, \ell_2), \qquad
\hat D _0(\ell_1, \ell_2)= \hat W(\ell_1+\ell_2)
}
or, after a  Laplace transformation,
\eqn\convV{
\label{convV}
D _{L} (z_1,z_2) = \oint {dz\over 2\pi i} \ {W(z_1)-W(z)\over z_1-z}\
D _{L-1}(-z, z_2)} 
where the contour of integration encircles the cut
$-\infty<z<-M$.  This integral equation implies a condition on the
discontinuity of $D _L(z)$ along the cut:
\eqn\discV{\Im 
D _{L}(z, z_2)
= \Im W(z) \cdot 
 D _{L-1}(-z, z_2),\qquad
z<-M.}
In terms of the   variable $\t$, this condition is equivalent to a
finite-difference equation  \cite{Ibliou}
\eqn\fdeqV{
\label{fdeqV}
 D _{L}(\t_1+i\pi , \t_2)-D _L(\t_1-i\pi , \t_2)
=[W(\t_1+i\pi )- W(\t_1-i\pi)]
 D _{L-1}(\t_1, \t_2)
}
or, taking into account the explicit expression (\ref{Wottau})
for $W(\t)$
\eqn\EQD{
\label{EQN}
\sin \(\pi {\p/\p \t_1}\) \
 D _{L}(\t_1  , \t_2) 
=  M^g \  \sinh g\t_1\
 D _{L-1}(\t_1, \t_2).
}
Together with  the initial condition
$D _{0}(\t_1, \t_2)\equiv W(\t_1, \t_2)$, eq.(\ref{Wtwocc}), 
this recurrence equation has the unique solution  
\eqal\ddid{
\label{ddid}
 D_L (z_1, z_2)=
  {c_L \over z_1-  (-)^Lz_2}\ \prod_{k=0}^L \Big( 
 W\big(\t_1+i\pi (L-2k)\big)-W(\t_2)  \Big)
 }
  where the normalization constants are given by
\eqal\normcsts{
c_L= (-)^L  \ \prod_{k=1}^L\frac{\sin\pi g}{\sin(k+1)\pi g}.
 \nonumber}
It  is a rational function of the  variables 
$z_a = M\cosh\t_a$ and   $W_a = M^g\cosh g\t_a  \ \ (a=1,2)$:
\eqal\SLsols{
\label{SLsols}
 D _0(z_1, z_2)&=& {W_1-W_2\over z_1-z_2}
 \cr
& & \cr
 D _1(z_1, z_2)&=&  {c_1\over z_1+z_2}\
 \( W_1^2 + W_2^2- 2\cos\pi g\ W_1W_2 - M^{2g}\) 
 \cr 
 & &... \cr
 D _{2m}(z_1, z_2)&=& c_{2m}\
  {W_1-W_2\over z_1-z_2}\ 
 \prod _{k=1} ^{m} 
 \( W_1^2 + W_2^2 -
 2\cos 2k\pi g\ W_1W_2 - M^{2g} {\sin^2 2k\pi g \over \sin^2\pi g}\)
\cr
 & & \cr
D _{2m+1}(z_1, z_2)&=&
  { c_{2m+1}\over z_1+z_2}\ 
 \prod _{k=0} ^{m} \( W_1^2 + W_2^2- 2\cos (2k+1) \pi g\ W_1W_2 
 - M^{2g}{\sin^2 (2k+1)\pi g \over \sin^2\pi g}
 \). \nonumber 
 }

 \bigskip
 
 Let us compare (\ref{fdeqV})  with the difference equation  
 obtained in Liouville gravity. First,  notice that this  requires the choice 
 (\ref{kapp})
of the constant $\k$.

 \bigskip

 \noindent
 a)  $g>1$.
 
 Then the operator   $S_L$ should be identified with the boundary field
$\CB_{L+1, 1} $ with  $P_{L+1, 1}=  \hf e_0 + L{1\over 2b}$.
 Then $|P|> {1\over 2b}$ and  eq. (\ref{fdeqV})    reproduces
the difference equation      (\ref{FDEa}) in Liouville gravity.

  \bigskip

 \noindent
 b)  $\hf <g<1$.
 
Then  the  operator $S_L$ should be identified with the boundary field
 $\CB_{1, L+1} $ with  $P_{1,L+1}=  \hf e_0 - L{b\over 2}$.
   Then  $|P|\ge {b\over 2} $ for all $L\ge 1$
  and   eq. (\ref{fdeqV})   coincides with the difference equation 
     (\ref{FDEb}).
      
   In  the interval $0<g<\hf$ the statistical interpretation of the
     loop gas partition function does not exists because the
      fugacity of the loops is negative:
  $-2\cos(\pi g)<0$.

\subsection{Boundary three-point  functions of 
magnetic operators}

The same techniques can be used to write difference 
equations for correlators of more than two operators.
Consider the   correlation function of  three boundary  
magnetic operators
${\bf S}_{L_1}$,  ${ \bf S}_{L_2}$ and ${\bf S}_{L_3}$
\eqn\SScorf{C_{L_1,L_2,L_3}(\tau_1,\tau_2,\tau_3 )
  =\< {\bf S}_{L_1}^{\t_2\t_3}{\bf S}_{L_2}^{\t_3\t_1}{ \bf 
S}_{L_3}^{\t_1\t_2}\>
}
or pictorially

 \epsfxsize=200pt
\eqn\tripF{
\epsfbox{   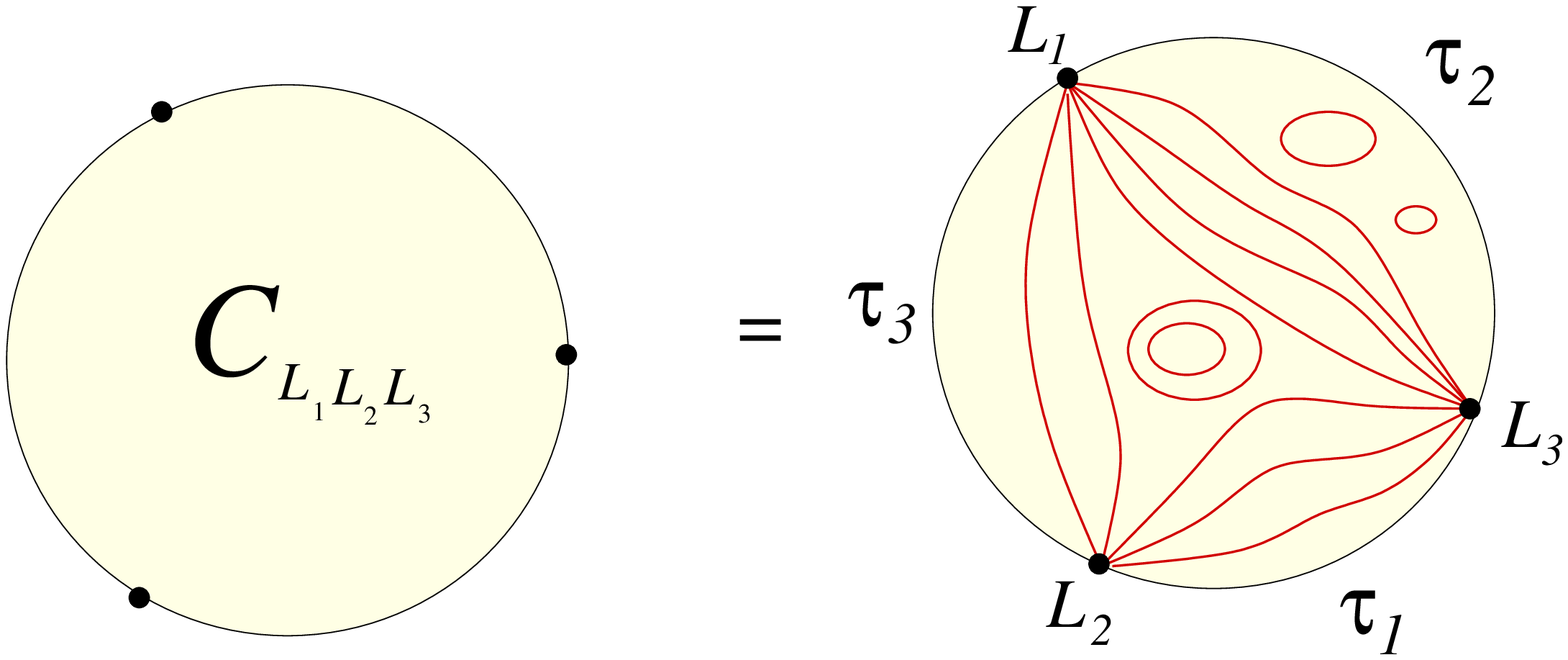 }
 }

 Let us first evaluate the three-point function of the identity 
boundary operator  with $ L=0$.  This is the partition function on the 
disc with three different cosmological constants $z_1, z_2, z_3$ 
along three segments on the boundary.
  It is equal to  the integral
$$C_{000} (z_1,  z_2,z_3)=
\int _0^\infty d\ell_1 ... d\ell_3\ 
e^{- \sum z_i\ell_i} W\(  \ell_1+\ell_2+\ell_3\)=\oint {dz \over 2\pi 
i} 
{W(z) \over\prod _i( z-z_i) }
$$
or, after performing the    integration,
\eqn\tripbbb{
\label{tripbbb}
 C_{000}  (z_1,z_2,z_3)
= {(z_2-z_3)W_1+(z_3-z_1)W_2+(z_1-z_2)W_3
\over (z_1-z_2)(z_2-z_3)(z_3-z_1)} .
}
This expression solves the Liouville difference equation (\ref{3fe})
and coincides, up to a multiplicative constant, with the
Liouville three-point function (\ref{Csss}).

    It is straightforward to write a recurrence equation relating the
correlation functions $C_{L_1L_2L_3}$ and $C_{L_1-1, L_2-1, L_3}$ by
using the factorization property associated with the most external
line relating the points 1 and 2: \eqn \CLLeps{ \epsfxsize=180pt
\epsfbox{ 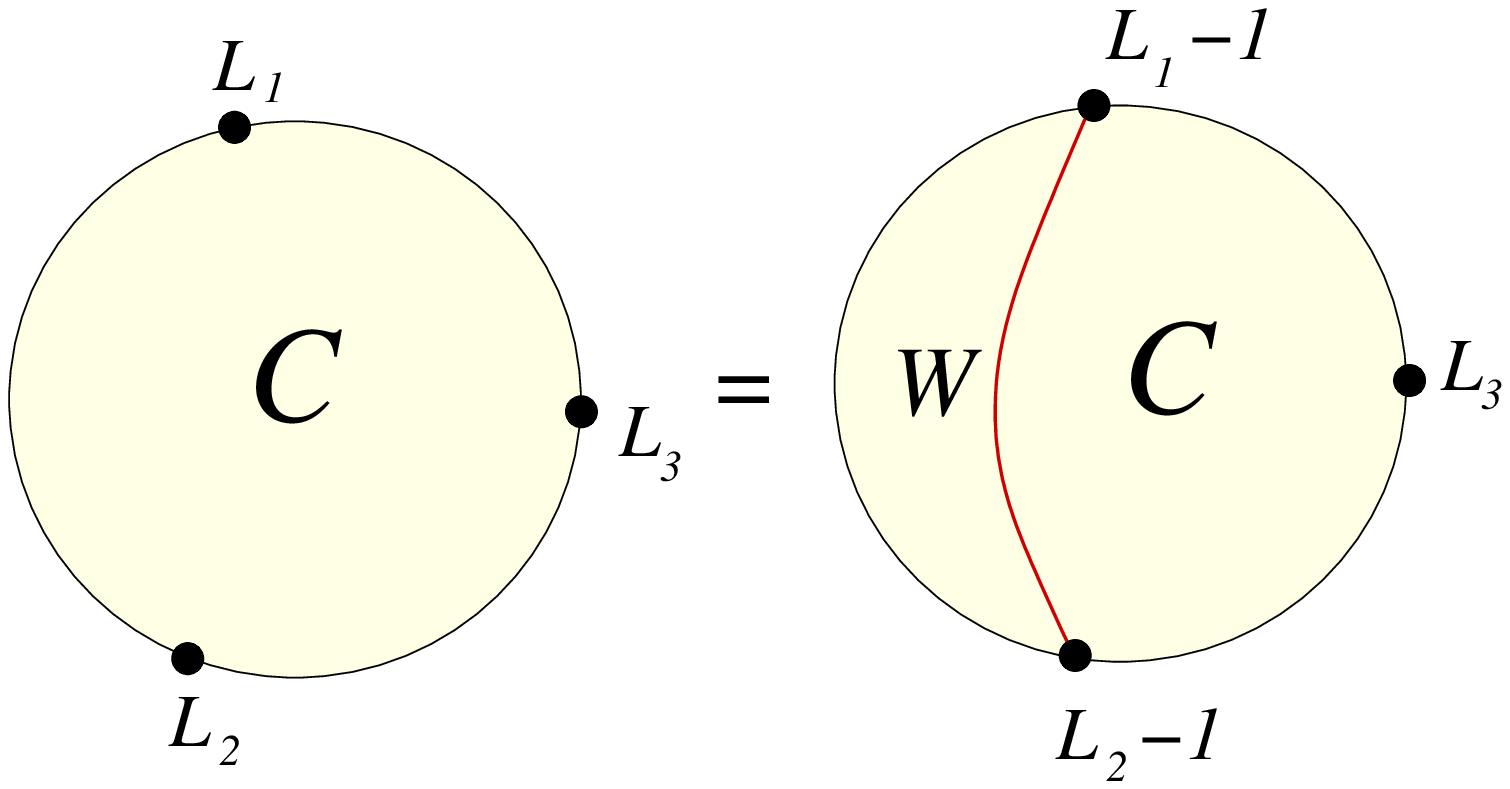 } }
  Skipping the derivation, which is
the same as in the case of the two-point function, we 
  write   
\eqal\steq{
\label{steq}
&C_{L_1,L_2,L_3}(\tau_3+i\pi ,\tau_1,\tau_2 )-
C_{L_1,L_2,L_3}(\tau_3-i\pi ,\tau_1,\tau_2 ) =\cr &
[W(\t_3+i\pi)-W(\t_3-i\pi)] \ C_{L_1-1,L_2-1,L_3}(\tau_3
,\tau_1,\tau_2 ).  } 
Knowing the explicit form of the function for $C_{000}$, given by eq. 
(\ref{tripbbb}), we  can give the following Ansatz for the general form
of the solution
\eqal\CLLL{
C_{L_1,L_2 ,L_3}=
{\sum_k z_k \ P_{L_1,L_2 L_3}^k (W_1,W_2,W_3) \over \prod _{i<j}
(z_i-(-1)^{n_i+n_j}z_j) }
}
where    $L_1=n_2+n_3, L_2=n_3+n_1, L_3=n_1+n_2$\ and 
  $P_{L_1,L_2, L_3} $ is a polynomial of degree 
$\hf(L_1+L_2+L_3) +1=n_1+n_2+n_3+1$.

    \bigskip
    
Let us compare eqn. (\ref{steq})  with the   difference equations
for the three-point function obtained   in Liouville  gravity
in sect. 4:
 
 \bigskip
 
 \noindent
  a)  $g>1$.
 
  \noindent
 Then the operator   $S_L$ should be identified with the boundary field
$\CB_{L+1, 1} $ with  $P_{L+1, 1}=  \hf e_0 + L{1\over 2b}$.
  In this case eq. (\ref{steq})     coincides with 
the difference equation      (\ref{FDEa})  
  and   $|P|> {1\over 2b}$.

    \bigskip
 \noindent

 a)  $\hf <g<1$.
 
  \noindent
Then  the  operator $S_L$ should be identified with the boundary field
 $\CB_{1, L+1} $ with  $P_{1,L+1}=  \hf e_0 - L{b\over 2}$.
   In this case eq. (\ref{fdeqV})   coincides with the difference equation 
     (\ref{FDEb})  and   $|P|\ge {b\over 2} $.
      
\paragraph{\bf Remark 1:}
 Eqn.  (\ref{steq}) was derived in the  SOS model, where the loop amplitude 
 $W(\t)$ does not depend on the height of the boundary.
 For a general ADE model this  equation is written as follows.
Let $\hat C_{L_1-1,L_2-1,L_3;}^{h_1,h_2,h_3}(\tau_1,\tau_2, \tau_3)$
be the boundary 3pt function with Liouville and matter boundary conditions 
$\s_k, h_k\ \ (k=1,2,3)$. Then the difference equation 
for $\beta_k = Q- [(L_k+1)g-1] $ takes the form
\eqal\steqa{
\label{steqa}
&\hat C_{L_1,L_2,L_3}^{h_1,h_2,h_3
}(\tau_1,\tau_2, \tau_3+i\pi )-
\hat C_{L_1,L_2,L_3}^{h_1,h_2,h_3}(\tau_1,\tau_2, \tau_3-i\pi )
 =\cr &
 =[W_{h_3}(\t_3+i\pi)-W_{h_3}
 (\t_3-i\pi)] \  A_{h_3 h'_3} 
\ \hat C_{L_1-1,L_2-1,L_3}^{
  h_1,h_2,h'_3}(\tau_1,\tau_2, \tau_3)
}
where  $W_h(\t)= S_h  M^{g}\cosh g\t$, $S_h$ is the
 Perron-Frobenius vector and 
$A_{hh'}$ is the adjacency matrix of the Dynkin graph.
 Writing the solution in a factorized form
  $$\hat C_{L_1-1,L_2-1,L_3}^{  h_1,h_2,h_3}(
\tau_1,\tau_2, \tau_3)=\check C_{L_1-1,L_2-1,L_3}^{h_1,h_2,h_3 }
C_{L_1-1,L_2-1,L_3}(
\tau_1,\tau_2, \tau_3)
$$
we get eqn.  (\ref{steq}) as well as another equation for the
matter factor.

\paragraph{\bf Remark 2:}
We find it interesting to propose the following {formal} 
representation of the three-point 
function  as the integrated product of three two-point functions and 
one loop amplitude
\eqal\didinaR{
\label{didinaR}
\epsfxsize=200pt
\epsfbox{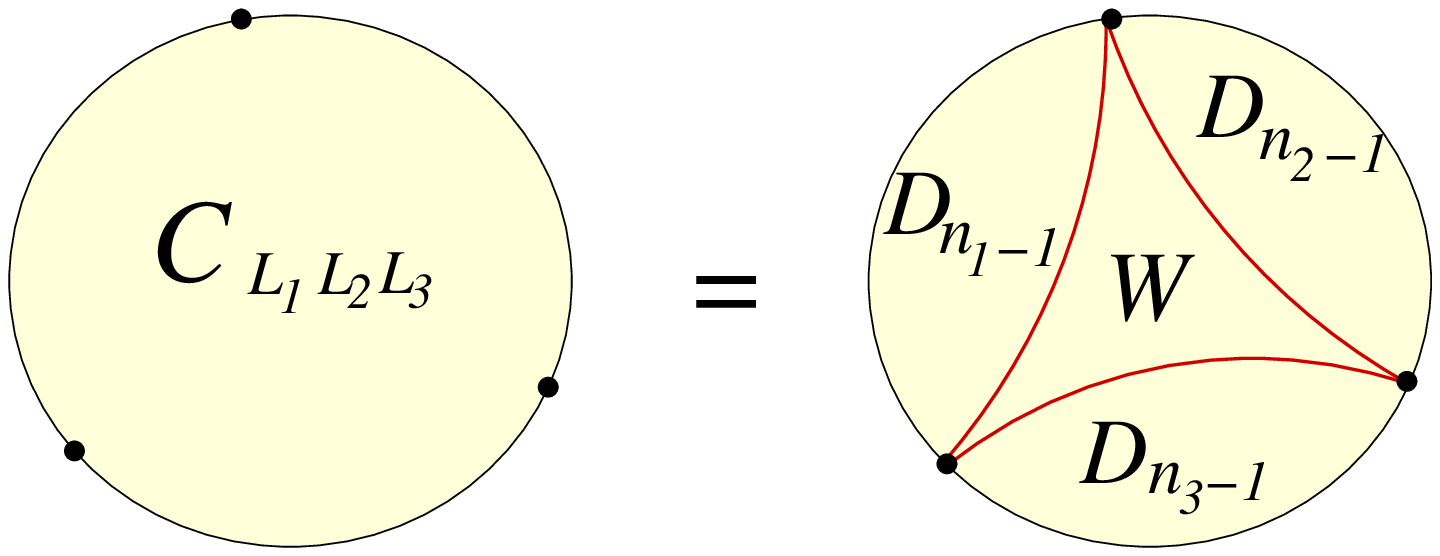 }
}
\eqal\eqdidina{
\label{eqdidina}
C_{L_1,L_2,L_3}(z_1, z_2,z_3)
= \oint {dz\over 2\pi i} \  W(z)\  D_{n_1-1}(-z, z_1) 
 D_{n_2-1}(-z, z_2) 
 D_{n_3-1}(-z, z_3) .
}
where
$ n_1= \hf(L_1+L_2-L_3) ,\
n_2= \hf(L_1+L_3-L_2) , \
n_3= \hf(L_2+L_3-L_1) 
$.
Written in terms of  $\t$-variables,
\eqn\didinaT{
\label{didinaT}
C_{L_1,L_2,L_3}(\t _1, \t _2,\t _3)
= {M^{g+1}  \over 2\pi} 
\int _{-\infty} ^\infty d\t  \sinh \t  \sinh g\t  \  D_{n_1-1}(\t , \t _1) 
 D_{n_2-1}(\t , \t _2) 
 D_{n_3-1}(\t , \t _3) .
}
This expression satisfies the difference equations on
 each of the variables $\t_1,\t_2, \t_3$ and for
  $L_1=L_2=L_3=0$ ({\it i.e.} $ n_1=n_2=n_3=-1$) 
  it reduces to the expression (\ref{tripbbb}).  
 
Although the three-point function in Liouville theory has an integral representation
(\ref{ff3p}),  this   seems very different from  (\ref{didinaT}).  
In particular, we were not able to evaluate directly the three point
function $C_{bbb}$  in LFT from its integral representation
(\ref{ff3p}); we managed
to give its dependence with respect to the boundary parameters
only,thanks to
 the functional equation (\ref{3fe}). We find it remarkable that its
equivalent in 2D gravity is so simple to compute.

It is natural to expect   that eq. (\ref{eqdidina})  has its 
counterpart in Liouville theory, namely the difference equation that follows from the  
 fusion rules with a higher  degenerate field $B_{1, L+1}$ or $B_{L+1, 1}$.

\subsection{Relation between the bulk-boundary and the 
boundary-boundary 
two-point  function of star operators}

Let $ R_L(z)$ be the structure constant for the fusion of a bulk and
boundary $L$-star operators $R_L(\t)= \< {\bf S}^{(L)} \ {\bf S}_L\>$:
   
\eqn\btripF{\epsfxsize=200pt
\epsfbox{ 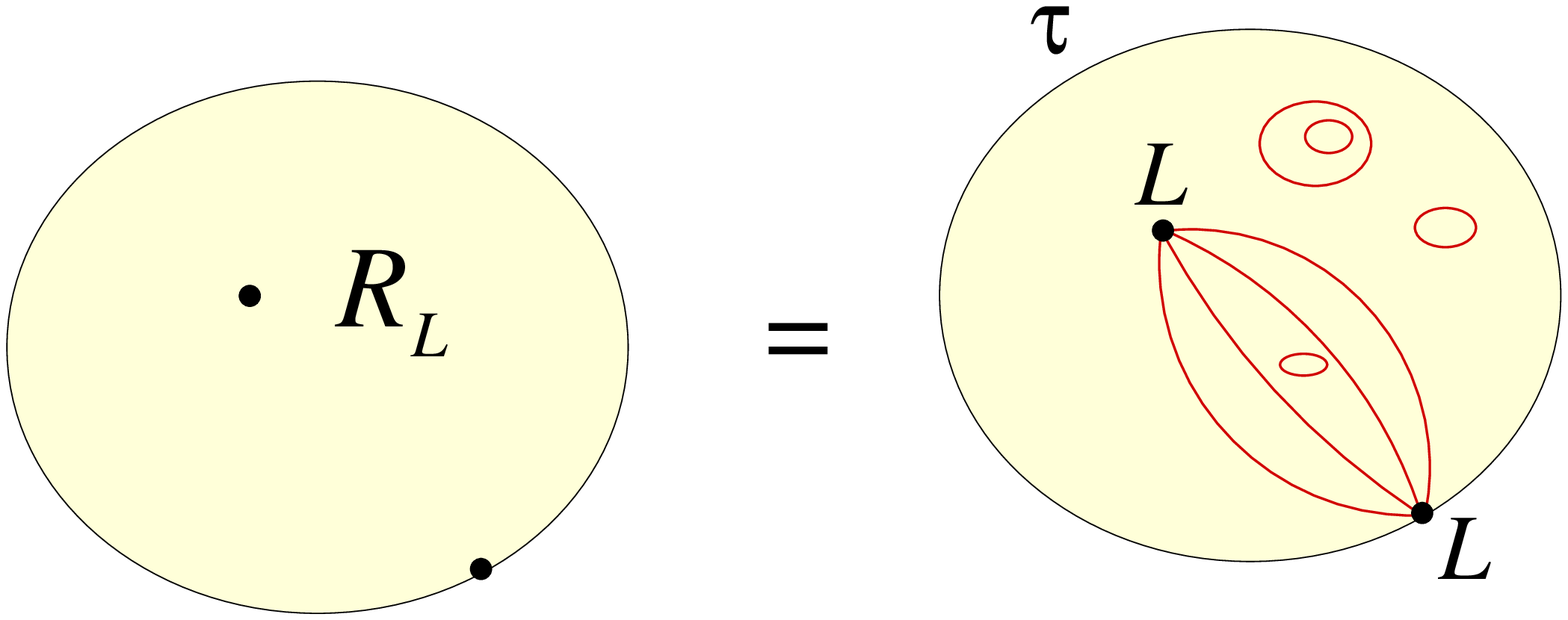 }
 }
The factorization of the measure applied for the   union of the two 
most external lines leads to the  identity
\eqn\rldeq{\epsfxsize=200pt
\epsfbox{ 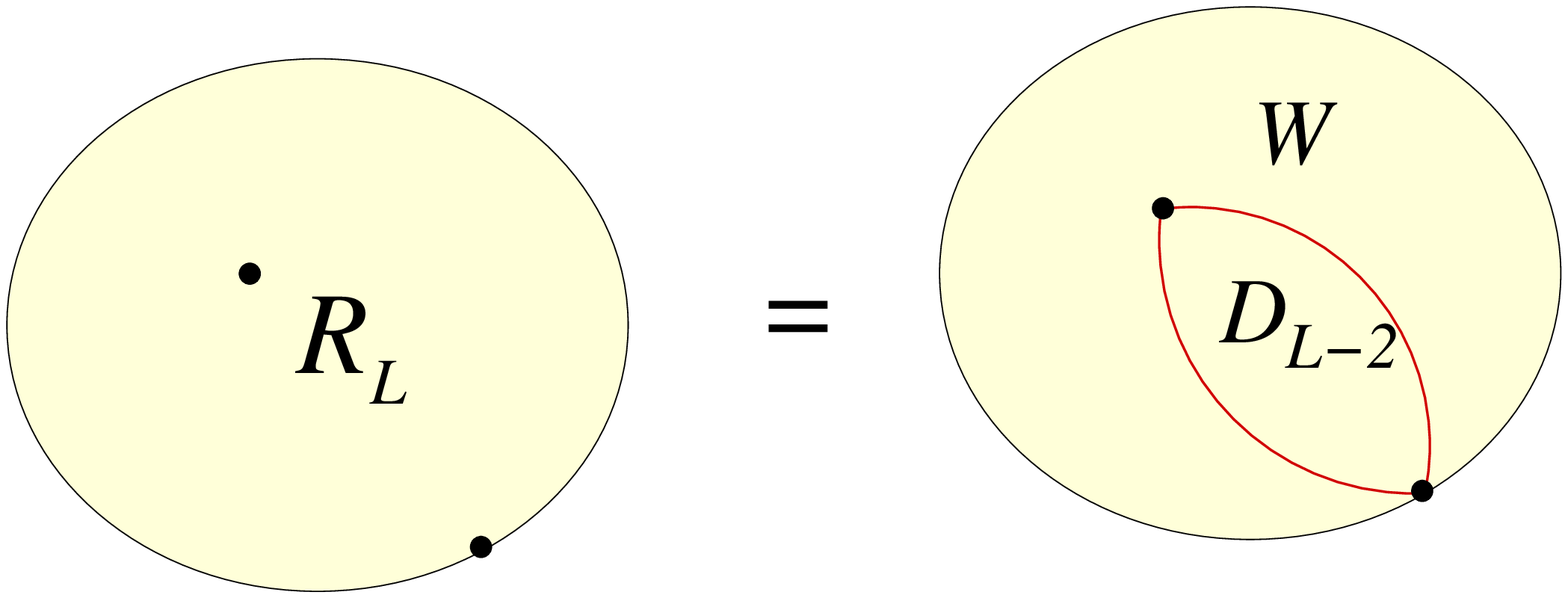 }
 }
which reads, in $z$-representation,
\eqn\RLDL{
R_L(z)=\oint dz' W(z,z') \ D_{L-2}(-z', -z').
}
  The condition on the cut
 $$ \Im R(z)= \Im W(z)\  D_{L-2}(-z,-z), \qquad z<-M
 $$
 gives the following finite-difference equation
 \eqn\eqnRLD{
 \label{eqnRLD}
R_L(\tau+i\pi) - R_L(\tau -i\pi) = [W(\tau+i\pi) - W(\tau -i\pi)] \
D_{L-2}(\t, \t) } The function $R_L(\t) $ should be proportional to
the Liouville bulk-boundary structure constant $R_\s(\a,\b)$ with $\a
= Q/2 - L/4b$ and $\b = Q/2 - (L+1)/2b+ b/2= b - L/2b$.  Indeed, with
the help of the identity (\ref{RvD}), eq.  (\ref{eqnRLD}) can be
identified as a particular case of the Liouville functional equation
(\ref{toto}).

\subsection{The boundary two-point function of twist operators}
 
 We have checked that the functional equations in the SOS string and in
Liouville theory are compatible for two infinite discrete sets of
equally spaced Liouville charges.  Now we are going to consider the
case when some parts of the boundary have Neumann boundary condition. 
This will allow to compare the two approaches for situations involving
a continuous spectrum of Liouville charges for some of the operators.

The simplest correlation function involving both Dirichlet and Neumann
boundary conditions has been calculated in \cite{KKloop} .  This is
the correlation function $\Omega(z, \tilde z)$ of two boundary
changing operators that intertwine between the Dirichlet and Neumann
boundary conditions (see Fig.  10 of ref.\cite{KKloop} ).  Recently it has
been checked \cite{Ibliou}\ that this correlation function coincide
(again up to a numerical factor) with the boundary Liouville
correlation function (\ref{Dbeta}) for \eqn\btwist{
 \label{btwist}
 \b= b/2 + 1/4b.
 }
 For completeness we review here the observation of \cite{Ibliou}.
We denote by  
\eqn\couplND{z = M\cosh \tau, \quad
\tilde z = M^g \cosh  g\tilde \t}
the cosmological constants along the
Dirichlet and  Neumann type 
boundaries, correspondingly
 \epsfxsize=200pt
\eqn\btripF{
\epsfbox{   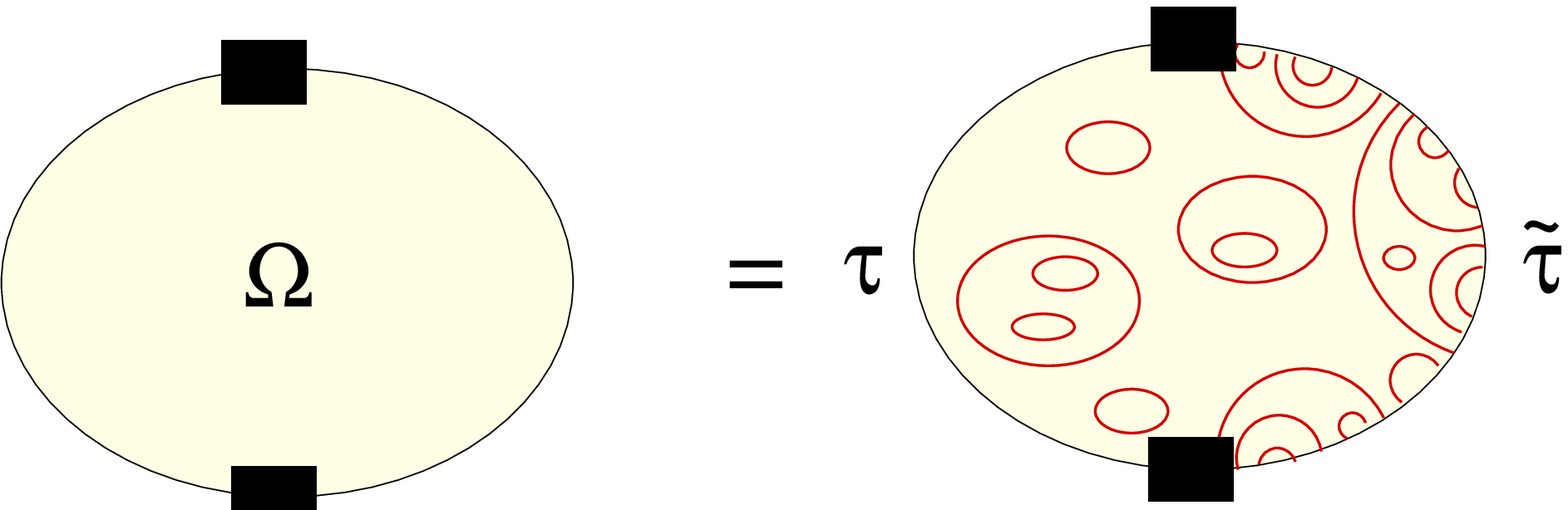 }
 }
Applying the factorization property to the maximal chords we get the
following integral equation
 \epsfxsize=200pt \eqn\bF{ \epsfbox{
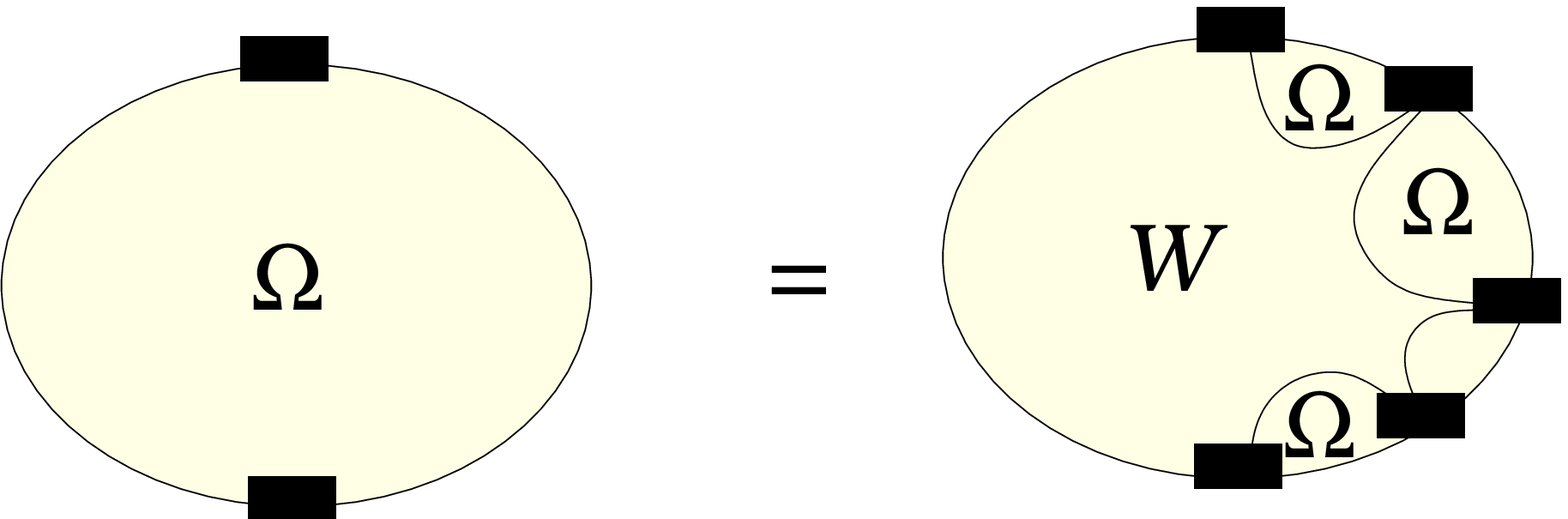 } } whose simplest analytic form is in the $(\ell, \tilde
z)$ representation, where $\ell$ is the length of the Dirichlet
boundary and $\tilde z$ is the cosmological constant associated with
the Neumann boundary:
\def\nb{ 2\cos(\hf \pi p_0)}
\eqn\intOm{
\hat  \Omega(\ell, \tilde z)=1+\sum_{n=0}^\infty  \hat n^n
\int_0^\infty d\ell_1...d\ell_n\
 \hat W(\ell+ \ell_1+...+\ell_n, \tilde z)\ 
 \hat \Omega(\ell_1, \tilde z)..
.\tilde \Omega (\ell_n, \tilde z), 
}
where
$\hat n = \nb= 2\sin (\pi g/2)$ is the fugacity of the open lines.
From here we  find for the Laplace transform

\noindent
 $\Omega(z, \tilde z) =1-
\int_0^\infty d\ell e^{-\ell z} \hat 
 \Omega(\ell, \tilde z)$
\eqn%
\eqnzOm{
\Omega(z, \tilde z)=- {1\over 2\sin (\pi g/2)}
\oint {d z'\over 2\pi i} {1\over z-z'}\
{W(z')\over  \ \Omega(-z', \tilde z)}.
}

Taking the imaginary part along the cut, we get the 
functional equation  of \cite{KKloop} 
\eqn\funcDN{
\label{funcDN}
\Im \Omega(z, \tilde z)  =- {1\over 2\sin (\pi g/2)}\ 
{\Im W(z) \over \Omega(-z, \tilde z)}
}
or, in terms of $\tau$,
\eqn\fctDN{
\label{fctDN}
\Omega(\tau +i\pi, \tilde\tau) -
 \Omega(\tau - i \pi, \tilde\tau)
= -{ W(\tau+i\pi) - W(\tau - i\pi)\over 
 2\sin (\pi g/2) \  \Omega(\tau, \tilde\tau)
}.}
   
Let us compare eq. (\ref{fctDN})    with 
   the difference equation 
(\ref{WEW}) for the boundary two-point function  in Liouville gravity.
  The two equations are compatible for $P={ 1\over 4b}$ and
$$  \Omega (\t,\tilde \t)= {i b\over \sqrt{2}}   \ 
W_{1\over 4b^2}(\t,\tilde \t).
$$
 This confirms the identification of 
the gravitationally dressed twist operator  as the 
boundary state $\CB_{P}$  with $P={ 1\over 4b}$:
 \eqn\twistKPZ{{\bf T}=   (ib)^{1/2}  2^{-1/4}  \ {
\CB}_{1\over 4b}.} 
This  is a ``half-degenerate" boundary  field with $r=\hf$ and $s=0$
 and gravitational  conformal dimension
  $$\delta_{\bf T} = {2b^2-1\over 4b^2}.$$
The flat dimension of this operator is given by  (\ref{gtwist}).

 We can consider, more generally, the excited  boundary twist operators
 defined as a juxtraposition of a boundary twist
and magnetic operators \eqn\TSL{{\bf T}_L = {\bf S}_{L} {\bf T}.  }
 Geometrically the operator ${\bf T}_L$ is constructed as the source of $L$
nonintersecting lines inserted between at the point where Dirichlet
and Neumann boundaries meet.  The two-point function $$\Omega_L(\t,
\tilde \t)=\left\langle {\bf T}_L^{\t\tilde \t}{\bf
T}_L^{\tilde \t\t} \right\rangle $$
  satisfies the same recurrence equation   (\ref{convV})
\eqn\EQDO{
 \Omega_L(\t+i\pi , \tilde \t)-\Omega_L(\t-i\pi , \tilde\t)
= [W(\t+i\pi)- W(\t-i\pi)]
 \Omega_{L-1}(\t, \tilde \t),
}
 but with 
different initial condition
$ \Omega_0 (\t,\tilde  \t)=\Omega (\t,\tilde  \t)$.
Comparing with the difference equations  in Liouville quantum gravity we
 identify  the  excited twist operators as fields   ``half-degenerate"
boundary KPZ fields $r= L+\hf, s=0$ with $P= \left(L +\hf\right)/b-b
$.

\subsection{ The boundary correlation function of two twist operators and one 
vertex  operator.}


Here we will consider the 
correlation function of two twist operators
 and  an  electric operator with arbitrary target space momentum $p$:
\eqn\defPsi{\Psi_p(\t, h|\tilde \t_1, \tilde \t_2)=
\<   {\bf T}^{\t \tilde\t_1}  {\CV }_p ^{ \tilde\t_1\tilde\t_2} {\bf 
T}^{\tilde\t_2\t}
\>.
}
In this case the correlation function depends explicitly on the 
height $h$ of the Dirichlet piece of the boundary 
between the two twist operators.
%
 The corresponding     pattern of  domain lines  and loops is:
   \eqal\patrnPsi{  
   \label{patrnPsi}
     \epsfxsize=  200pt \epsfbox{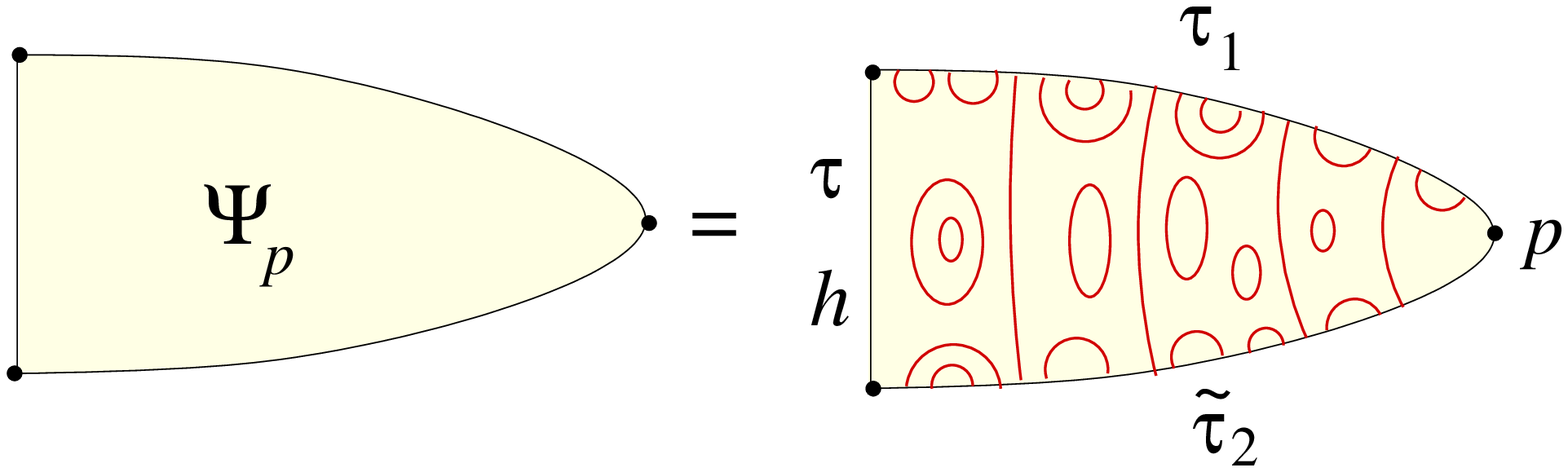}
   }

   
\bigskip \noindent There are two kinds of open lines: the contractible
ones whose ends belong to the same Neumann boundary and the
non-contractible ones that connect two different Neumann boundaries. 
The non-contractible lines have additional weight factor $\cos\pi
p/\cos\pi (\hf p_0)$.
    
    Due to the momentum introduced by the vertex operator  $\CV_p$
    the correlation function has nontrivial dependence on the height $h$:  
    \eqn\psih{
    \label{psih}
    \Psi_p(\t, h|\tilde \t_1, \tilde \t_2)= e^{2\pi i ph}\
   \Psi_p(\t|\tilde \t_1, \tilde \t_2).
   }
The function $\Psi_p(z, h|\tilde z_1, \tilde z_2)$ satisfies a linear integral equation, obtained by
factorizing with respect to the non-contractible line closest to the
Dirichlet boundary.  The equation involves the 4-point function
$\G(z, z'|\tilde z_1\tilde z_2)$, which 
is the partition function on a disc whose boundary is divided into
four segments, with alternating Dirichlet and Neumann boundary
conditions and no open  lines connecting the two Neumann 
boundaries:\footnote{ 
The function $\G$ is actually the   correlation function of four 
boundary twist operators,  $\G(z, z'|\tilde z _1\tilde z_2)= \< [{\bf T} ]^{z \tilde z_2}{\bf T} ^{\tilde 
z_2  z'} {\bf T}^{z'\tilde z_1}{\bf T} ^{\tilde z_1 z} \> $,
with target-space momenta $p=1/2$ and $p=-1/2$ associated with the 
two Dirichlet boundaries (with boundary parameters $\t$ and $\t'$).
The effect of  inserting the momentum $1/2$ is that all lines connecting the 
two Neumann boundaries will acquire a factor $2\cos (\hf \pi) =0$.
 }

\eqn\ptnTTTT{
 \label{ptnTTTT}
\epsfxsize= 150pt \epsfbox{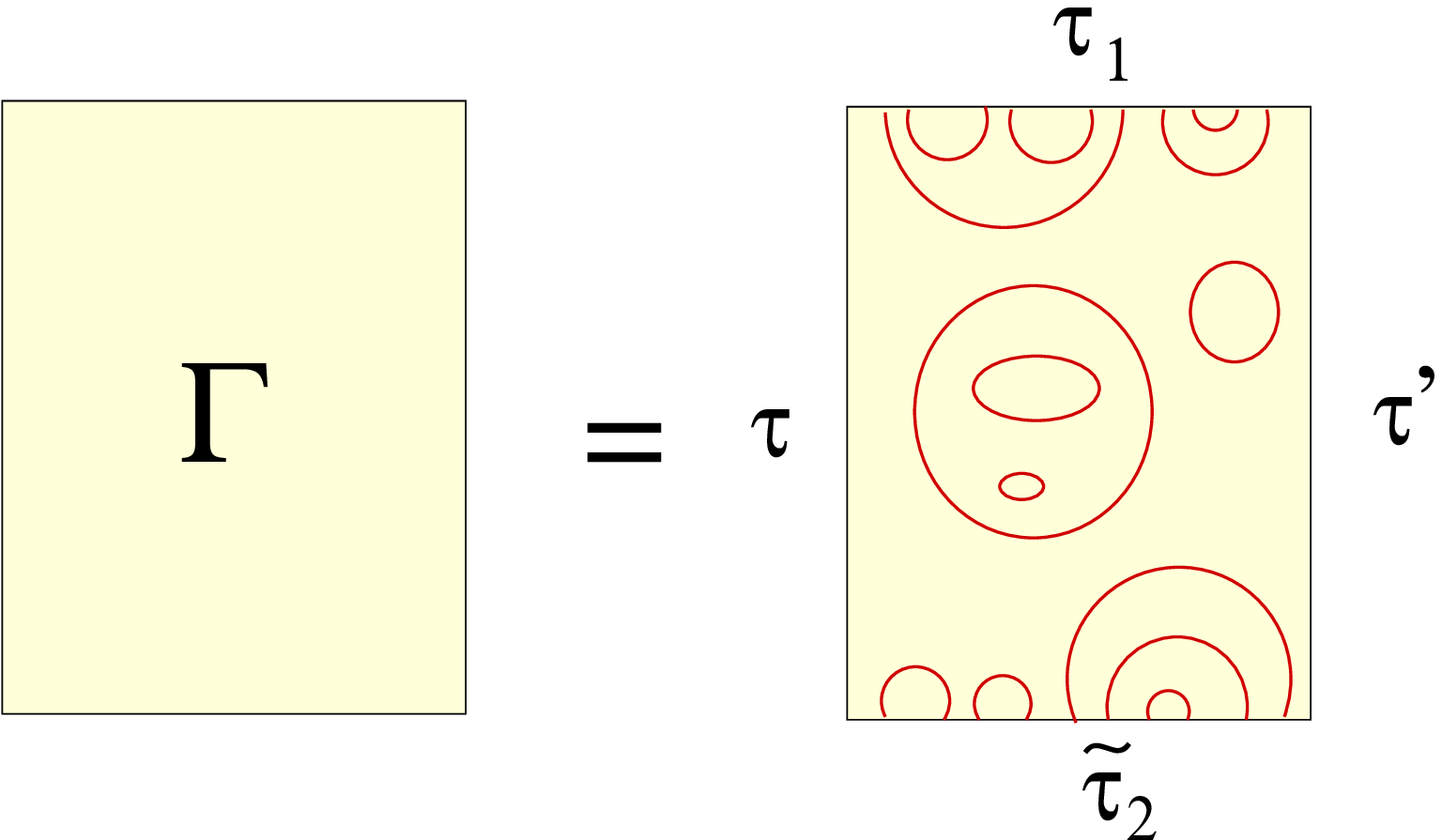} } 

\noindent
We can write (\ref{patrnPsi})
as the sum of a term containing no non-contractible lines, which we
denote by $\Psi_{1/2}$, and a term that factorizes into $\G$ and
$\Psi_p$:
\bigskip
   \eqn\eqnPsi{  
    \label{eqnPsi}
     \epsfxsize=  280pt \epsfbox{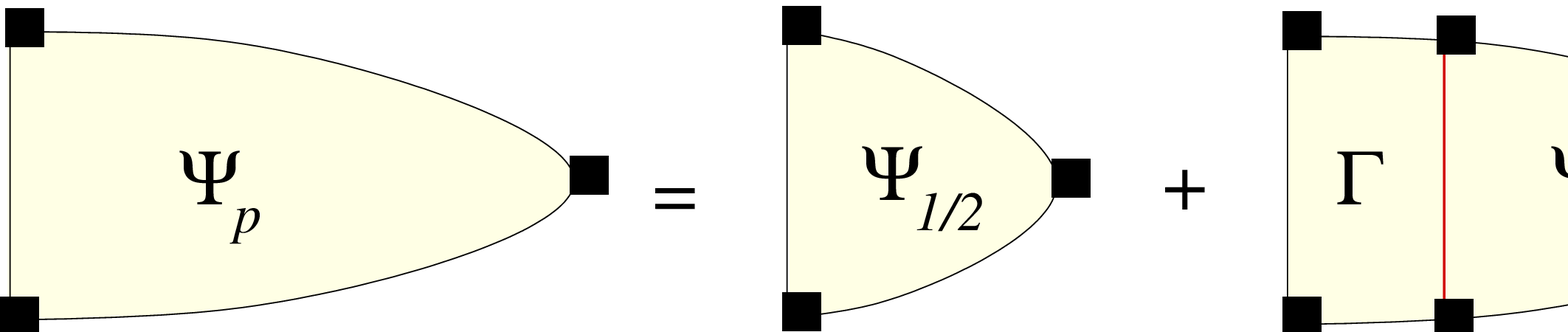}
   }
  
     \bigskip
     \noindent
This yields the integral equation
\eqn\BEeq{
\label{BEeq}
\Psi_p(\t|\tilde \t_1, \tilde \t_2)= \Psi_{1/2}(\t |\tilde 
\t_1,{\tilde \tau}_2) + \cos\pi p 
 \ \oint {dz'\over 2\pi i}\G (z, z'|\tilde z_1 \tilde z_2)
\Psi_p(z'|\tilde z_1 \tilde z_2).
}
 The 3-point function
$ \Psi_{1/2}(\t |\tilde \t_1,{\tilde \t}_2) $
is defined in the same way as $\Psi_p$, but 
 since $\cos(\hf \pi)=0$ there are no lines connecting the 
sides $\tilde \t_1$ and $\tilde \t_2$.
It is not difficult to see that the  function $\G$  is 
 expressed through  $ \Psi_{1/2} $  as
\eqn\Snot{\G(z, z'|\tilde z_1\tilde z_2)=
{\Psi_{1/2}(z|\tilde z_1 , \tilde z_2 )-\Psi_{1/2}(z'|\tilde z_1 , 
\tilde 
z_2) \over z- z'}.
}
The  3-point function $\Psi_{1/2}$ itself  is expressed 
in  terms of the  two-point 
correlator of twist operators $\Omega(z, \tilde z)  $ (see sect. 6 of \cite{KKloop} )
 \eqal\Gmma{ 
 \label{Gmma}
 \Psi_{1/2}(z|\tilde z_1,\tilde z_2)= {1\over [\nb]^2}
\oint {d\z\over 
2\pi i} {W(z)-W(\z)\over z-\z} {1\over \Omega(\z, 
\tilde z_1) \Omega(\z, \tilde z_2)}.
 }
   Again integral equations  (\ref{BEeq})   and    (\ref{Gmma}) can be turned into
finite-difference equations
for the $\t$-variables:

 \eqal\loopeqS{ 
 \label{loopeqS}
 \sin\pi \p_\t \ \Psi_p(\t |{\tilde \tau}_1,{\tilde \tau}_2)=
 [ \sin\pi \p_\t \ \Psi_{1/2}(\t|{\tilde \tau}_1,{\tilde \tau}_2) 
]\ \[ 1+ \cos\pi p  \
 \Psi_p(\t|{\tilde \tau}_1,{\tilde \tau}_2) \] }

 \eqal\Gmmabs{ \sin\pi \p_\t \ \Psi_{1/2} (\t |{\tilde 
\tau}_1,{\tilde \tau}_2)=
 - {1\over [\nb]^2}\ 
 {\sin g\t \over \Omega(\t , {\tilde \tau}_1) \Omega(\t, {\tilde 
\tau}_2)}.
 }
The non-homogeneous term on the rhs can be eliminated by shifting 
$\Psi_p\to 
\Psi_p- 1/\cos\pi p $. As a result we finally arrive at the 
homogeneous equation
 \eqal\loopeqPsi{ 
 \label{loopeqPsi}
 \sin\pi \p_\t \ \Psi_p(\t |{\tilde \tau}_1,{\tilde \tau}_2)=- 
{ \cos\pi p  \over [\sin\hf \pi g]^2}\
{\sin g\t \over \Omega(\t , {\tilde \tau}_1) \Omega(\t, {\tilde 
\tau}_2)}
  \
 \Psi_p(\t|{\tilde \tau}_1,{\tilde \tau}_2)  }

This equation cannot be compared directly to the difference equations  
obtained in Liouville  quantum gravity, which were written for too special situation
(different signs of the two adjacent momenta and the neutrality condition imposed.)
Nevertheless  is we consider the formal  expression
\eqn\BTPF{
\label{BTPF}
\< [\CB_{1/4b} ]^{\t\tilde \t} [ \CB_{bp}]^{\tilde \t \tilde \t}
 [\CB_{1/4b} ]^{\tilde \t \t}\>_{\rm disc},
}
the corresponding difference equation
(\ref{FDEab}) is almost identical with (\ref{loopeqPsi}).
The extra factor $\sin(\pi g/2) = \sin (\pi/2b^2)$ in the denominator 
can be explained with the fact that here both fields have positive momentum
$P=1/2b$. The factor $\cos\pi p$  has its origin in the matter field.
In the second term on the r.h.s. of eq. (\ref{eqnPsi}) the height 
of the Dirichlet boundary of $\Psi_p$ is either $h+\hf$  or $h-\hf$, 
because it is separated by one domain wall from the height $h$.
The sum over the two possibilities gives, taking account of (\ref{psih}),
the factor $2\cos \pi p$.

Note that we can replace the twist operators ${\bf T}$ by excited
twist operators ${\bf T}_L = {\bf S}_L {\bf T}$ and apply the
factorization condition to one of the lines of the star operators. 
This will   produce eq. (\ref{FDEa})  with $P_1=(L+\hf){1\over 2b} - {b\over 2}$
and $P-3 = bp$
for all $L\ge 1$.

 \newsec{Conclusions}
 
 In this paper showed that the  the continuous and the discrete approaches to \QG\  
 lead to the same boundary correlation functions.
First we considered pure  Liouville  theory and  showed that  all 
 Liouville boundary structure constants satisfy linear difference equations 
 with respect to the boundary parameters.

  We  interpreted  these equations in the context of 
   Liouville quantum gravity  with  gaussian matter field.
   We observed that  after rescaling  the boundary fields by the    leg 
  factors and taking into account the charge neutrality condition,  
  the charge dependence of the difference operator disappears.  
   The difference equations   have the form of recurrence  relations for the charges. 
  After  repeated application of the difference operator, one obtains a 
   correlation function in which some of the matter fields have 
  ``wrong"  Liouville dressing.   If this is the case, we get, using the 
  reflection property, nonlinear difference equations for the physical fields.
  
  Then we considered the microscopic realization of  \QG\ 
  and  defined    the boundary correlators as  expectation 
  values in the ensemble of non-intersecting   loops  and lines   on a 
  randomly triangulated disc.  We   derived    difference equations  of the same form using 
  the  factorization   of the measure over random surfaces  
  after ``cutting  it open" along a non-intersecting line on the world sheet.
  
  Comparing the difference equations obtained in the two approaches 
  we see that they become identical for certain normalization of the boundary one-point
  function.  We consider this  as  an important
``experimental" confirmation of the Liouville boundary bootstrap
approach.  
It is remarkable that the above equations are sufficient to reproduce 
the correlation functions.   These are much simpler than the standard loop equations,
which can be obtained as Ward identities in
the corresponding matrix model, and which have   
contact terms associated with degenerate world sheets.   

In Liouville theory the difference equations follow from the fusion rules 
with the lowest degenerate boundary fields which   shift   the boundary parameter.
In the microscopic picture, the difference equations are associated with
boundary operators, whose geometrical meaning is to create an open line 
on the world sheet  starting at the boundary.  These operators are 
associated with the lowest degenerate {\it matter}  fields.   
 It is therefore natural to expect that there is an underlying algebraic structure 
involving operators which are products of Liouville and matter degenerate fields.
In the case of vanishing boundary cosmological constant such a structure, 
the `boundary ground ring', has been  discovered in \cite{bershkut}.
Its generalization to the case of finite boundary parameters seems 
quite straightforward.  The ground ring structure should lead to difference equations for
 the boundary $ n$-point function for any $n\ge 2$, which is also the 
 case for the difference equations derived here in the microscopic approach. 
   
   In this paper we considered only the gaussian field realization of 
   the matter field (without screening operators), whose lattice analog is the 
   non-restricted SOS model. The difference equations can be of course derived 
  for  the  $O(n)$ matrix model and the matrix models associated with ADE Dynkin graphs.

 \smallskip\smallskip\smallskip
 \bigskip
\noindent
{\bf Acknowledgments}
\smallskip

\noindent
We thank D. Kutasov, H. Saleur, V. Schomerus, J. Teschner, and especially  V. Petkova  and Al. 
Zamolodchikov 
for valuable discussions.  This research is supported in
part by the European network EUCLID, HPRN-CT-2002-00325.

 \appendix

  \newsec{Special functions}
\begin{itemize}
\item{$\Ga(x)$ function}\\
The Double Gamma function introduced by Barnes \cite{Barnes} is
defined by:
\begin{eqnarray}
\nonumber \\
&&
\text{log}\Gamma_{2}(s|\omega_1,\omega_2)=\left(\frac{\partial}{\partial
t}
\sum_{n_1,n_2=0}^{\infty}(s+n_1\omega_1+n_2\omega_2)^{-t}\right)_{t=0}
\nonumber
\end{eqnarray}
Definition: $\Gamma_b(x) \equiv 
\frac{\Gamma_2(x|b,b^{-1})}{\Gamma_2(Q/2|b,b^{-1})}$.\\
Functional relations:
\begin{eqnarray}
\nonumber \\
&&\Ga(x+b)= \frac{\sqrt{2\pi}b^{bx-\frac{1}{2}}}{\Gamma(bx)}\Ga(x), 
\nonumber \\
&&\Ga(x+1/b)=
\frac{\sqrt{2\pi}b^{-\frac{x}{b}+\frac{1}{2}}}{\Gamma(x/b)}\Ga(x).
\nonumber
\end{eqnarray}
$\Ga(x)$ is a meromorphic function of $x$, whose poles are located
at
$x=-nb-mb^{-1}, n,m \in \mathbb{N}$.\\
Integral representation convergent for $0<\mathrm{Re}x$
\begin{eqnarray}
&&\text{log}\Ga(x)=\int_{0}^{\infty}\frac{dt}{t}
\left\lbrack\frac{e^{-xt}-e^{-Qt/2}}{(1-e^{-bt})(1-e^{-t/b})}
-\frac{(Q/2-x)^{2}}{2}e^{-t}-\frac{Q/2-x}{t}\right\rbrack\nonumber
\end{eqnarray}
\item{$S_b(x)$ function}\\
Definition: $S_b(x)\equiv \frac{\Ga(x)}{\Ga(Q-x)}$ \nonumber\\
Functional relations:
\begin{eqnarray}
&& S_b(x+b) = 2\text{sin}(\pi bx)S_b(x),  \nonumber \\
&& S_b(x+1/b) = 2\text{sin}(\pi x/b)S_b(x). \nonumber
\end{eqnarray}
$S_b(x)$ is a meromorphic function of $x$, whose poles are located
at $x=-nb-mb^{-1}, n,m \in \mathbb{N}$,
and whose zeros are located at $x=Q+nb+mb^{-1}, n,m \in \mathbb{N}$.\\
Integral representation convergent in the strip $0<\mathrm{Re}x<Q$
\begin{eqnarray}
&&\text{log}S_b(x)=\int_{0}^{\infty}\frac{dt}{t}
\left\lbrack\frac{\text{sinh}(\frac{Q}{2}-x)t}
{2\text{sinh}(\frac{bt}{2})\text{sinh}(\frac{t}{2b})}-
\frac{(Q-2x)}{t}\right\rbrack\nonumber
\end{eqnarray}
\end{itemize}

\newsec{Functional relations for the boundary three 
point function}
 The LFT boundary three point function satisfies the
pentagonal equation \cite{PTtwo}:
\begin{equation}
\int
d\beta_{21}C_{\beta_{3},\beta_{21}}^{(\sigma_{4}\sigma_{3}\sigma_{1})\beta_{4}}
C_{\beta_{2}\beta_{1}}^{(\sigma_{3}\sigma_{2}\sigma_{1})\beta_{21}}%
F_{\beta_{21}\beta_{32}}\left[
\begin{array}
[c]{cc}%
\beta_{3} & \beta_{2}\\
\beta_{4} & \beta_{1}%
\end{array}
\right]
 =%
C_{\beta_{32},\beta_{1}}^{(\sigma_{4}\sigma_{2}\sigma_{1})\beta_{4}}
C_{\beta_{3}\beta_{2}}^{(\sigma_{4}\sigma_{3}\sigma_{2})\beta_{32}}
\label{pb}
\end{equation}
where $ F_{\beta_{21}\beta_{32}}\left[
\begin{array}
[c]{cc}%
\beta_{3} & \beta_{2}\\
\beta_{4} & \beta_{1}%
\end{array}
\right] $ is the fusion coefficients (its explicit expression can
be found in \cite{PT1}) that express the invertible fusion
transformation between the s- and t-channel Liouville conformal
blocks.
\begin{itemize}
\item
In the case where the boundary field
$B_{\beta_2}^{\sigma_3,\sigma_2}$ is replaced by the degenerate
boundary field $B_{-\frac{b}{2}}^{\sigma_3,\sigma_3+\frac{b}{2}}$,
then the associativity condition (\ref{pb}) gets replaced by
\begin{align}
\sum_{s=\pm}C_{\beta_{3},\beta_{1}-s\frac{b}{2}}^{(\sigma_{4},
\sigma_{3},\sigma_{1})\beta_{4}}
C_{-\frac{b}{2},\beta_{1}}^{(\sigma_{3},\sigma_{3}+\frac{b}{2},
\sigma_{1})\beta_{1}-s\frac{b}{2}}
\Fus{\beta_{1}}{-\frac{b}{2}}{\beta_3}{\beta_4}{\beta_{1}
-s\frac{b}{2},}{\beta_{32}}
\nonumber \\
= C_{\beta_{32},\beta_{1}}^{(\sigma_{4},\sigma_{3}+\frac{b}{2},
\sigma_{1})\beta_{4}}
C_{\beta_{3},-\frac{b}{2}}^{(\sigma_{4},\sigma_{3},
\sigma_{3}+\frac{b}{2})\beta_{32}}
\end{align}
and $\beta_{32}$ can take the two values $\beta_3\pm \frac{b}{2}$.  We
make the choice $\beta_{32}=\beta_3 -\frac{b}{2}$.  We now use the
particular values for the boundary three point functions \cite{FZZb}:
$$C_{-\frac{b}{2},\beta_{1}}^{(\sigma_{3},
\sigma_{3}+\frac{b}{2},\sigma_{1})\beta_{1}-\frac{b}{2}}=
C_{\beta_{3},-\frac{b}{2}}^{(\sigma_{4},
\sigma_{3},\sigma_{3}+\frac{b}{2})\beta_{3}-\frac{b}{2}} \equiv 1 ,$$
and $$C_{-\frac{b}{2},\beta_{1}}^{(\sigma_{3},
\sigma_{3}+\frac{b}{2},\sigma_{1})\beta_{1}+\frac{b}{2}}
=c_{-}^{-}(\beta_1,\sigma_1,\sigma_3).
$$
This leads to the equation
\begin{align}
C_{\beta_{3},\beta_{1}-\frac{b}{2}}^{(\sigma_{4},\sigma_{3},\sigma_{1})\beta_{4}}
\Fus{\beta_{1}}{-\frac{b}{2}}{\beta_3}{\beta_4}{+}{+} +
C_{\beta_{3},\beta_{1}+\frac{b}{2}}^{(\sigma_{4},\sigma_{3},\sigma_{1})\beta_{4}}
c_{-}^{-}(\beta_1,\sigma_1,\sigma_3)
\Fus{\beta_{1}}{-\frac{b}{2}}{\beta_3}{\beta_4}{-}{+} \nonumber \\
 =
C_{\beta_{3}-\frac{b}{2},\beta_{1}}^{(\sigma_{4},\sigma_{3}+\frac{b}{2},\sigma_{1})\beta_{4}}
\end{align}
A similar equation is obtained by considering the degenerate
boundary field $B_{-\frac{b}{2}}^{\sigma_3,\sigma_3-\frac{b}{2}}$
in (\ref{pb}):
\begin{align}
C_{\beta_{3},\beta_{1}-\frac{b}{2}}^{(\sigma_{4},\sigma_{3},\sigma_{1})\beta_{4}}
\Fus{\beta_{1}}{-\frac{b}{2}}{\beta_3}{\beta_4}{+}{+} +
C_{\beta_{3},\beta_{1}+\frac{b}{2}}^{(\sigma_{4},\sigma_{3},\sigma_{1})\beta_{4}}
c_{-}^{+}(\beta_1,\sigma_1,\sigma_3)
\Fus{\beta_{1}}{-\frac{b}{2}}{\beta_3}{\beta_4}{-}{+}
\nonumber \\
 =
C_{\beta_{3}-\frac{b}{2},\beta_{1}}^{(\sigma_{4},\sigma_{3}-\frac{b}{2},\sigma_{1})\beta_{4}}
\end{align}
where we introduced
$C_{-\frac{b}{2},\beta_{1}}^{(\sigma_{3},\sigma_{3}-\frac{b}{2},\sigma_{1})\beta_{1}+\frac{b}{2}}=
c_{-}^{+}(\beta_1,\sigma_1,\sigma_3)$.\\
 Subtracting these two
equations gives
\begin{eqnarray}
\lefteqn{C_{\beta_{3}-\frac{b}{2},\beta_{1}}^{(\sigma_{4},\sigma_{3}-\frac{b}{2},\sigma_{1})\beta_{4}}-
C_{\beta_{3}-\frac{b}{2},\beta_{1}}^{(\sigma_{4},\sigma_{3}+\frac{b}{2},\sigma_{1})\beta_{4}}
=}\nonumber \\
&&
(c_{-}^{+}(\beta_1,\sigma_1,\sigma_3)-c_{-}^{-}(\beta_1,\sigma_1,\sigma_3))
\Fus{\beta_1}{-\frac{b}{2}}{\beta_3}{\beta_{4}}{-}{+}
C_{\beta_{3},\beta_{1}+\frac{b}{2}}^{(\sigma_{4},\sigma_{3},\sigma_{1})\beta_{4}}.
\end{eqnarray}
From the results of \cite{FZZb} we derive
 $$ c_{-}^+(\beta_1,\sigma_1,\sigma_3)-
c_{-}^-(\beta_1,\sigma_1,\sigma_3) =
\left(-\frac{\mu}{\pi\gamma(-b^2)}\right)^{1/2}
\frac{-2\pi\Gamma(1-2b\beta_1)}{\Gamma(2+b^2-2b\beta_1)} \sin
 2\pi b(\sigma_3-Q/2),
$$
and it is well known \cite{BPZ} that the fusion coefficient
$F_{-+}$ is equal to
$$
\Fus{\beta_1}{-\frac{b}{2}}{\beta_3}{\beta_{4}}{-}{+}
=\frac{\Gamma(2-b(2\beta_1-b))\Gamma(b(b-2\beta_3)+1)}
{\Gamma(2-b(\beta_1+\beta_3+\beta_4-3\frac{b}{2}))
\Gamma(1-b(\beta_1+\beta_3-\beta_4-\frac{b}{2}))}\quad
.
$$
We finally get the following functional relation for the boundary
three point function :
\begin{eqnarray}
\lefteqn{C_{\beta_{3}-\frac{b}{2},\beta_{1}}^{(\sigma_{4},\sigma_{3}-\frac{b}{2},\sigma_{1})\beta_{4}}-
C_{\beta_{3}-\frac{b}{2},\beta_{1}}^{(\sigma_{4},\sigma_{3}+\frac{b}{2},\sigma_{1})\beta_{4}}
= -2\pi \left(-\frac{\mu}{\pi\gamma(-b^2)}\right)^{1/2} \sin\pi
b(2\sigma_3-Q)\times}
\nonumber \\
&& \times \frac{\Gamma(1-2b\beta_1) \Gamma(b(b-2\beta_3)+1)}
{\Gamma(2-b(\beta_1+\beta_3+\beta_4-\frac{3b}{2}))\Gamma(1-b(\beta_1+\beta_3-\beta_4-\frac{b}{2}))}
C_{\beta_{3},\beta_{1}+\frac{b}{2}}^{(\sigma_{4},\sigma_{3},\sigma_{1})\beta_{4}}.\nonumber
\\
\end{eqnarray}

\item
Another insertion of the spin $-b/2$ gives:
\begin{eqnarray}
\lefteqn{C_{\beta_1,\sigma_1}^{(\sigma_2,\sigma_1+b/2,-b/2)\sigma_2-b/2}C_{\beta_2,\sigma_2-b/2}^
{(\sigma_3,\sigma_2,-b/2)\sigma_3-b/2}
\Fus{\sigma_1}{\beta_1}{\beta_2}{\sigma_3-b/2}{\sigma_2-b/2,}{\beta_{3}}}
\nonumber \\
&&+
C_{\beta_1\sigma_1}^{(\sigma_2,\sigma_1+b/2,-b/2)\sigma_2+b/2}C_{\beta_2,\sigma_2+b/2}^
{(\sigma_3,\sigma_2,-b/2)\sigma_3-b/2}
\Fus{\sigma_1}{\beta_1}{\beta_2}{\sigma_3-b/2}{\sigma_2+b/2,}{\beta_{3}}
\nonumber \\
&&=
C_{\beta_3,\sigma_1+b/2}^{(\sigma_3,\sigma_1+b/2,-b/2)\sigma_3-b/2}
\fus{\sigma_1}{\beta_1}{\beta_2}{\sigma_3}{\sigma_{2},}{\beta_{3}}\quad
.
\end{eqnarray}

Now we use the relation between the boundary three point function
and the fusion matrix \cite{PTtwo}
\begin{equation}
C_{\beta_{2}\beta_{1}}^{(\sigma_{3}\sigma_{2}\sigma_{1})\beta_{3}}=%
\frac{g(\beta_{3},\sigma_{3},\sigma_{1})}{g(\beta_{2},\sigma_{3},\sigma_{2})g(\beta_{1},\sigma_{2},\sigma_{1})}
F_{\sigma_{2}\beta_{3}}\left[
\begin{array}
[c]{cc}%
\beta_{2}     & \beta_{1}\\
\sigma_{3}    &   \sigma_{1}%
\end{array}
\right],
\end{equation}
where $g(\beta_{1},\sigma_{2},\sigma_{1})$ is the normalization of
the boundary operators computed in \cite{PTtwo}; this leads to the
following functional relation for the boundary three point
function, that produces shifts on the boundary conditions {\it
only}:
\begin{eqnarray}
\lefteqn{-\sin\pi b(\sigma_1+\sigma_2-\beta_1-b/2)\sin\pi
b(\sigma_3+\sigma_2-\beta_2-b)
C_{\beta_2,\beta_1}^{(\sigma_3-b/2,\sigma_2-b/2,\sigma_1)\beta_3}}\nonumber
\\
&& + \sin\pi b(\sigma_1-\sigma_2-\beta_1+b/2)\sin\pi
b(\sigma_3-\sigma_2-\beta_2)
C_{\beta_2,\beta_1}^{(\sigma_3-b/2,\sigma_2+b/2,\sigma_1)\beta_3}\nonumber
\\
&& =\sin\pi b(\sigma_1+\sigma_3-\beta_3-b/2)\sin\pi b(2\sigma_2-Q)
C_{\beta_2,\beta_1}^{(\sigma_3,\sigma_2,\sigma_1+b/2)\beta_3}
\quad.
\end{eqnarray}
\end{itemize}

\end{document}